\documentclass[longauth]{aa}

\usepackage{csquotes}

\usepackage{graphicx}
\usepackage{natbib}
\usepackage{scalerel}
\usepackage{lastpage}
\usepackage{dcolumn} 
\usepackage{tablefootnote} 
\usepackage{multirow}
\usepackage{threeparttable}
\usepackage[table]{xcolor}
\usepackage{tikz}
\usepackage{natbib}

\usepackage{tikz}
\usetikzlibrary{shapes.geometric, arrows, positioning}
\tikzstyle{startstop} = [rectangle, rounded corners, text centered, draw=black, align=center]
\tikzstyle{process} = [rectangle, text width=4cm, minimum height=1cm, text centered, draw=black, align=center]
\tikzstyle{arrow} = [thick,->,>=stealth]
\usepackage{xcolor}

\usepackage{txfonts}
\usepackage[pdfencoding=auto,psdextra]{hyperref}
\hypersetup{
    colorlinks=true,
    linkcolor=blue,
    filecolor=magenta,      
    urlcolor=blue,
    citecolor=blue
}
\makeatletter
\renewcommand*\aa@pageof{, page \thepage{} of \pageref*{LastPage}}
\makeatother

\usepackage[utf8]{inputenc}

\usepackage[switch, modulo]{lineno}

\usepackage[nameinlink,capitalise]{cleveref}
\crefname{section}{Sect.}{Sects.}
\Crefname{section}{Section}{Sections}
\crefname{figure}{Fig.}{Figs.}
\Crefname{figure}{Figure}{Figures}
\crefname{equation}{Eq.}{Eqs.}
\Crefname{equation}{Equation}{Equations}
\crefname{table}{Table}{Tables}
\crefname{appendix}{Appendix}{Appendices}

\usepackage{orcidlink} 
\newcommand{\orcid}[1]{\orcidlink{#1}}

\usepackage{euclid}
\newcommand*{\gaia}{\textit{Gaia}\xspace}
\newcommand\gdr[1]{\gaia DR#1}

\newcommand\gbp{\ensuremath{G_\mathrm{BP}}}
\newcommand\grp{\ensuremath{G_\mathrm{RP}}}

\begin{document}

\defcitealias{EP-Bisigello}{BL24}
\defcitealias{Q1-SP003}{RW25}
\defcitealias{Q1-SP009}{SG25}
\defcitealias{Q1-SP015}{MB25}
\defcitealias{Q1-SP013}{LM25}
\defcitealias{Q1-TP005}{TM25}
\defcitealias{Q1-TP004}{RE25}
\defcitealias{Assef_2018_2018ApJS..234...23A}{A18}
\defcitealias{Q1-SP011}{BL25}
\defcitealias{EP-Selwood}{SM25}

%
%
\title{Euclid Quick Data Release (Q1)} \subtitle{The active galaxies of  \Euclid}

\author{Euclid Collaboration: T.~Matamoro~Zatarain\orcid{0009-0007-2976-293X}\thanks{\email{ky23883@bristol.ac.uk}}\inst{\ref{aff1}}
\and S.~Fotopoulou\orcid{0000-0002-9686-254X}\inst{\ref{aff1}}
\and F.~Ricci\orcid{0000-0001-5742-5980}\inst{\ref{aff2},\ref{aff3}}
\and M.~Bolzonella\orcid{0000-0003-3278-4607}\inst{\ref{aff4}}
\and F.~La~Franca\orcid{0000-0002-1239-2721}\inst{\ref{aff2}}
\and A.~Viitanen\orcid{0000-0001-9383-786X}\inst{\ref{aff5},\ref{aff3}}
\and G.~Zamorani\orcid{0000-0002-2318-301X}\inst{\ref{aff4}}
\and M.~B.~Taylor\orcid{0000-0002-4209-1479}\inst{\ref{aff1}}
\and M.~Mezcua\orcid{0000-0003-4440-259X}\inst{\ref{aff6},\ref{aff7}}
\and B.~Laloux\orcid{0000-0001-9996-9732}\inst{\ref{aff8},\ref{aff9}}
\and A.~Bongiorno\orcid{0000-0002-0101-6624}\inst{\ref{aff3}}
\and K.~Jahnke\orcid{0000-0003-3804-2137}\inst{\ref{aff10}}
\and G.~Stevens\orcid{0000-0002-8885-4443}\inst{\ref{aff1}}
\and R.~A.~Shaw\orcid{0000-0003-4318-0737}\inst{\ref{aff1}}
\and L.~Bisigello\orcid{0000-0003-0492-4924}\inst{\ref{aff11}}
\and W.~Roster\orcid{0000-0002-9149-6528}\inst{\ref{aff9}}
\and Y.~Fu\orcid{0000-0002-0759-0504}\inst{\ref{aff12},\ref{aff13}}
\and B.~Margalef-Bentabol\orcid{0000-0001-8702-7019}\inst{\ref{aff14}}
\and A.~La~Marca\orcid{0000-0002-7217-5120}\inst{\ref{aff14},\ref{aff13}}
\and F.~Tarsitano\orcid{0000-0002-5919-0238}\inst{\ref{aff15}}
\and A.~Feltre\orcid{0000-0001-6865-2871}\inst{\ref{aff16}}
\and J.~Calhau\orcid{0000-0003-1803-6899}\inst{\ref{aff8}}
\and X.~López~López\orcid{0009-0008-5194-5908}\inst{\ref{aff17},\ref{aff4}}
\and M.~Scialpi\orcid{0009-0006-5100-4986}\inst{\ref{aff18},\ref{aff19},\ref{aff16}}
\and M.~Salvato\orcid{0000-0001-7116-9303}\inst{\ref{aff9}}
\and V.~Allevato\orcid{0000-0001-7232-5152}\inst{\ref{aff8}}
\and M.~Siudek\orcid{0000-0002-2949-2155}\inst{\ref{aff20},\ref{aff6}}
\and C.~Saulder\orcid{0000-0002-0408-5633}\inst{\ref{aff9},\ref{aff21}}
\and D.~Vergani\orcid{0000-0003-0898-2216}\inst{\ref{aff4}}
\and M.~N.~Bremer\inst{\ref{aff1}}
\and L.~Wang\orcid{0000-0002-6736-9158}\inst{\ref{aff14},\ref{aff13}}
\and M.~Giulietti\orcid{0000-0002-1847-4496}\inst{\ref{aff22}}
\and D.~M.~Alexander\orcid{0000-0002-5896-6313}\inst{\ref{aff23}}
\and D.~Sluse\orcid{0000-0001-6116-2095}\inst{\ref{aff24}}
\and F.~Shankar\orcid{0000-0001-8973-5051}\inst{\ref{aff25}}
\and L.~Spinoglio\orcid{0000-0001-8840-1551}\inst{\ref{aff26}}
\and D.~Scott\orcid{0000-0002-6878-9840}\inst{\ref{aff27}}
\and R.~Shirley\orcid{0000-0002-1114-0135}\inst{\ref{aff9}}
\and H.~Landt\inst{\ref{aff23}}
\and M.~Selwood\inst{\ref{aff1}}
\and Y.~Toba\orcid{0000-0002-3531-7863}\inst{\ref{aff28},\ref{aff29},\ref{aff30}}
\and P.~Dayal\orcid{0000-0001-8460-1564}\inst{\ref{aff13}}
\and N.~Aghanim\orcid{0000-0002-6688-8992}\inst{\ref{aff31}}
\and B.~Altieri\orcid{0000-0003-3936-0284}\inst{\ref{aff32}}
\and A.~Amara\inst{\ref{aff33}}
\and S.~Andreon\orcid{0000-0002-2041-8784}\inst{\ref{aff34}}
\and N.~Auricchio\orcid{0000-0003-4444-8651}\inst{\ref{aff4}}
\and H.~Aussel\orcid{0000-0002-1371-5705}\inst{\ref{aff35}}
\and C.~Baccigalupi\orcid{0000-0002-8211-1630}\inst{\ref{aff36},\ref{aff37},\ref{aff38},\ref{aff39}}
\and M.~Baldi\orcid{0000-0003-4145-1943}\inst{\ref{aff40},\ref{aff4},\ref{aff41}}
\and S.~Bardelli\orcid{0000-0002-8900-0298}\inst{\ref{aff4}}
\and A.~Basset\inst{\ref{aff42}}
\and P.~Battaglia\orcid{0000-0002-7337-5909}\inst{\ref{aff4}}
\and A.~Biviano\orcid{0000-0002-0857-0732}\inst{\ref{aff37},\ref{aff36}}
\and A.~Bonchi\orcid{0000-0002-2667-5482}\inst{\ref{aff43}}
\and E.~Branchini\orcid{0000-0002-0808-6908}\inst{\ref{aff44},\ref{aff45},\ref{aff34}}
\and M.~Brescia\orcid{0000-0001-9506-5680}\inst{\ref{aff46},\ref{aff8}}
\and J.~Brinchmann\orcid{0000-0003-4359-8797}\inst{\ref{aff47},\ref{aff48}}
\and S.~Camera\orcid{0000-0003-3399-3574}\inst{\ref{aff49},\ref{aff50},\ref{aff51}}
\and G.~Ca\~nas-Herrera\orcid{0000-0003-2796-2149}\inst{\ref{aff52},\ref{aff53},\ref{aff12}}
\and V.~Capobianco\orcid{0000-0002-3309-7692}\inst{\ref{aff51}}
\and C.~Carbone\orcid{0000-0003-0125-3563}\inst{\ref{aff54}}
\and J.~Carretero\orcid{0000-0002-3130-0204}\inst{\ref{aff55},\ref{aff56}}
\and S.~Casas\orcid{0000-0002-4751-5138}\inst{\ref{aff57}}
\and F.~J.~Castander\orcid{0000-0001-7316-4573}\inst{\ref{aff6},\ref{aff7}}
\and M.~Castellano\orcid{0000-0001-9875-8263}\inst{\ref{aff3}}
\and G.~Castignani\orcid{0000-0001-6831-0687}\inst{\ref{aff4}}
\and S.~Cavuoti\orcid{0000-0002-3787-4196}\inst{\ref{aff8},\ref{aff58}}
\and K.~C.~Chambers\orcid{0000-0001-6965-7789}\inst{\ref{aff59}}
\and A.~Cimatti\inst{\ref{aff60}}
\and C.~Colodro-Conde\inst{\ref{aff61}}
\and G.~Congedo\orcid{0000-0003-2508-0046}\inst{\ref{aff62}}
\and C.~J.~Conselice\orcid{0000-0003-1949-7638}\inst{\ref{aff63}}
\and L.~Conversi\orcid{0000-0002-6710-8476}\inst{\ref{aff64},\ref{aff32}}
\and Y.~Copin\orcid{0000-0002-5317-7518}\inst{\ref{aff65}}
\and F.~Courbin\orcid{0000-0003-0758-6510}\inst{\ref{aff66},\ref{aff67}}
\and H.~M.~Courtois\orcid{0000-0003-0509-1776}\inst{\ref{aff68}}
\and M.~Cropper\orcid{0000-0003-4571-9468}\inst{\ref{aff69}}
\and A.~Da~Silva\orcid{0000-0002-6385-1609}\inst{\ref{aff70},\ref{aff71}}
\and H.~Degaudenzi\orcid{0000-0002-5887-6799}\inst{\ref{aff15}}
\and G.~De~Lucia\orcid{0000-0002-6220-9104}\inst{\ref{aff37}}
\and A.~M.~Di~Giorgio\orcid{0000-0002-4767-2360}\inst{\ref{aff26}}
\and C.~Dolding\orcid{0009-0003-7199-6108}\inst{\ref{aff69}}
\and H.~Dole\orcid{0000-0002-9767-3839}\inst{\ref{aff31}}
\and F.~Dubath\orcid{0000-0002-6533-2810}\inst{\ref{aff15}}
\and C.~A.~J.~Duncan\orcid{0009-0003-3573-0791}\inst{\ref{aff63}}
\and X.~Dupac\inst{\ref{aff32}}
\and S.~Dusini\orcid{0000-0002-1128-0664}\inst{\ref{aff72}}
\and A.~Ealet\orcid{0000-0003-3070-014X}\inst{\ref{aff65}}
\and S.~Escoffier\orcid{0000-0002-2847-7498}\inst{\ref{aff73}}
\and M.~Fabricius\orcid{0000-0002-7025-6058}\inst{\ref{aff9},\ref{aff21}}
\and M.~Farina\orcid{0000-0002-3089-7846}\inst{\ref{aff26}}
\and R.~Farinelli\inst{\ref{aff4}}
\and F.~Faustini\orcid{0000-0001-6274-5145}\inst{\ref{aff43},\ref{aff3}}
\and S.~Ferriol\inst{\ref{aff65}}
\and F.~Finelli\orcid{0000-0002-6694-3269}\inst{\ref{aff4},\ref{aff74}}
\and N.~Fourmanoit\orcid{0009-0005-6816-6925}\inst{\ref{aff73}}
\and M.~Frailis\orcid{0000-0002-7400-2135}\inst{\ref{aff37}}
\and E.~Franceschi\orcid{0000-0002-0585-6591}\inst{\ref{aff4}}
\and S.~Galeotta\orcid{0000-0002-3748-5115}\inst{\ref{aff37}}
\and K.~George\orcid{0000-0002-1734-8455}\inst{\ref{aff21}}
\and B.~Gillis\orcid{0000-0002-4478-1270}\inst{\ref{aff62}}
\and C.~Giocoli\orcid{0000-0002-9590-7961}\inst{\ref{aff4},\ref{aff41}}
\and P.~G\'omez-Alvarez\orcid{0000-0002-8594-5358}\inst{\ref{aff75},\ref{aff32}}
\and J.~Gracia-Carpio\inst{\ref{aff9}}
\and B.~R.~Granett\orcid{0000-0003-2694-9284}\inst{\ref{aff34}}
\and A.~Grazian\orcid{0000-0002-5688-0663}\inst{\ref{aff11}}
\and F.~Grupp\inst{\ref{aff9},\ref{aff21}}
\and S.~Gwyn\orcid{0000-0001-8221-8406}\inst{\ref{aff76}}
\and S.~V.~H.~Haugan\orcid{0000-0001-9648-7260}\inst{\ref{aff77}}
\and H.~Hoekstra\orcid{0000-0002-0641-3231}\inst{\ref{aff12}}
\and W.~Holmes\inst{\ref{aff78}}
\and I.~M.~Hook\orcid{0000-0002-2960-978X}\inst{\ref{aff79}}
\and F.~Hormuth\inst{\ref{aff80}}
\and A.~Hornstrup\orcid{0000-0002-3363-0936}\inst{\ref{aff81},\ref{aff82}}
\and P.~Hudelot\inst{\ref{aff83}}
\and M.~Jhabvala\inst{\ref{aff84}}
\and E.~Keih\"anen\orcid{0000-0003-1804-7715}\inst{\ref{aff5}}
\and S.~Kermiche\orcid{0000-0002-0302-5735}\inst{\ref{aff73}}
\and A.~Kiessling\orcid{0000-0002-2590-1273}\inst{\ref{aff78}}
\and B.~Kubik\orcid{0009-0006-5823-4880}\inst{\ref{aff65}}
\and M.~K\"ummel\orcid{0000-0003-2791-2117}\inst{\ref{aff21}}
\and M.~Kunz\orcid{0000-0002-3052-7394}\inst{\ref{aff85}}
\and H.~Kurki-Suonio\orcid{0000-0002-4618-3063}\inst{\ref{aff86},\ref{aff87}}
\and Q.~Le~Boulc'h\inst{\ref{aff88}}
\and A.~M.~C.~Le~Brun\orcid{0000-0002-0936-4594}\inst{\ref{aff89}}
\and D.~Le~Mignant\orcid{0000-0002-5339-5515}\inst{\ref{aff90}}
\and P.~Liebing\inst{\ref{aff69}}
\and S.~Ligori\orcid{0000-0003-4172-4606}\inst{\ref{aff51}}
\and P.~B.~Lilje\orcid{0000-0003-4324-7794}\inst{\ref{aff77}}
\and V.~Lindholm\orcid{0000-0003-2317-5471}\inst{\ref{aff86},\ref{aff87}}
\and I.~Lloro\orcid{0000-0001-5966-1434}\inst{\ref{aff91}}
\and G.~Mainetti\orcid{0000-0003-2384-2377}\inst{\ref{aff88}}
\and D.~Maino\inst{\ref{aff92},\ref{aff54},\ref{aff93}}
\and E.~Maiorano\orcid{0000-0003-2593-4355}\inst{\ref{aff4}}
\and O.~Mansutti\orcid{0000-0001-5758-4658}\inst{\ref{aff37}}
\and S.~Marcin\inst{\ref{aff94}}
\and O.~Marggraf\orcid{0000-0001-7242-3852}\inst{\ref{aff95}}
\and M.~Martinelli\orcid{0000-0002-6943-7732}\inst{\ref{aff3},\ref{aff96}}
\and N.~Martinet\orcid{0000-0003-2786-7790}\inst{\ref{aff90}}
\and F.~Marulli\orcid{0000-0002-8850-0303}\inst{\ref{aff17},\ref{aff4},\ref{aff41}}
\and R.~Massey\orcid{0000-0002-6085-3780}\inst{\ref{aff97}}
\and D.~C.~Masters\orcid{0000-0001-5382-6138}\inst{\ref{aff98}}
\and S.~Maurogordato\inst{\ref{aff99}}
\and E.~Medinaceli\orcid{0000-0002-4040-7783}\inst{\ref{aff4}}
\and S.~Mei\orcid{0000-0002-2849-559X}\inst{\ref{aff100},\ref{aff101}}
\and M.~Melchior\inst{\ref{aff102}}
\and Y.~Mellier\inst{\ref{aff103},\ref{aff83}}
\and M.~Meneghetti\orcid{0000-0003-1225-7084}\inst{\ref{aff4},\ref{aff41}}
\and E.~Merlin\orcid{0000-0001-6870-8900}\inst{\ref{aff3}}
\and G.~Meylan\inst{\ref{aff104}}
\and A.~Mora\orcid{0000-0002-1922-8529}\inst{\ref{aff105}}
\and M.~Moresco\orcid{0000-0002-7616-7136}\inst{\ref{aff17},\ref{aff4}}
\and L.~Moscardini\orcid{0000-0002-3473-6716}\inst{\ref{aff17},\ref{aff4},\ref{aff41}}
\and R.~Nakajima\orcid{0009-0009-1213-7040}\inst{\ref{aff95}}
\and C.~Neissner\orcid{0000-0001-8524-4968}\inst{\ref{aff106},\ref{aff56}}
\and S.-M.~Niemi\inst{\ref{aff52}}
\and J.~W.~Nightingale\orcid{0000-0002-8987-7401}\inst{\ref{aff107}}
\and C.~Padilla\orcid{0000-0001-7951-0166}\inst{\ref{aff106}}
\and S.~Paltani\orcid{0000-0002-8108-9179}\inst{\ref{aff15}}
\and F.~Pasian\orcid{0000-0002-4869-3227}\inst{\ref{aff37}}
\and K.~Pedersen\inst{\ref{aff108}}
\and W.~J.~Percival\orcid{0000-0002-0644-5727}\inst{\ref{aff109},\ref{aff110},\ref{aff111}}
\and V.~Pettorino\inst{\ref{aff52}}
\and S.~Pires\orcid{0000-0002-0249-2104}\inst{\ref{aff35}}
\and G.~Polenta\orcid{0000-0003-4067-9196}\inst{\ref{aff43}}
\and M.~Poncet\inst{\ref{aff42}}
\and L.~A.~Popa\inst{\ref{aff112}}
\and L.~Pozzetti\orcid{0000-0001-7085-0412}\inst{\ref{aff4}}
\and F.~Raison\orcid{0000-0002-7819-6918}\inst{\ref{aff9}}
\and R.~Rebolo\orcid{0000-0003-3767-7085}\inst{\ref{aff61},\ref{aff113},\ref{aff114}}
\and A.~Renzi\orcid{0000-0001-9856-1970}\inst{\ref{aff115},\ref{aff72}}
\and J.~Rhodes\orcid{0000-0002-4485-8549}\inst{\ref{aff78}}
\and G.~Riccio\inst{\ref{aff8}}
\and E.~Romelli\orcid{0000-0003-3069-9222}\inst{\ref{aff37}}
\and M.~Roncarelli\orcid{0000-0001-9587-7822}\inst{\ref{aff4}}
\and E.~Rossetti\orcid{0000-0003-0238-4047}\inst{\ref{aff40}}
\and H.~J.~A.~Rottgering\orcid{0000-0001-8887-2257}\inst{\ref{aff12}}
\and B.~Rusholme\orcid{0000-0001-7648-4142}\inst{\ref{aff116}}
\and R.~Saglia\orcid{0000-0003-0378-7032}\inst{\ref{aff21},\ref{aff9}}
\and Z.~Sakr\orcid{0000-0002-4823-3757}\inst{\ref{aff117},\ref{aff118},\ref{aff119}}
\and A.~G.~S\'anchez\orcid{0000-0003-1198-831X}\inst{\ref{aff9}}
\and D.~Sapone\orcid{0000-0001-7089-4503}\inst{\ref{aff120}}
\and B.~Sartoris\orcid{0000-0003-1337-5269}\inst{\ref{aff21},\ref{aff37}}
\and J.~A.~Schewtschenko\orcid{0000-0002-4913-6393}\inst{\ref{aff62}}
\and P.~Schneider\orcid{0000-0001-8561-2679}\inst{\ref{aff95}}
\and T.~Schrabback\orcid{0000-0002-6987-7834}\inst{\ref{aff121}}
\and M.~Scodeggio\inst{\ref{aff54}}
\and A.~Secroun\orcid{0000-0003-0505-3710}\inst{\ref{aff73}}
\and G.~Seidel\orcid{0000-0003-2907-353X}\inst{\ref{aff10}}
\and M.~Seiffert\orcid{0000-0002-7536-9393}\inst{\ref{aff78}}
\and S.~Serrano\orcid{0000-0002-0211-2861}\inst{\ref{aff7},\ref{aff122},\ref{aff6}}
\and P.~Simon\inst{\ref{aff95}}
\and C.~Sirignano\orcid{0000-0002-0995-7146}\inst{\ref{aff115},\ref{aff72}}
\and G.~Sirri\orcid{0000-0003-2626-2853}\inst{\ref{aff41}}
\and J.~Skottfelt\orcid{0000-0003-1310-8283}\inst{\ref{aff123}}
\and L.~Stanco\orcid{0000-0002-9706-5104}\inst{\ref{aff72}}
\and J.~Steinwagner\orcid{0000-0001-7443-1047}\inst{\ref{aff9}}
\and P.~Tallada-Cresp\'{i}\orcid{0000-0002-1336-8328}\inst{\ref{aff55},\ref{aff56}}
\and A.~N.~Taylor\inst{\ref{aff62}}
\and I.~Tereno\inst{\ref{aff70},\ref{aff124}}
\and S.~Toft\orcid{0000-0003-3631-7176}\inst{\ref{aff125},\ref{aff126}}
\and R.~Toledo-Moreo\orcid{0000-0002-2997-4859}\inst{\ref{aff127}}
\and F.~Torradeflot\orcid{0000-0003-1160-1517}\inst{\ref{aff56},\ref{aff55}}
\and I.~Tutusaus\orcid{0000-0002-3199-0399}\inst{\ref{aff118}}
\and L.~Valenziano\orcid{0000-0002-1170-0104}\inst{\ref{aff4},\ref{aff74}}
\and J.~Valiviita\orcid{0000-0001-6225-3693}\inst{\ref{aff86},\ref{aff87}}
\and T.~Vassallo\orcid{0000-0001-6512-6358}\inst{\ref{aff21},\ref{aff37}}
\and G.~Verdoes~Kleijn\orcid{0000-0001-5803-2580}\inst{\ref{aff13}}
\and A.~Veropalumbo\orcid{0000-0003-2387-1194}\inst{\ref{aff34},\ref{aff45},\ref{aff44}}
\and Y.~Wang\orcid{0000-0002-4749-2984}\inst{\ref{aff98}}
\and J.~Weller\orcid{0000-0002-8282-2010}\inst{\ref{aff21},\ref{aff9}}
\and A.~Zacchei\orcid{0000-0003-0396-1192}\inst{\ref{aff37},\ref{aff36}}
\and F.~M.~Zerbi\inst{\ref{aff34}}
\and I.~A.~Zinchenko\orcid{0000-0002-2944-2449}\inst{\ref{aff21}}
\and E.~Zucca\orcid{0000-0002-5845-8132}\inst{\ref{aff4}}
\and M.~Ballardini\orcid{0000-0003-4481-3559}\inst{\ref{aff128},\ref{aff129},\ref{aff4}}
\and E.~Bozzo\orcid{0000-0002-8201-1525}\inst{\ref{aff15}}
\and C.~Burigana\orcid{0000-0002-3005-5796}\inst{\ref{aff22},\ref{aff74}}
\and R.~Cabanac\orcid{0000-0001-6679-2600}\inst{\ref{aff118}}
\and A.~Cappi\inst{\ref{aff4},\ref{aff99}}
\and D.~Di~Ferdinando\inst{\ref{aff41}}
\and J.~A.~Escartin~Vigo\inst{\ref{aff9}}
\and G.~Fabbian\orcid{0000-0002-3255-4695}\inst{\ref{aff130}}
\and L.~Gabarra\orcid{0000-0002-8486-8856}\inst{\ref{aff131}}
\and J.~Mart\'{i}n-Fleitas\orcid{0000-0002-8594-569X}\inst{\ref{aff105}}
\and S.~Matthew\orcid{0000-0001-8448-1697}\inst{\ref{aff62}}
\and N.~Mauri\orcid{0000-0001-8196-1548}\inst{\ref{aff60},\ref{aff41}}
\and R.~B.~Metcalf\orcid{0000-0003-3167-2574}\inst{\ref{aff17},\ref{aff4}}
\and A.~Pezzotta\orcid{0000-0003-0726-2268}\inst{\ref{aff132},\ref{aff9}}
\and M.~P\"ontinen\orcid{0000-0001-5442-2530}\inst{\ref{aff86}}
\and C.~Porciani\orcid{0000-0002-7797-2508}\inst{\ref{aff95}}
\and I.~Risso\orcid{0000-0003-2525-7761}\inst{\ref{aff133}}
\and V.~Scottez\inst{\ref{aff103},\ref{aff134}}
\and M.~Sereno\orcid{0000-0003-0302-0325}\inst{\ref{aff4},\ref{aff41}}
\and M.~Tenti\orcid{0000-0002-4254-5901}\inst{\ref{aff41}}
\and M.~Viel\orcid{0000-0002-2642-5707}\inst{\ref{aff36},\ref{aff37},\ref{aff39},\ref{aff38},\ref{aff135}}
\and M.~Wiesmann\orcid{0009-0000-8199-5860}\inst{\ref{aff77}}
\and Y.~Akrami\orcid{0000-0002-2407-7956}\inst{\ref{aff136},\ref{aff137}}
\and I.~T.~Andika\orcid{0000-0001-6102-9526}\inst{\ref{aff138},\ref{aff139}}
\and S.~Anselmi\orcid{0000-0002-3579-9583}\inst{\ref{aff72},\ref{aff115},\ref{aff140}}
\and M.~Archidiacono\orcid{0000-0003-4952-9012}\inst{\ref{aff92},\ref{aff93}}
\and F.~Atrio-Barandela\orcid{0000-0002-2130-2513}\inst{\ref{aff141}}
\and C.~Benoist\inst{\ref{aff99}}
\and K.~Benson\inst{\ref{aff69}}
\and D.~Bertacca\orcid{0000-0002-2490-7139}\inst{\ref{aff115},\ref{aff11},\ref{aff72}}
\and M.~Bethermin\orcid{0000-0002-3915-2015}\inst{\ref{aff142}}
\and A.~Blanchard\orcid{0000-0001-8555-9003}\inst{\ref{aff118}}
\and L.~Blot\orcid{0000-0002-9622-7167}\inst{\ref{aff143},\ref{aff140}}
\and H.~B\"ohringer\orcid{0000-0001-8241-4204}\inst{\ref{aff9},\ref{aff144},\ref{aff145}}
\and M.~L.~Brown\orcid{0000-0002-0370-8077}\inst{\ref{aff63}}
\and S.~Bruton\orcid{0000-0002-6503-5218}\inst{\ref{aff146}}
\and A.~Calabro\orcid{0000-0003-2536-1614}\inst{\ref{aff3}}
\and B.~Camacho~Quevedo\orcid{0000-0002-8789-4232}\inst{\ref{aff7},\ref{aff6}}
\and F.~Caro\inst{\ref{aff3}}
\and C.~S.~Carvalho\inst{\ref{aff124}}
\and T.~Castro\orcid{0000-0002-6292-3228}\inst{\ref{aff37},\ref{aff38},\ref{aff36},\ref{aff135}}
\and F.~Cogato\orcid{0000-0003-4632-6113}\inst{\ref{aff17},\ref{aff4}}
\and T.~Contini\orcid{0000-0003-0275-938X}\inst{\ref{aff118}}
\and A.~R.~Cooray\orcid{0000-0002-3892-0190}\inst{\ref{aff147}}
\and O.~Cucciati\orcid{0000-0002-9336-7551}\inst{\ref{aff4}}
\and S.~Davini\orcid{0000-0003-3269-1718}\inst{\ref{aff45}}
\and F.~De~Paolis\orcid{0000-0001-6460-7563}\inst{\ref{aff148},\ref{aff149},\ref{aff150}}
\and G.~Desprez\orcid{0000-0001-8325-1742}\inst{\ref{aff13}}
\and A.~D\'iaz-S\'anchez\orcid{0000-0003-0748-4768}\inst{\ref{aff151}}
\and J.~J.~Diaz\inst{\ref{aff61}}
\and S.~Di~Domizio\orcid{0000-0003-2863-5895}\inst{\ref{aff44},\ref{aff45}}
\and J.~M.~Diego\orcid{0000-0001-9065-3926}\inst{\ref{aff152}}
\and P.-A.~Duc\orcid{0000-0003-3343-6284}\inst{\ref{aff142}}
\and A.~Enia\orcid{0000-0002-0200-2857}\inst{\ref{aff40},\ref{aff4}}
\and Y.~Fang\inst{\ref{aff21}}
\and A.~G.~Ferrari\orcid{0009-0005-5266-4110}\inst{\ref{aff41}}
\and A.~Finoguenov\orcid{0000-0002-4606-5403}\inst{\ref{aff86}}
\and A.~Fontana\orcid{0000-0003-3820-2823}\inst{\ref{aff3}}
\and F.~Fontanot\orcid{0000-0003-4744-0188}\inst{\ref{aff37},\ref{aff36}}
\and A.~Franco\orcid{0000-0002-4761-366X}\inst{\ref{aff149},\ref{aff148},\ref{aff150}}
\and K.~Ganga\orcid{0000-0001-8159-8208}\inst{\ref{aff100}}
\and J.~Garc\'ia-Bellido\orcid{0000-0002-9370-8360}\inst{\ref{aff136}}
\and T.~Gasparetto\orcid{0000-0002-7913-4866}\inst{\ref{aff37}}
\and V.~Gautard\inst{\ref{aff153}}
\and E.~Gaztanaga\orcid{0000-0001-9632-0815}\inst{\ref{aff6},\ref{aff7},\ref{aff154}}
\and F.~Giacomini\orcid{0000-0002-3129-2814}\inst{\ref{aff41}}
\and F.~Gianotti\orcid{0000-0003-4666-119X}\inst{\ref{aff4}}
\and G.~Gozaliasl\orcid{0000-0002-0236-919X}\inst{\ref{aff155},\ref{aff86}}
\and A.~Gregorio\orcid{0000-0003-4028-8785}\inst{\ref{aff156},\ref{aff37},\ref{aff38}}
\and M.~Guidi\orcid{0000-0001-9408-1101}\inst{\ref{aff40},\ref{aff4}}
\and C.~M.~Gutierrez\orcid{0000-0001-7854-783X}\inst{\ref{aff157}}
\and A.~Hall\orcid{0000-0002-3139-8651}\inst{\ref{aff62}}
\and W.~G.~Hartley\inst{\ref{aff15}}
\and S.~Hemmati\orcid{0000-0003-2226-5395}\inst{\ref{aff116}}
\and C.~Hern\'andez-Monteagudo\orcid{0000-0001-5471-9166}\inst{\ref{aff114},\ref{aff61}}
\and H.~Hildebrandt\orcid{0000-0002-9814-3338}\inst{\ref{aff158}}
\and J.~Hjorth\orcid{0000-0002-4571-2306}\inst{\ref{aff108}}
\and J.~J.~E.~Kajava\orcid{0000-0002-3010-8333}\inst{\ref{aff159},\ref{aff160}}
\and Y.~Kang\orcid{0009-0000-8588-7250}\inst{\ref{aff15}}
\and V.~Kansal\orcid{0000-0002-4008-6078}\inst{\ref{aff161},\ref{aff162}}
\and D.~Karagiannis\orcid{0000-0002-4927-0816}\inst{\ref{aff128},\ref{aff163}}
\and K.~Kiiveri\inst{\ref{aff5}}
\and C.~C.~Kirkpatrick\inst{\ref{aff5}}
\and S.~Kruk\orcid{0000-0001-8010-8879}\inst{\ref{aff32}}
\and L.~Legrand\orcid{0000-0003-0610-5252}\inst{\ref{aff164},\ref{aff165}}
\and M.~Lembo\orcid{0000-0002-5271-5070}\inst{\ref{aff128},\ref{aff129}}
\and F.~Lepori\orcid{0009-0000-5061-7138}\inst{\ref{aff166}}
\and G.~Leroy\orcid{0009-0004-2523-4425}\inst{\ref{aff23},\ref{aff97}}
\and G.~F.~Lesci\orcid{0000-0002-4607-2830}\inst{\ref{aff17},\ref{aff4}}
\and J.~Lesgourgues\orcid{0000-0001-7627-353X}\inst{\ref{aff57}}
\and L.~Leuzzi\orcid{0009-0006-4479-7017}\inst{\ref{aff17},\ref{aff4}}
\and T.~I.~Liaudat\orcid{0000-0002-9104-314X}\inst{\ref{aff167}}
\and A.~Loureiro\orcid{0000-0002-4371-0876}\inst{\ref{aff168},\ref{aff169}}
\and J.~Macias-Perez\orcid{0000-0002-5385-2763}\inst{\ref{aff170}}
\and G.~Maggio\orcid{0000-0003-4020-4836}\inst{\ref{aff37}}
\and M.~Magliocchetti\orcid{0000-0001-9158-4838}\inst{\ref{aff26}}
\and E.~A.~Magnier\orcid{0000-0002-7965-2815}\inst{\ref{aff59}}
\and F.~Mannucci\orcid{0000-0002-4803-2381}\inst{\ref{aff16}}
\and R.~Maoli\orcid{0000-0002-6065-3025}\inst{\ref{aff171},\ref{aff3}}
\and C.~J.~A.~P.~Martins\orcid{0000-0002-4886-9261}\inst{\ref{aff172},\ref{aff47}}
\and L.~Maurin\orcid{0000-0002-8406-0857}\inst{\ref{aff31}}
\and M.~Miluzio\inst{\ref{aff32},\ref{aff173}}
\and P.~Monaco\orcid{0000-0003-2083-7564}\inst{\ref{aff156},\ref{aff37},\ref{aff38},\ref{aff36}}
\and C.~Moretti\orcid{0000-0003-3314-8936}\inst{\ref{aff39},\ref{aff135},\ref{aff37},\ref{aff36},\ref{aff38}}
\and G.~Morgante\inst{\ref{aff4}}
\and C.~Murray\inst{\ref{aff100}}
\and K.~Naidoo\orcid{0000-0002-9182-1802}\inst{\ref{aff154}}
\and A.~Navarro-Alsina\orcid{0000-0002-3173-2592}\inst{\ref{aff95}}
\and S.~Nesseris\orcid{0000-0002-0567-0324}\inst{\ref{aff136}}
\and F.~Passalacqua\orcid{0000-0002-8606-4093}\inst{\ref{aff115},\ref{aff72}}
\and K.~Paterson\orcid{0000-0001-8340-3486}\inst{\ref{aff10}}
\and L.~Patrizii\inst{\ref{aff41}}
\and A.~Pisani\orcid{0000-0002-6146-4437}\inst{\ref{aff73},\ref{aff174}}
\and D.~Potter\orcid{0000-0002-0757-5195}\inst{\ref{aff166}}
\and S.~Quai\orcid{0000-0002-0449-8163}\inst{\ref{aff17},\ref{aff4}}
\and M.~Radovich\orcid{0000-0002-3585-866X}\inst{\ref{aff11}}
\and P.-F.~Rocci\inst{\ref{aff31}}
\and G.~Rodighiero\orcid{0000-0002-9415-2296}\inst{\ref{aff115},\ref{aff11}}
\and S.~Sacquegna\orcid{0000-0002-8433-6630}\inst{\ref{aff148},\ref{aff149},\ref{aff150}}
\and M.~Sahl\'en\orcid{0000-0003-0973-4804}\inst{\ref{aff175}}
\and D.~B.~Sanders\orcid{0000-0002-1233-9998}\inst{\ref{aff59}}
\and E.~Sarpa\orcid{0000-0002-1256-655X}\inst{\ref{aff39},\ref{aff135},\ref{aff38}}
\and A.~Schneider\orcid{0000-0001-7055-8104}\inst{\ref{aff166}}
\and M.~Schultheis\inst{\ref{aff99}}
\and D.~Sciotti\orcid{0009-0008-4519-2620}\inst{\ref{aff3},\ref{aff96}}
\and E.~Sellentin\inst{\ref{aff176},\ref{aff12}}
\and A.~Shulevski\orcid{0000-0002-1827-0469}\inst{\ref{aff177},\ref{aff13},\ref{aff178},\ref{aff179}}
\and L.~C.~Smith\orcid{0000-0002-3259-2771}\inst{\ref{aff180}}
\and S.~A.~Stanford\orcid{0000-0003-0122-0841}\inst{\ref{aff181}}
\and K.~Tanidis\orcid{0000-0001-9843-5130}\inst{\ref{aff131}}
\and G.~Testera\inst{\ref{aff45}}
\and R.~Teyssier\orcid{0000-0001-7689-0933}\inst{\ref{aff174}}
\and S.~Tosi\orcid{0000-0002-7275-9193}\inst{\ref{aff44},\ref{aff133}}
\and A.~Troja\orcid{0000-0003-0239-4595}\inst{\ref{aff115},\ref{aff72}}
\and M.~Tucci\inst{\ref{aff15}}
\and C.~Valieri\inst{\ref{aff41}}
\and A.~Venhola\orcid{0000-0001-6071-4564}\inst{\ref{aff182}}
\and G.~Verza\orcid{0000-0002-1886-8348}\inst{\ref{aff183}}
\and P.~Vielzeuf\orcid{0000-0003-2035-9339}\inst{\ref{aff73}}
\and N.~A.~Walton\orcid{0000-0003-3983-8778}\inst{\ref{aff180}}
\and E.~Soubrie\orcid{0000-0001-9295-1863}\inst{\ref{aff31}}}
										   
\institute{School of Physics, HH Wills Physics Laboratory, University of Bristol, Tyndall Avenue, Bristol, BS8 1TL, UK\label{aff1}
\and
Department of Mathematics and Physics, Roma Tre University, Via della Vasca Navale 84, 00146 Rome, Italy\label{aff2}
\and
INAF-Osservatorio Astronomico di Roma, Via Frascati 33, 00078 Monteporzio Catone, Italy\label{aff3}
\and
INAF-Osservatorio di Astrofisica e Scienza dello Spazio di Bologna, Via Piero Gobetti 93/3, 40129 Bologna, Italy\label{aff4}
\and
Department of Physics and Helsinki Institute of Physics, Gustaf H\"allstr\"omin katu 2, 00014 University of Helsinki, Finland\label{aff5}
\and
Institute of Space Sciences (ICE, CSIC), Campus UAB, Carrer de Can Magrans, s/n, 08193 Barcelona, Spain\label{aff6}
\and
Institut d'Estudis Espacials de Catalunya (IEEC),  Edifici RDIT, Campus UPC, 08860 Castelldefels, Barcelona, Spain\label{aff7}
\and
INAF-Osservatorio Astronomico di Capodimonte, Via Moiariello 16, 80131 Napoli, Italy\label{aff8}
\and
Max Planck Institute for Extraterrestrial Physics, Giessenbachstr. 1, 85748 Garching, Germany\label{aff9}
\and
Max-Planck-Institut f\"ur Astronomie, K\"onigstuhl 17, 69117 Heidelberg, Germany\label{aff10}
\and
INAF-Osservatorio Astronomico di Padova, Via dell'Osservatorio 5, 35122 Padova, Italy\label{aff11}
\and
Leiden Observatory, Leiden University, Einsteinweg 55, 2333 CC Leiden, The Netherlands\label{aff12}
\and
Kapteyn Astronomical Institute, University of Groningen, PO Box 800, 9700 AV Groningen, The Netherlands\label{aff13}
\and
SRON Netherlands Institute for Space Research, Landleven 12, 9747 AD, Groningen, The Netherlands\label{aff14}
\and
Department of Astronomy, University of Geneva, ch. d'Ecogia 16, 1290 Versoix, Switzerland\label{aff15}
\and
INAF-Osservatorio Astrofisico di Arcetri, Largo E. Fermi 5, 50125, Firenze, Italy\label{aff16}
\and
Dipartimento di Fisica e Astronomia "Augusto Righi" - Alma Mater Studiorum Universit\`a di Bologna, via Piero Gobetti 93/2, 40129 Bologna, Italy\label{aff17}
\and
Dipartimento di Fisica e Astronomia, Universit\`{a} di Firenze, via G. Sansone 1, 50019 Sesto Fiorentino, Firenze, Italy\label{aff18}
\and
University of Trento, Via Sommarive 14, I-38123 Trento, Italy\label{aff19}
\and
Instituto de Astrof\'isica de Canarias (IAC); Departamento de Astrof\'isica, Universidad de La Laguna (ULL), 38200, La Laguna, Tenerife, Spain\label{aff20}
\and
Universit\"ats-Sternwarte M\"unchen, Fakult\"at f\"ur Physik, Ludwig-Maximilians-Universit\"at M\"unchen, Scheinerstrasse 1, 81679 M\"unchen, Germany\label{aff21}
\and
INAF, Istituto di Radioastronomia, Via Piero Gobetti 101, 40129 Bologna, Italy\label{aff22}
\and
Department of Physics, Centre for Extragalactic Astronomy, Durham University, South Road, Durham, DH1 3LE, UK\label{aff23}
\and
STAR Institute, University of Li{\`e}ge, Quartier Agora, All\'ee du six Ao\^ut 19c, 4000 Li\`ege, Belgium\label{aff24}
\and
School of Physics \& Astronomy, University of Southampton, Highfield Campus, Southampton SO17 1BJ, UK\label{aff25}
\and
INAF-Istituto di Astrofisica e Planetologia Spaziali, via del Fosso del Cavaliere, 100, 00100 Roma, Italy\label{aff26}
\and
Department of Physics and Astronomy, University of British Columbia, Vancouver, BC V6T 1Z1, Canada\label{aff27}
\and
Department of Physical Sciences, Ritsumeikan University, Kusatsu, Shiga 525-8577, Japan\label{aff28}
\and
National Astronomical Observatory of Japan, 2-21-1 Osawa, Mitaka, Tokyo 181-8588, Japan\label{aff29}
\and
Academia Sinica Institute of Astronomy and Astrophysics (ASIAA), 11F of ASMAB, No.~1, Section 4, Roosevelt Road, Taipei 10617, Taiwan\label{aff30}
\and
Universit\'e Paris-Saclay, CNRS, Institut d'astrophysique spatiale, 91405, Orsay, France\label{aff31}
\and
ESAC/ESA, Camino Bajo del Castillo, s/n., Urb. Villafranca del Castillo, 28692 Villanueva de la Ca\~nada, Madrid, Spain\label{aff32}
\and
School of Mathematics and Physics, University of Surrey, Guildford, Surrey, GU2 7XH, UK\label{aff33}
\and
INAF-Osservatorio Astronomico di Brera, Via Brera 28, 20122 Milano, Italy\label{aff34}
\and
Universit\'e Paris-Saclay, Universit\'e Paris Cit\'e, CEA, CNRS, AIM, 91191, Gif-sur-Yvette, France\label{aff35}
\and
IFPU, Institute for Fundamental Physics of the Universe, via Beirut 2, 34151 Trieste, Italy\label{aff36}
\and
INAF-Osservatorio Astronomico di Trieste, Via G. B. Tiepolo 11, 34143 Trieste, Italy\label{aff37}
\and
INFN, Sezione di Trieste, Via Valerio 2, 34127 Trieste TS, Italy\label{aff38}
\and
SISSA, International School for Advanced Studies, Via Bonomea 265, 34136 Trieste TS, Italy\label{aff39}
\and
Dipartimento di Fisica e Astronomia, Universit\`a di Bologna, Via Gobetti 93/2, 40129 Bologna, Italy\label{aff40}
\and
INFN-Sezione di Bologna, Viale Berti Pichat 6/2, 40127 Bologna, Italy\label{aff41}
\and
Centre National d'Etudes Spatiales -- Centre spatial de Toulouse, 18 avenue Edouard Belin, 31401 Toulouse Cedex 9, France\label{aff42}
\and
Space Science Data Center, Italian Space Agency, via del Politecnico snc, 00133 Roma, Italy\label{aff43}
\and
Dipartimento di Fisica, Universit\`a di Genova, Via Dodecaneso 33, 16146, Genova, Italy\label{aff44}
\and
INFN-Sezione di Genova, Via Dodecaneso 33, 16146, Genova, Italy\label{aff45}
\and
Department of Physics "E. Pancini", University Federico II, Via Cinthia 6, 80126, Napoli, Italy\label{aff46}
\and
Instituto de Astrof\'isica e Ci\^encias do Espa\c{c}o, Universidade do Porto, CAUP, Rua das Estrelas, PT4150-762 Porto, Portugal\label{aff47}
\and
Faculdade de Ci\^encias da Universidade do Porto, Rua do Campo de Alegre, 4150-007 Porto, Portugal\label{aff48}
\and
Dipartimento di Fisica, Universit\`a degli Studi di Torino, Via P. Giuria 1, 10125 Torino, Italy\label{aff49}
\and
INFN-Sezione di Torino, Via P. Giuria 1, 10125 Torino, Italy\label{aff50}
\and
INAF-Osservatorio Astrofisico di Torino, Via Osservatorio 20, 10025 Pino Torinese (TO), Italy\label{aff51}
\and
European Space Agency/ESTEC, Keplerlaan 1, 2201 AZ Noordwijk, The Netherlands\label{aff52}
\and
Institute Lorentz, Leiden University, Niels Bohrweg 2, 2333 CA Leiden, The Netherlands\label{aff53}
\and
INAF-IASF Milano, Via Alfonso Corti 12, 20133 Milano, Italy\label{aff54}
\and
Centro de Investigaciones Energ\'eticas, Medioambientales y Tecnol\'ogicas (CIEMAT), Avenida Complutense 40, 28040 Madrid, Spain\label{aff55}
\and
Port d'Informaci\'{o} Cient\'{i}fica, Campus UAB, C. Albareda s/n, 08193 Bellaterra (Barcelona), Spain\label{aff56}
\and
Institute for Theoretical Particle Physics and Cosmology (TTK), RWTH Aachen University, 52056 Aachen, Germany\label{aff57}
\and
INFN section of Naples, Via Cinthia 6, 80126, Napoli, Italy\label{aff58}
\and
Institute for Astronomy, University of Hawaii, 2680 Woodlawn Drive, Honolulu, HI 96822, USA\label{aff59}
\and
Dipartimento di Fisica e Astronomia "Augusto Righi" - Alma Mater Studiorum Universit\`a di Bologna, Viale Berti Pichat 6/2, 40127 Bologna, Italy\label{aff60}
\and
Instituto de Astrof\'{\i}sica de Canarias, V\'{\i}a L\'actea, 38205 La Laguna, Tenerife, Spain\label{aff61}
\and
Institute for Astronomy, University of Edinburgh, Royal Observatory, Blackford Hill, Edinburgh EH9 3HJ, UK\label{aff62}
\and
Jodrell Bank Centre for Astrophysics, Department of Physics and Astronomy, University of Manchester, Oxford Road, Manchester M13 9PL, UK\label{aff63}
\and
European Space Agency/ESRIN, Largo Galileo Galilei 1, 00044 Frascati, Roma, Italy\label{aff64}
\and
Universit\'e Claude Bernard Lyon 1, CNRS/IN2P3, IP2I Lyon, UMR 5822, Villeurbanne, F-69100, France\label{aff65}
\and
Institut de Ci\`{e}ncies del Cosmos (ICCUB), Universitat de Barcelona (IEEC-UB), Mart\'{i} i Franqu\`{e}s 1, 08028 Barcelona, Spain\label{aff66}
\and
Instituci\'o Catalana de Recerca i Estudis Avan\c{c}ats (ICREA), Passeig de Llu\'{\i}s Companys 23, 08010 Barcelona, Spain\label{aff67}
\and
UCB Lyon 1, CNRS/IN2P3, IUF, IP2I Lyon, 4 rue Enrico Fermi, 69622 Villeurbanne, France\label{aff68}
\and
Mullard Space Science Laboratory, University College London, Holmbury St Mary, Dorking, Surrey RH5 6NT, UK\label{aff69}
\and
Departamento de F\'isica, Faculdade de Ci\^encias, Universidade de Lisboa, Edif\'icio C8, Campo Grande, PT1749-016 Lisboa, Portugal\label{aff70}
\and
Instituto de Astrof\'isica e Ci\^encias do Espa\c{c}o, Faculdade de Ci\^encias, Universidade de Lisboa, Campo Grande, 1749-016 Lisboa, Portugal\label{aff71}
\and
INFN-Padova, Via Marzolo 8, 35131 Padova, Italy\label{aff72}
\and
Aix-Marseille Universit\'e, CNRS/IN2P3, CPPM, Marseille, France\label{aff73}
\and
INFN-Bologna, Via Irnerio 46, 40126 Bologna, Italy\label{aff74}
\and
FRACTAL S.L.N.E., calle Tulip\'an 2, Portal 13 1A, 28231, Las Rozas de Madrid, Spain\label{aff75}
\and
NRC Herzberg, 5071 West Saanich Rd, Victoria, BC V9E 2E7, Canada\label{aff76}
\and
Institute of Theoretical Astrophysics, University of Oslo, P.O. Box 1029 Blindern, 0315 Oslo, Norway\label{aff77}
\and
Jet Propulsion Laboratory, California Institute of Technology, 4800 Oak Grove Drive, Pasadena, CA, 91109, USA\label{aff78}
\and
Department of Physics, Lancaster University, Lancaster, LA1 4YB, UK\label{aff79}
\and
Felix Hormuth Engineering, Goethestr. 17, 69181 Leimen, Germany\label{aff80}
\and
Technical University of Denmark, Elektrovej 327, 2800 Kgs. Lyngby, Denmark\label{aff81}
\and
Cosmic Dawn Center (DAWN), Denmark\label{aff82}
\and
Institut d'Astrophysique de Paris, UMR 7095, CNRS, and Sorbonne Universit\'e, 98 bis boulevard Arago, 75014 Paris, France\label{aff83}
\and
NASA Goddard Space Flight Center, Greenbelt, MD 20771, USA\label{aff84}
\and
Universit\'e de Gen\`eve, D\'epartement de Physique Th\'eorique and Centre for Astroparticle Physics, 24 quai Ernest-Ansermet, CH-1211 Gen\`eve 4, Switzerland\label{aff85}
\and
Department of Physics, P.O. Box 64, 00014 University of Helsinki, Finland\label{aff86}
\and
Helsinki Institute of Physics, Gustaf H{\"a}llstr{\"o}min katu 2, University of Helsinki, Helsinki, Finland\label{aff87}
\and
Centre de Calcul de l'IN2P3/CNRS, 21 avenue Pierre de Coubertin 69627 Villeurbanne Cedex, France\label{aff88}
\and
Laboratoire d'etude de l'Univers et des phenomenes eXtremes, Observatoire de Paris, Universit\'e PSL, Sorbonne Universit\'e, CNRS, 92190 Meudon, France\label{aff89}
\and
Aix-Marseille Universit\'e, CNRS, CNES, LAM, Marseille, France\label{aff90}
\and
SKA Observatory, Jodrell Bank, Lower Withington, Macclesfield, Cheshire SK11 9FT, UK\label{aff91}
\and
Dipartimento di Fisica "Aldo Pontremoli", Universit\`a degli Studi di Milano, Via Celoria 16, 20133 Milano, Italy\label{aff92}
\and
INFN-Sezione di Milano, Via Celoria 16, 20133 Milano, Italy\label{aff93}
\and
University of Applied Sciences and Arts of Northwestern Switzerland, School of Computer Science, 5210 Windisch, Switzerland\label{aff94}
\and
Universit\"at Bonn, Argelander-Institut f\"ur Astronomie, Auf dem H\"ugel 71, 53121 Bonn, Germany\label{aff95}
\and
INFN-Sezione di Roma, Piazzale Aldo Moro, 2 - c/o Dipartimento di Fisica, Edificio G. Marconi, 00185 Roma, Italy\label{aff96}
\and
Department of Physics, Institute for Computational Cosmology, Durham University, South Road, Durham, DH1 3LE, UK\label{aff97}
\and
Infrared Processing and Analysis Center, California Institute of Technology, Pasadena, CA 91125, USA\label{aff98}
\and
Universit\'e C\^{o}te d'Azur, Observatoire de la C\^{o}te d'Azur, CNRS, Laboratoire Lagrange, Bd de l'Observatoire, CS 34229, 06304 Nice cedex 4, France\label{aff99}
\and
Universit\'e Paris Cit\'e, CNRS, Astroparticule et Cosmologie, 75013 Paris, France\label{aff100}
\and
CNRS-UCB International Research Laboratory, Centre Pierre Binetruy, IRL2007, CPB-IN2P3, Berkeley, USA\label{aff101}
\and
University of Applied Sciences and Arts of Northwestern Switzerland, School of Engineering, 5210 Windisch, Switzerland\label{aff102}
\and
Institut d'Astrophysique de Paris, 98bis Boulevard Arago, 75014, Paris, France\label{aff103}
\and
Institute of Physics, Laboratory of Astrophysics, Ecole Polytechnique F\'ed\'erale de Lausanne (EPFL), Observatoire de Sauverny, 1290 Versoix, Switzerland\label{aff104}
\and
Aurora Technology for European Space Agency (ESA), Camino bajo del Castillo, s/n, Urbanizacion Villafranca del Castillo, Villanueva de la Ca\~nada, 28692 Madrid, Spain\label{aff105}
\and
Institut de F\'{i}sica d'Altes Energies (IFAE), The Barcelona Institute of Science and Technology, Campus UAB, 08193 Bellaterra (Barcelona), Spain\label{aff106}
\and
School of Mathematics, Statistics and Physics, Newcastle University, Herschel Building, Newcastle-upon-Tyne, NE1 7RU, UK\label{aff107}
\and
DARK, Niels Bohr Institute, University of Copenhagen, Jagtvej 155, 2200 Copenhagen, Denmark\label{aff108}
\and
Waterloo Centre for Astrophysics, University of Waterloo, Waterloo, Ontario N2L 3G1, Canada\label{aff109}
\and
Department of Physics and Astronomy, University of Waterloo, Waterloo, Ontario N2L 3G1, Canada\label{aff110}
\and
Perimeter Institute for Theoretical Physics, Waterloo, Ontario N2L 2Y5, Canada\label{aff111}
\and
Institute of Space Science, Str. Atomistilor, nr. 409 M\u{a}gurele, Ilfov, 077125, Romania\label{aff112}
\and
Consejo Superior de Investigaciones Cientificas, Calle Serrano 117, 28006 Madrid, Spain\label{aff113}
\and
Universidad de La Laguna, Departamento de Astrof\'{\i}sica, 38206 La Laguna, Tenerife, Spain\label{aff114}
\and
Dipartimento di Fisica e Astronomia "G. Galilei", Universit\`a di Padova, Via Marzolo 8, 35131 Padova, Italy\label{aff115}
\and
Caltech/IPAC, 1200 E. California Blvd., Pasadena, CA 91125, USA\label{aff116}
\and
Institut f\"ur Theoretische Physik, University of Heidelberg, Philosophenweg 16, 69120 Heidelberg, Germany\label{aff117}
\and
Institut de Recherche en Astrophysique et Plan\'etologie (IRAP), Universit\'e de Toulouse, CNRS, UPS, CNES, 14 Av. Edouard Belin, 31400 Toulouse, France\label{aff118}
\and
Universit\'e St Joseph; Faculty of Sciences, Beirut, Lebanon\label{aff119}
\and
Departamento de F\'isica, FCFM, Universidad de Chile, Blanco Encalada 2008, Santiago, Chile\label{aff120}
\and
Universit\"at Innsbruck, Institut f\"ur Astro- und Teilchenphysik, Technikerstr. 25/8, 6020 Innsbruck, Austria\label{aff121}
\and
Satlantis, University Science Park, Sede Bld 48940, Leioa-Bilbao, Spain\label{aff122}
\and
Centre for Electronic Imaging, Open University, Walton Hall, Milton Keynes, MK7~6AA, UK\label{aff123}
\and
Instituto de Astrof\'isica e Ci\^encias do Espa\c{c}o, Faculdade de Ci\^encias, Universidade de Lisboa, Tapada da Ajuda, 1349-018 Lisboa, Portugal\label{aff124}
\and
Cosmic Dawn Center (DAWN)\label{aff125}
\and
Niels Bohr Institute, University of Copenhagen, Jagtvej 128, 2200 Copenhagen, Denmark\label{aff126}
\and
Universidad Polit\'ecnica de Cartagena, Departamento de Electr\'onica y Tecnolog\'ia de Computadoras,  Plaza del Hospital 1, 30202 Cartagena, Spain\label{aff127}
\and
Dipartimento di Fisica e Scienze della Terra, Universit\`a degli Studi di Ferrara, Via Giuseppe Saragat 1, 44122 Ferrara, Italy\label{aff128}
\and
Istituto Nazionale di Fisica Nucleare, Sezione di Ferrara, Via Giuseppe Saragat 1, 44122 Ferrara, Italy\label{aff129}
\and
School of Physics and Astronomy, Cardiff University, The Parade, Cardiff, CF24 3AA, UK\label{aff130}
\and
Department of Physics, Oxford University, Keble Road, Oxford OX1 3RH, UK\label{aff131}
\and
INAF - Osservatorio Astronomico di Brera, via Emilio Bianchi 46, 23807 Merate, Italy\label{aff132}
\and
INAF-Osservatorio Astronomico di Brera, Via Brera 28, 20122 Milano, Italy, and INFN-Sezione di Genova, Via Dodecaneso 33, 16146, Genova, Italy\label{aff133}
\and
ICL, Junia, Universit\'e Catholique de Lille, LITL, 59000 Lille, France\label{aff134}
\and
ICSC - Centro Nazionale di Ricerca in High Performance Computing, Big Data e Quantum Computing, Via Magnanelli 2, Bologna, Italy\label{aff135}
\and
Instituto de F\'isica Te\'orica UAM-CSIC, Campus de Cantoblanco, 28049 Madrid, Spain\label{aff136}
\and
CERCA/ISO, Department of Physics, Case Western Reserve University, 10900 Euclid Avenue, Cleveland, OH 44106, USA\label{aff137}
\and
Technical University of Munich, TUM School of Natural Sciences, Physics Department, James-Franck-Str.~1, 85748 Garching, Germany\label{aff138}
\and
Max-Planck-Institut f\"ur Astrophysik, Karl-Schwarzschild-Str.~1, 85748 Garching, Germany\label{aff139}
\and
Laboratoire Univers et Th\'eorie, Observatoire de Paris, Universit\'e PSL, Universit\'e Paris Cit\'e, CNRS, 92190 Meudon, France\label{aff140}
\and
Departamento de F{\'\i}sica Fundamental. Universidad de Salamanca. Plaza de la Merced s/n. 37008 Salamanca, Spain\label{aff141}
\and
Universit\'e de Strasbourg, CNRS, Observatoire astronomique de Strasbourg, UMR 7550, 67000 Strasbourg, France\label{aff142}
\and
Center for Data-Driven Discovery, Kavli IPMU (WPI), UTIAS, The University of Tokyo, Kashiwa, Chiba 277-8583, Japan\label{aff143}
\and
Ludwig-Maximilians-University, Schellingstrasse 4, 80799 Munich, Germany\label{aff144}
\and
Max-Planck-Institut f\"ur Physik, Boltzmannstr. 8, 85748 Garching, Germany\label{aff145}
\and
California Institute of Technology, 1200 E California Blvd, Pasadena, CA 91125, USA\label{aff146}
\and
Department of Physics \& Astronomy, University of California Irvine, Irvine CA 92697, USA\label{aff147}
\and
Department of Mathematics and Physics E. De Giorgi, University of Salento, Via per Arnesano, CP-I93, 73100, Lecce, Italy\label{aff148}
\and
INFN, Sezione di Lecce, Via per Arnesano, CP-193, 73100, Lecce, Italy\label{aff149}
\and
INAF-Sezione di Lecce, c/o Dipartimento Matematica e Fisica, Via per Arnesano, 73100, Lecce, Italy\label{aff150}
\and
Departamento F\'isica Aplicada, Universidad Polit\'ecnica de Cartagena, Campus Muralla del Mar, 30202 Cartagena, Murcia, Spain\label{aff151}
\and
Instituto de F\'isica de Cantabria, Edificio Juan Jord\'a, Avenida de los Castros, 39005 Santander, Spain\label{aff152}
\and
CEA Saclay, DFR/IRFU, Service d'Astrophysique, Bat. 709, 91191 Gif-sur-Yvette, France\label{aff153}
\and
Institute of Cosmology and Gravitation, University of Portsmouth, Portsmouth PO1 3FX, UK\label{aff154}
\and
Department of Computer Science, Aalto University, PO Box 15400, Espoo, FI-00 076, Finland\label{aff155}
\and
Dipartimento di Fisica - Sezione di Astronomia, Universit\`a di Trieste, Via Tiepolo 11, 34131 Trieste, Italy\label{aff156}
\and
Instituto de Astrof\'\i sica de Canarias, c/ Via Lactea s/n, La Laguna 38200, Spain. Departamento de Astrof\'\i sica de la Universidad de La Laguna, Avda. Francisco Sanchez, La Laguna, 38200, Spain\label{aff157}
\and
Ruhr University Bochum, Faculty of Physics and Astronomy, Astronomical Institute (AIRUB), German Centre for Cosmological Lensing (GCCL), 44780 Bochum, Germany\label{aff158}
\and
Department of Physics and Astronomy, Vesilinnantie 5, 20014 University of Turku, Finland\label{aff159}
\and
Serco for European Space Agency (ESA), Camino bajo del Castillo, s/n, Urbanizacion Villafranca del Castillo, Villanueva de la Ca\~nada, 28692 Madrid, Spain\label{aff160}
\and
ARC Centre of Excellence for Dark Matter Particle Physics, Melbourne, Australia\label{aff161}
\and
Centre for Astrophysics \& Supercomputing, Swinburne University of Technology,  Hawthorn, Victoria 3122, Australia\label{aff162}
\and
Department of Physics and Astronomy, University of the Western Cape, Bellville, Cape Town, 7535, South Africa\label{aff163}
\and
DAMTP, Centre for Mathematical Sciences, Wilberforce Road, Cambridge CB3 0WA, UK\label{aff164}
\and
Kavli Institute for Cosmology Cambridge, Madingley Road, Cambridge, CB3 0HA, UK\label{aff165}
\and
Department of Astrophysics, University of Zurich, Winterthurerstrasse 190, 8057 Zurich, Switzerland\label{aff166}
\and
IRFU, CEA, Universit\'e Paris-Saclay 91191 Gif-sur-Yvette Cedex, France\label{aff167}
\and
Oskar Klein Centre for Cosmoparticle Physics, Department of Physics, Stockholm University, Stockholm, SE-106 91, Sweden\label{aff168}
\and
Astrophysics Group, Blackett Laboratory, Imperial College London, London SW7 2AZ, UK\label{aff169}
\and
Univ. Grenoble Alpes, CNRS, Grenoble INP, LPSC-IN2P3, 53, Avenue des Martyrs, 38000, Grenoble, France\label{aff170}
\and
Dipartimento di Fisica, Sapienza Universit\`a di Roma, Piazzale Aldo Moro 2, 00185 Roma, Italy\label{aff171}
\and
Centro de Astrof\'{\i}sica da Universidade do Porto, Rua das Estrelas, 4150-762 Porto, Portugal\label{aff172}
\and
HE Space for European Space Agency (ESA), Camino bajo del Castillo, s/n, Urbanizacion Villafranca del Castillo, Villanueva de la Ca\~nada, 28692 Madrid, Spain\label{aff173}
\and
Department of Astrophysical Sciences, Peyton Hall, Princeton University, Princeton, NJ 08544, USA\label{aff174}
\and
Theoretical astrophysics, Department of Physics and Astronomy, Uppsala University, Box 515, 751 20 Uppsala, Sweden\label{aff175}
\and
Mathematical Institute, University of Leiden, Einsteinweg 55, 2333 CA Leiden, The Netherlands\label{aff176}
\and
ASTRON, the Netherlands Institute for Radio Astronomy, Postbus 2, 7990 AA, Dwingeloo, The Netherlands\label{aff177}
\and
Anton Pannekoek Institute for Astronomy, University of Amsterdam, Postbus 94249, 1090 GE Amsterdam, The Netherlands\label{aff178}
\and
Center for Advanced Interdisciplinary Research, Ss. Cyril and Methodius University in Skopje, Macedonia\label{aff179}
\and
Institute of Astronomy, University of Cambridge, Madingley Road, Cambridge CB3 0HA, UK\label{aff180}
\and
Department of Physics and Astronomy, University of California, Davis, CA 95616, USA\label{aff181}
\and
Space physics and astronomy research unit, University of Oulu, Pentti Kaiteran katu 1, FI-90014 Oulu, Finland\label{aff182}
\and
Center for Computational Astrophysics, Flatiron Institute, 162 5th Avenue, 10010, New York, NY, USA\label{aff183}}    
   

%
%
\abstract
{

We present a catalogue of candidate active galactic nuclei (AGN) in the \Euclid Quick Release (Q1) fields. For each \Euclid source we collect multi-wavelength photometry and spectroscopy information from Galaxy Evolution Explorer (GALEX), \gaia, Dark Energy Survey (DES), Wise-field Infrared Survey Explorer (WISE), \textit{Spitzer}, Dark Energy Survey (DESI), and Sloan Digital Sky Survey (SDSS), including spectroscopic redshift from public compilations.
We investigate the AGN contents of the Q1 fields by applying selection criteria using \Euclid colours and WISE-AllWISE cuts finding respectively 292\,222 and 65\,131 candidates. We also create a high-purity QSO catalogue based on \gaia DR3 information containing 1971 candidates. Furthermore, we utilise the collected spectroscopic information from DESI to perform broad-line and narrow-line AGN selections, leading to a total of 4392 AGN candidates in the Q1 field. We investigate and refine the Q1 probabilistic random forest QSO population, selecting a total of  180\,666 candidates. Additionally, we perform SED fitting on a subset of sources with available $z_{\text{spec}}$, and by utilizing the derived AGN fraction, we identify a total of 7766 AGN candidates. We discuss purity and completeness of the selections and define two new colour selection criteria ($JH$\_$I_{\text{E}}Y$ and $I_{\text{E}}H$\_$gz$) to improve on purity, finding 313\,714 and 267\,513 candidates respectively in the Q1 data. We find a total of 229\,779 AGN candidates equivalent to an AGN surface density of 3641 deg$^{-2}$ for $18<\IE\leq 24.5$, and a subsample of 30\,422 candidates corresponding to an AGN surface density of 482 deg$^{-2}$ when limiting the depth to $18<\IE\leq 22$. The surface density of AGN recovered from this work is in line with predictions based on the AGN X-ray luminosity functions.
}
    \keywords{  Galaxies: active, Catalogues, Surveys}

   \titlerunning{ The active galaxies of  \Euclid}
   \authorrunning{Euclid Collaboration: T.\ Matamoro Zatarain et al.}
   
   \maketitle
%
%
%
%


\section{\label{sc:Intro}Introduction}

Active galactic nuclei (AGN) are some of the most powerful sources in the Universe. With bolometric luminosities up to $L_{\text{bol}} = 10^{48}$\,erg\,s$^{-1}$ \citep{Padovani_2017_2017A&ARv..25....2P}, these objects exist at the centres of massive galaxies and emit immense amounts of non-stellar radiation \citep{Peterson_1997_1997iagn.book.....P, Netzer_2015_2015ARA&A..53..365N, Alexander_2012,Combes_2021_2021agnf.book.....C} 
due to the accretion of matter onto a super-massive black hole (SMBH) and its surrounding accretion disc
embedded in a dusty, clumpy, obscuring torus \citep{Shakura_1973_1973A&A....24..337S, Zel'dovich_1964_1964SPhD....9..246Z, Rees_1984_1984ARA&A..22..471R, Peterson_1997_1997iagn.book.....P,Antonucci_1993_1993ARA&A..31..473A,Netzer_2015_2015ARA&A..53..365N}.

The activity of a SMBH is closely related to the properties of its host galaxy through energetic winds providing valuable feedback, which gives rise to relationships such as the $M$--$\sigma$ relation \citep{Silk-Rees-1998A&A...331L...1S,Merritt_2000_2000ASPC..197..221M, Haehnelt_2000_2000MNRAS.318L..35H,2019ApJ...887...10S}, the black hole mass to bulge mass relation \citep{Magorrian_1998_1998AJ....115.2285M, Haring_2004_2004ApJ...604L..89H, Kormendy_2013_2013arXiv1308.6483K}, and even the black hole mass to host galaxy stellar mass relation \citep{Bandara_2009_2009ApJ...704.1135B,2016MNRAS.460.3119S}, indicating that understanding the many types of AGN is key to deciphering the origin and evolution of galaxies. This is why identifying AGN in their different states of accretion and obscuration is fundamental to build a full picture of the evolution and properties of their host galaxies \citep{Harrison_2024_2024Galax..12...17H}.

Our current census of AGN is incomplete, partly because we lack a universal diagnostic tool to identify the overall population of these objects \citep{Lacy_2004_2004ApJS..154..166L, Stern_2005_2005ApJ...631..163S, Stern_2012_2012ApJ...753...30S, Kirkpatrick_2012_2012ApJ...759..139K}, leading to samples whose properties are strongly biased by their selection methods \citep{Hickox_2019_2019HEAD...1710648H, Cann_2019_2019ApJ...870L...2C, Hviding_2024_2024AJ....167..169H}. AGN diagnostics have been developed for most wavelength ranges. Some of the most common techniques involve using radio observations \citep{Mushotzky_2004_2004ASSL..308...53M, Smolcic_2017_2017A&A...602A...2S, Hickox_2018_2018ARA&A..56..625H}, X-ray emission \citep{Pounds_1979_1979RSPSA.366..375P, Brandt_2015, Lusso_2016_2016ApJ...819..154L}, emission line diagnostics \citep{Baldwin1981, Veilleux_1987_1987ApJS...63..295V, Osmer_1991_1991ApJS...75..273O, Greene_2005_2005ApJ...630..122G}, variability diagnostics \citep{Ulrich_1997_1997ARA&A..35..445U, Kawaguchi_1998_1998ApJ...504..671K, Paolillo_2004_2004ApJ...611...93P}, or colour criteria \citep{Sandage_1971_1971swng.conf..271S, Koo_1988_1988ApJ...325...92K, Richards_2001_2001AJ....121.2308R, Stern_2005_2005ApJ...631..163S, Wang_2016_2016ApJ...819...24W}, as well as machine-learning methods \citep{Fotopoulou_2018_2018A&A...619A..14F}. However, all of these techniques have their own limitations. For instance, AGN selection in the ultra-violet (UV), optical, and soft X-rays, are affected by dust and gas obscuration, creating a bias against obscured AGN, also known as Type II AGN \citep{Gilli_2007_2007A&A...463...79G, Treister_2009_2009ApJ...706..535T, Bornancini_2022_2022A&A...664A.110B}. 

The infrared (IR) regime is a powerful alternative for AGN identification, particularly for obscured sources \citep{Hickox_2018_2018ARA&A..56..625H, Bornancini_2022_2022A&A...664A.110B, Calabro_2023_2023A&A...679A..80C}. IR radiation is created by UV and optical accretion disc photons that are absorbed by a surrounding dusty torus and re-emitted in the IR \citep{Antonucci_1993_1993ARA&A..31..473A, Urry_1995_1995PASP..107..803U, Mor_2012_2012MNRAS.420..526M}. This theoretically means that by using IR diagnostics, one should be able to detect a sizeable population of obscured AGN \citep{Calabro_2023_2023A&A...679A..80C}. Previous works have already developed both spectroscopic and photometric selection approaches for IR surveys \citep{deGrijp_1987_1987A&AS...70...95D, Clavel_2000_2000A&A...357..839C, Lacy_2004_2004ApJS..154..166L,Stern_2005_2005ApJ...631..163S, Stern_2012_2012ApJ...753...30S, Assef_2013_2013ApJ...772...26A}. In particular, colour criteria present a rapid and inexpensive approach for cataloguing these sources. Nonetheless, these techniques still have limitations, and not only are some AGN types still missed, but contaminants also play a role in this selection regime \citep{Bornancini_2022_2022A&A...664A.110B}.

\Euclid is an optical and near-IR (NIR) European Space Agency (ESA) mission \citep{Laureijs_2011_2011arXiv1110.3193L}. The details of the instruments and its scientific goals can be found in \citet{EuclidSkyOverview}. Briefly, \Euclid will observe approximately 14\,000 deg$^2$ of the extra-galactic sky while undertaking two surveys during its expected 6-year lifetime. The Euclid Wide Survey \citep[EWS,][]{Scaramella-EP1}, which will observe about 14\,000 deg$^2$ with a visible depth of $\IE=26.2$, and the Euclid Deep Survey (EDS), which will concentrate on three different areas of the sky covering over 53\,deg$^2$ with a visible depth of $\IE=28.2$ \citep{EuclidSkyOverview}. It is expected that \Euclid will be able to detect billions of sources, of which at least $10$ million are anticipated to be AGN identified through a combination of its Visible Camera \citep[VIS,][]{EuclidSkyVIS} and Near-Infrared Spectrometer and Photometer \citep[NISP,][]{EuclidSkyNISP} instruments \citep{EP-Selwood, EP-Bisigello,EuclidCollaboration_2024_2024A&A...685A.108E}. This will increase our known number of AGN dramatically, meaning that -- with the right target selection tools and strategic overlap with other multi-wavelength data sets --  \Euclid will play a crucial role in creating a more complete AGN census.

For this reason, and in anticipation of \Euclid's first Quick Data Release \citep{Q1cite}, which constitutes a first visit to the Euclid Deep Fields (EDFs), covering a total area of 63.1 deg$^2$,  various approaches to identify AGN have already been developed. In particular, \citet{EP-Bisigello} carried out a systematic study to find the best colour-selection criteria for AGN based on \Euclid's photometry. That paper emphasizes distinct selection methods that may be more appropriate for either the EWS or EDS. Although the purity of these diagnostics could be enhanced, they provide an excellent foundation for examining the populations present in the Q1 data. 

The Q1 data provides observations of the EDF-North (EDF-N), EDF-South (EDF-S), and EDF-Fornax (EDF-F), at the depth of the EWS \citep{Q1-TP001}. The three EDF regions were selected primarily due to the nearly perennial visibility of ecliptic poles under the survey strategy \citep{EuclidSkyOverview}. The overlap of the EDF regions with multi-waveband external surveys provides an excellent opportunity to investigate the multi-wavelength properties of \Euclid's sources. Detailed catalogues of detected sources have been produced for various different missions, and collectively, over 20 million AGN candidates have been identified \citep{Assef_2013_2013ApJ...772...26A, Assef_2018_2018ApJS..234...23A, Storey-Fisher2024, Fu_2024_2024ApJS..271...54F}. Studies based on the combination of external catalogues with those created from the \Euclid data sets will play a crucial role in advancing our understanding of AGN demography and evolution.

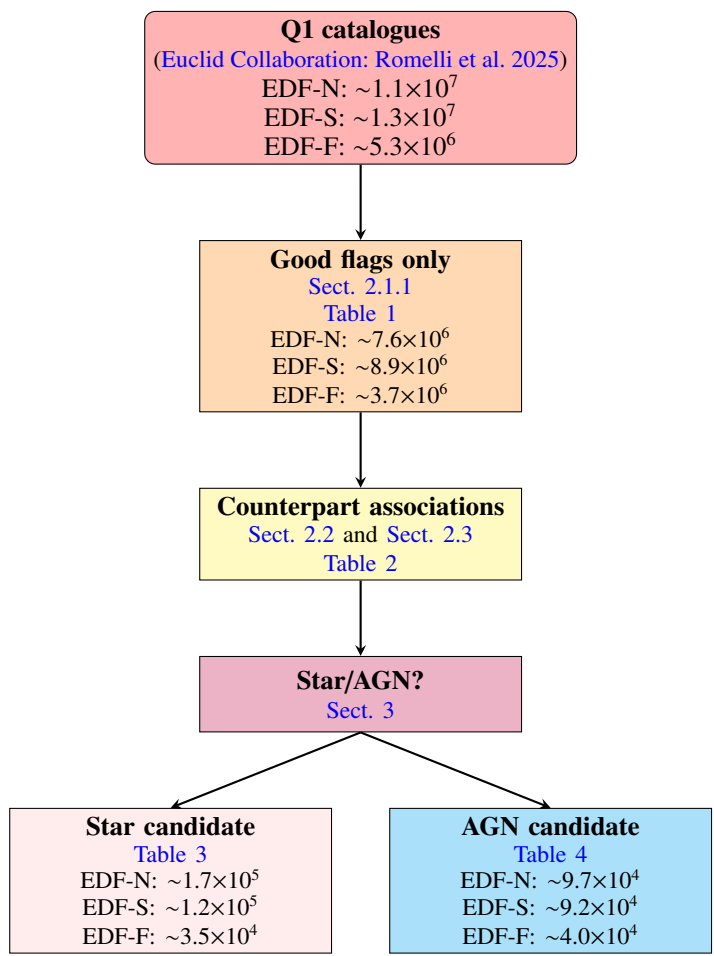
\begin{figure}[ht]
\centering
\begin{tikzpicture}[node distance=1cm]

\node (start) [startstop, fill=red!30] {\textbf{Q1 catalogues} \\ \small{\citep{Q1-TP004}} \\ EDF-N: $\sim$1.1$\times$10$^7$ \\
EDF-S: $\sim$1.3$\times$10$^7$ \\
EDF-F: $\sim$5.3$\times$10$^6$ };
\node (step1) [process, below=of start, fill=orange!30] {\textbf{Good flags only} \\ \small{\cref{sc:quality} \\ \cref{Flags} \\
\footnotesize{EDF-N: $\sim$7.6$\times$10$^6$ \\
EDF-S: $\sim$8.9$\times$10$^6$ \\
EDF-F: $\sim$3.7$\times$10$^6$ }}};
\node (step2) [process, below=of step1, fill=yellow!30] {\textbf{Counterpart associations} \\ \small{\cref{sc:matchedphot} and \cref{sc:spec} \\ \cref{Table:matches_breakdown}}};
\node (step4) [process, below=of step2, fill=purple!30] {\textbf{Star/AGN?} \\ \small{\cref{sc:qso} }};
\node (step5) [process, below=of step4, xshift=-2.5cm, fill=pink!30] {\textbf{Star candidate} \\ \small{\cref{Table:star} \\ 
\footnotesize{EDF-N: $\sim$1.7$\times$10$^5$ \\
EDF-S: $\sim$1.2$\times$10$^5$ \\
EDF-F: $\sim$3.5$\times$10$^4$ }}};
\node (step6) [process, below=of step4, xshift=2.5cm, fill=cyan!30] {\textbf{AGN candidate} \\ \small{\cref{Table:AGN_candidates_photometry} \\ 
\footnotesize{EDF-N: $\sim$9.7$\times$10$^4$ \\
EDF-S: $\sim$9.2$\times$10$^4$ \\
EDF-F: $\sim$4.0$\times$10$^4$ }}};
\draw [arrow] (start.south) -- (step1.north);
\draw [arrow] (step1.south) -- (step2.north);
\draw [arrow] (step2.south) -- (step4.north);
\draw [arrow] (step4.south) -- (step5.north);
\draw [arrow] (step4.south) --  (step6.north);
\end{tikzpicture}
\caption{Sketch outlining the steps adopted in this work to attain
the AGN candidate catalogue. We report the number of stellar and AGN candidates for the magnitude range $18<\IE\leq$24.5.}\label{fig:flowchart}
\end{figure}

In this paper, we present a multi-wavelength AGN candidate catalogue derived from \Euclid's photometry in combination with external surveys. In \cref{sc:data}, we introduce and describe \Euclid's Q1 source catalogues, along with the external photometric and spectroscopic catalogues utilised in this work. Additionally, we explain how we perform counterpart (CTP) associations for each survey and provide the number of matches found. In \cref{sc:qso}, we explore the various source populations identified in the data, with a primary focus on stellar and AGN candidates. For the AGN candidates, we examine multiple selection methods, both spectroscopic and photometric, and we compare these diagnostic techniques to those used in other Q1 papers (\citealp{Q1-SP003, Q1-SP015, Q1-SP009, Q1-SP013}, Euclid Collaboration: Laloux et al., in prep). \Cref{sc:discussion} examines the AGN candidates obtained, compares them with expected results from the literature, and analyses the different AGN populations identified in this work. Finally, in \cref{sc:agn_catalogue}, the overall multi-wavelength AGN catalogue is presented. With \cref{fig:flowchart} we provide a diagram that illustrates the procedures we follow to compile the AGN candidate catalogue and we indicate the relevant sections of the paper associated with each step throughout. We adopt a $\Lambda$CDM cosmology with $H_0 = 70$\,km\,s$^{-1}$\,Mpc$^{-1}$, $\Omega_{\rm m} = 0.3$ and $\Omega_{\Lambda} = 0.7$. All magnitudes are in the AB system \citep{Oke_1983_1983ApJ...266..713O} unless stated otherwise.

\section{Data and counterpart associations}
\label{sc:data}

This section provides an overview of the data utilised in this study and the nearest-neighbour matching we perform to identify \Euclid's multi-wavelength counterparts. We begin with a concise summary of the Q1 catalogues, which serve as the foundation for our AGN catalogue. Subsequently, we split the external data into photometry and spectroscopy, detailing the respective instruments and surveys used.


\subsection{Euclid Q1 data}

\Euclid is scheduled to have three major data releases (DRs) over its 6-year nominal mission duration. Detailed descriptions of these releases, as well as information about the mission, are available in \citet{EuclidSkyOverview}. However, in addition to the primary releases, there are also interspersed quick releases of smaller volume planned between them. The first of these, the Q1, signifies the initial public release to the scientific community. Details on the  data available for the Q1 release can be found in \citet{Q1-TP001}, \citet{Q1-TP002}, \citet{Q1-TP003}, and Euclid Collaboration: Altieri et al., in prep.

Q1 encompasses a range of data products. Of particular significance for this study are the photometric catalogues generated by the Euclid MERge Processing Function \citep[MER,][]{Q1-TP004}, which include aperture flux measurements with the corresponding errors, quality flags, and morphological information, as well as template fit and Sérsic fit fluxes in each band for all sources detected in the EDFs. Moreover, Q1 includes imaging \citep{Q1-TP002,Q1-TP003} and spectroscopic data \citep{Q1-TP006,Q1-TP007}, as well as physical parameter estimations \citep{Q1-TP005}. 

All EDFs have been observed by the four \Euclid photometric bands, i.e.,\ \IE from VIS in the visible \citep{EuclidSkyVIS}, and \YE, \JE, and \HE from NISP in the NIR \citep{Schirmer-EP18,EuclidSkyNISP}. These measurements are accompanied by ground-based optical photometry taken with the $ugriz$ bands of various instruments, including the Ultraviolet Near-Infrared Optical Northern Survey (UNIONS, Gwyn et al. in prep.) and the Dark Energy Survey \citep{Abbott_2018ApJS..239...18A}, that are re-processed through the official \Euclid pipelines and homogenised by MER. The breakdown of the photometry available for each EDF and their corresponding instrument can be found in \cite{Q1-TP001}. 

Additional to the photometric information provided by the \Euclid catalogues, Q1 also provides spectroscopic catalogues. These data are obtained from NISP-S observations in two red grisms (RGS000 and RGS180) covering the 1206--1892\,nm wavelength range. The data reduction process, spectral extractions and data specifics for Q1 spectroscopy are described in \citet{Q1-TP006} and \citet{Q1-TP007}. In this work, we focus on the \Euclid MER photometric catalogues to investigate the source populations present in the Q1 data.

\subsubsection{Quality flags and data cleaning}\label{sc:quality}

\begin{table*}
  \caption{\label{Flags} Number of sources in each EDF and impact of quality cuts.} 
  \centering
  \newcolumntype{.}{D{.}{.}{.}{10}}
\begin{tabular}{c r r r r r}  
    \hline\hline
    \noalign{\vskip 1pt}

    Field & Q1 Catalogue&Good flags \tablefootmark{a} &\multicolumn{3}{c}{Magnitude bins}\\

    &&& \multicolumn{1}{c}{$18<$\IE $\leq 21$}&\multicolumn{1}{c}{$21<$ \IE $\leq22$}&\multicolumn{1}{c}{$22<$ \IE $\leq24.5$} \\
    \hline
    \noalign{\vskip 1pt}
    EDF-N & $11\,378\,352$ & $7\,573\,476$ & 115\,606  & 192\,722& 2\,568\,146  \\
    EDF-S & $13\,060\,965$& $8\,913\,816$  & 127\,649 & 217\,375& 3\,056\,362 \\
    EDF-F & $5\,328\,489$ & $3\,705\,597$ & 51\,294 & 88\,629 & 1\,326\,915\\
    \hline 
  \end{tabular} 
\tablefoot{
\tablefoottext{a}{These are the `quality-filtered' catalogues.}
}
\end{table*}
The Q1 photometric catalogues include a number of artefacts, easily identified through a series of flags that are provided as data models. For instance, the reference photometric measurement of a source is given by the \verb|FLUX_DETECTION_TOTAL| column, and the reliability of this measurement can be assessed using the binary \verb|DET_QUALITY_FLAG| column. With this flag, a source can be identified to be contaminated by close neighbours, bad pixels, blending with other sources, saturation, being close to a CCD border, being within the VIS or NIR bright star masks, being within an extended object area, or being skipped by the de-blending algorithm. The \verb|DET_QUALITY_FLAG| is the most informative flag we use to clean the data from the contaminants listed above. Nevertheless, several other flags can also be used to detect contamination in specific bands (i.e.,\ using \verb|<band>_FLAG|) or contamination by spurious sources (\verb|SPURIOUS_FLAG|).

In the process of constructing our AGN catalogue from the existing MER Q1 catalogues, we retain only those sources that meet our `good flags' criteria,
\begin{itemize}
    \item \verb|SPURIOUS_FLAG = 0|\;, 
    \item \verb|<band>_FLAG = 0|\;, 
    \item \verb|DET_QUALITY_FLAG  = 0|| \verb|2|| \verb|512|\;,
\end{itemize}
where \verb|DET_QUALITY_FLAG| values of $0$, $2$, and $512$ indicate no problems with the data, sources blended together, and sources within an extended object area, respectively.

By applying this `good flags' method, we exclude approximately $32\%$ of the data, resulting in what we from now on refer to as the `quality-filtered' catalogues.
Furthermore, considering the varying magnitude limits of the external catalogues we use, we refine the data by dividing them into three magnitude bins: $18<\IE\leq21$, $21<\IE\leq22$, and $22<\IE<24.5$. A detailed breakdown of the number of sources left after these cleaning steps and splitting of the data is provided in \cref{Flags}.


\subsection{Photometry}\label{sc:matchedphot}

In the following sections we discuss the multi-wavelength photometric data used to identify the different source populations, and the counterpart associations performed in this work, ordered by descending energy. Positional matches with the external surveys were performed using the STIL Tool Set \citep[{\it STILTS} version 3.5-1,][]{Taylor_2006_2006ASPC..351..666T}, which is a package for command-line processing of tabular data, such as astronomical tables. The matches for the three Q1 fields were tailored to account for their different survey coverages. \cref{Table:matches_breakdown} indicates the data sets matched to each EDF, the numbers of sources per data set that fall within the Q1 fields, and the number of counterparts found for the quality-filtered versions of the Q1 catalogues.

\begin{table*}
  \caption{\label{Table:matches_breakdown} Surveys cross-matched per \Euclid field. } 
  \centering
\begin{minipage}{0.63\textwidth}
  \newcolumntype{.}{D{.}{.}{.}}
\begin{tabular}{c c r r r r}  
    \hline\hline
    \noalign{\vskip 1pt}
    Field & Data set & Sources in field & No. matches & Reference \\
    \hline 
    \noalign{\vskip 1pt}
    EDF-N   &  GALEX & 225\,685 &  52\,663 & This work\\
            & \gaia & 192\,109 & 43\,253 & \citetalias{Q1-TP004}\\
            & WISE-AllWISE & 487\,397 & 266\,029 & This work\\
            & \textit{Spitzer} & 11\,378\,352 &  7\,573\,476 & \citetalias{Q1-SP011}\\
            & DESI & 110\,459 & 24\,922 & This work\\
            & SDSS & 326 & 18 & This work\\
        
    \hline
    \noalign{\vskip 1pt}
    EDF-S       & GALEX & 225\,268 &  58\,093 &This work\\
            & \gaia & 130\,647 &  35\,263 & \citetalias{Q1-TP004}\\
            & DES & 4\,258\,555 & 3\,197\,960 &This work\\
            & WISE-AllWISE & 352\,135 & 268\,281  &This work\\
            & \textit{Spitzer} & 11\,378\,352 &  8\,913\,816&\citetalias{Q1-SP011}\\

    \hline
    
    \noalign{\vskip 1pt}
    EDF-F  & GALEX & 778\,194 &  230\,466  & This work\\
        & \gaia & 46\,500 &  14\,682 &\citetalias{Q1-TP004}\\
        & DES & 4\,258\,555 &  1\,330\,109  &This work\\
        & WISE-AllWISE & 232\,079 &  147\,389  &This work\\
        & \textit{Spitzer} & 11\,378\,352 &  3\,705\,597  & \citetalias{Q1-SP011}\\

    \hline 
  \end{tabular}
\end{minipage}
\tablefoot{
The reported matches are between the external survey sources and the quality-filtered \Euclid catalogues (i.e.,\ only good quality flags).}
\end{table*}

\subsubsection{Ultra-Violet}\label{UV-dataset}

The UV regime offers insights into some of the most active processes in the Universe that are not observable with optical bands. In this energy range, the EDFs overlap with NASA's Galaxy Evolution Explorer \citep[GALEX,][]{Bianchi_1999_1999MmSAI..70..365B}. GALEX imaged the sky in two ultraviolet bands: the far-UV (FUV, $\lambda_\mathrm{eff} = 1528$\,\AA); and the near-UV (NUV, $\lambda_\mathrm{eff} \sim 2310$\,\AA). It provided the first UV sky surveys using two observing modes: direct imaging and grism field spectroscopy. GALEX achieved an image full width half maximum (FWHM) of 4\farcs2 in the FUV and 5\farcs3 in the NUV. The GALEX GR6/7 data release \citep{Bianchi_2017_2017ApJS..230...24B} includes millions of source measurements, mostly from the All-Sky Imaging Survey (AIS), with a 5$\sigma$ limiting magnitude of about 20 in FUV and $\sim$\,21 in NUV. In this work, we use the combined \texttt{photoobj} catalogue, which includes all GALEX programmes: AIS, Medium Imaging survey (MIS), and the Deep Imaging Survey (DIS). GALEX has a lower angular resolution compared to \Euclid's VIS point spread function (PSF) FWHM of 0\farcs13 \citep{EuclidSkyOverview}. Therefore, when matching between \Euclid and GALEX, we set the fixed error radius in \verb|STILTS| to a conservative value of 1\farcs5. This allows for a more flexible matching, which in itself is important because sources that might have appeared as blended for the GALEX survey, can potentially be disentangled with \Euclid's resolution. The total number of matches we obtain between GALEX and the quality-filtered catalogues is 341\,222 (see \cref{Table:matches_breakdown} for the breakdown of matches per EDF).

\subsubsection{Optical}

Historically, optical surveys have been significant for identifying and cataloguing a vast number of sources, therefore enhancing our knowledge of the Universe and the populations found within it.
In this energy range, all three EDFs overlap with ESA's \gaia mission \citep{GaiaCollaboration_2016_2016A&A...595A...1G}. \gaia, with its PSF FWHM of 0\farcs{1}, aims to measure the three-dimensional spatial and the three-dimensional velocity distribution of stars in order to map and understand the formation, structure, and evolution of our Galaxy. The most recent \gaia Data Release 3 \citep[DR3,][]{GaiaCollaboration_2023_2023A&A...674A...1G} provides comprehensive source lists that include celestial positions, proper motions, parallaxes, and broadband photometry in the \textit{G}, {\gbp} (330--680\,nm), and {\grp} (630--1050\,nm) passbands, with a limiting depth of $G \approx 21$. Additionally, it offers astrophysical parameters and source class probabilities, including stars, galaxies, and quasars (QSOs) over the entire sky. The Q1 catalogues include \gaia IDs from matches performed within the \Euclid pipeline, which are released as part of the overall Q1 products. The matching performed between these two surveys is explained in \citet{Q1-TP004}, hereafter referred to as \citetalias{Q1-TP004}. The number of identified matches between the quality-filtered \Euclid catalogues and \gaia is 93\,198, the breakdown of which is reported in \cref{Table:matches_breakdown}.

Moreover, the EDF-S and EDF-F share coverage with the Dark Energy Survey \citep[DES,][]{DES_2005}, which is a ground-based visible and near-infrared imaging survey, aiming to cover 5000\,deg$^2$ of the southern high Galactic latitude sky. The second DES large data release \citep[DR2,][]{Abbott_2021_2021ApJS..255...20A} contains co-added images and source catalogues, as well as calibrated single-epoch CCD images, from the processing of all six years of DES wide-area survey observations in five broad photometric bands, $grizY$ \citep{Kessler_2015_2015AJ....150..172K} and all five years of DES supernova survey observations in the $griz$ bands \citep{DES:2019rtl}, with a detection limit of $g<25$ and a PSF FWHM typically around 0\farcs{8} \citep{Abbott_2021_2021ApJS..255...20A}. To perform the counterpart associations between \Euclid and DES, after investigating different fixed error radii based on the PSF FWHM of both surveys, we set this parameter to 0\farcs55 and obtain a total of 4\,528\,069 matches (see \cref{Table:matches_breakdown} for the breakdown of matches per EDF).

\subsubsection{Infrared}
The infrared regime provides valuable insights into regions of the Universe that are obscured by dust. In this energy range, the Q1 photometry can be combined with various surveys to enhance our understanding of obscured sources.

In the mid-infrared (MIR), the Wide-field Infrared Survey Explorer \citep[WISE,][]{Wright_2010_2010AJ....140.1868W} is a survey mapping the entire sky in four infrared bands (i.e.,\ \rm{W}1, \rm{W}2, \rm{W}3, \rm{W}4) centred at 3.4, 4.6, 12, and 22\,\micron. Its AllWISE programme combined data from the WISE cryogenic and NEOWISE post-cryogenic survey \citep{Mainzer_2011_2011ApJ...731...53M} to form the most comprehensive view of the full mid-infrared sky currently available. The AllWISE Data Release, mapping the entire sky and therefore including all three EDFs, provides images with a pixel scale of 1\farcs375, source catalogues, multi-epoch photometry catalogues, and reject catalogues up to a detection limit of $W1$<17.1. Similar to GALEX, the WISE-AllWISE resolution is not as powerful as that of \Euclid, having a PSF with FWHM of 6\farcs1, 6\farcs8, 7\farcs4, and 12\arcsec, for its four \rm{W}1, \rm{W}2, \rm{W}3, and \rm{W}4 bands \citep{Wright_2010_2010AJ....140.1868W} Therefore, when matching between the two surveys, we decided to set the fixed error radius to a conservative value of 1\farcs5, allowing for a more flexible matching and resulting in a total number of 681\,699 matches, the breakdown of which is reported in \cref{Table:matches_breakdown}.

To further complement the \Euclid catalogues, \citet{Q1-SP011}, from now on referred to as \citetalias{Q1-SP011}, performed forced photometry on \text{Spitzer} IRAC images at the position of the \Euclid sources (i.e.,\ fixed positions). Briefly, starting from the public images by \citet{Moneti-EP17} in all four IRAC bands, which include the [3.6], [4.5], [5.6], and [8.0] filters, they first remove the sky background, using a 3\,$\times$\,3 pixel filter. Then, the extraction is performed using the position of all \Euclid sources, both VIS- and NISP- detected, using an aperture with $1\arcsec$ radius, resulting in IRAC aperture photometry, which they correct to total, for every \Euclid source, therefore making the counterpart association unnecessary.


\subsection{Spectroscopy}\label{sc:spec}

The following sections discuss the spectroscopic data used to identify the different source populations, and the counterpart associations performed in this work.

\subsubsection{DESI EDR}

The Dark Energy Spectroscopic Instrument (DESI) is a multi-object fibre spectrograph installed at the Mayall-Telescope at Kitt Peak \citep{DESI_inst}, capable of covering a 3$^{\circ}$.2 wide field of view \citep{DESI_focalplane}. In preparation for its ambitious main survey, a set of survey validation projects \citep{DESI_validation} were conducted with DESI to optimise the final target selection and explore the capabilities and limits of the instrument. These tests consisted of a commissioning data set and a series of survey validations (SV) 1, 2, and 3 \citep{Alexander_2023_2023AJ....165..124A, Brodzeller_2023_2023AJ....166...66B, Guy_2023_2023AJ....165..144G, Lan_2023}. The data collected during these pilot surveys were published as the early data release (EDR) of DESI \citep{DESI_EDR}. It contains spectra of 2\,847\,435 unique pointings (including sky), which yield 1\,202\,846 reliable extragalactic spectroscopic redshifts.

Since the DESI survey is based on optical ground-based data, it has a spatial resolution of roughly 1\arcsec. For its counterpart association we start by investigating the entire DESI EDR catalogue \citep{DESI_EDR}, to which we apply the following set of selection criteria to obtain a subsample of objects with robust spectroscopic redshifts: 
\begin{itemize}
    \item \texttt{objtype = TGT};
    \item \texttt{deltachi2 $>$ 10};
    \item \texttt{zcat\_primary = 1};
    \item \texttt{coadd\_fiberstatus = 0};
    \item \texttt{zwarn$<$ 4}. 
\end{itemize}

\noindent Furthermore, only spectra with good model fits and no serious issues with the redshift determination are used. 

Out of the three EDFs, only the EDF-N overlaps with the DESI EDR. To obtain the counterparts, we set the maximum error to 1\arcsec\, based on DESI's spatial resolution, and obtain a total number of 64\,039 matches with the raw MER catalogues. After applying the quality cuts specified in \cref{sc:quality}, this number is reduced to 24\,922 matches between DESI EDR and the quality-filtered EDF-N catalogue.  

All the matched DESI spectra belong to the SV3 sample of the DESI EDR, which was covered by a larger number of passes than the yet to be released main survey of DESI. This means that these regions have a higher completeness and at times, due to stacking, deeper observations than what we can expect from the DESI main survey.

\subsubsection{SDSS DR17}

The Sloan Digital Sky Survey \citep[SDSS,][]{York_2000_2000AJ....120.1579Y} is a large-scale imaging and multi-fibre spectroscopic redshift survey that has mapped millions of objects from our Galaxy to the distant Universe, including stars, galaxies, and quasars. SDSS' Data Release 17 (DR17) marks its fifth and final release from the fourth phase \citep{Abdurro'uf_2022_2022ApJS..259...35A}. DR17 contains the entire release of the Mapping Nearby Galaxies at APO Survey \citep[MaNGA,][]{Bundy_2015_2015ApJ...798....7B}, as well as the MaNGA Stellar Library, and the complete release of the Apache Point Observatory Galactic Evolution Experiment 2 survey \citep[APOGEE,][]{Majewski_2017_2017AJ....154...94M}. Moreover, DR17 also includes data from the SPectroscopic IDentification of ERosita Survey subsurvey \citep[SPIDERS,][]{Clerc_2016_2016MNRAS.463.4490C, Dwelly_2017_2017MNRAS.469.1065D} and the eBOSS-RM programme\citep{Shen_2015_2015ApJS..216....4S}, as well as $25$ new or updated value-added catalogues, covering a total of 14\,555\,deg$^2$ with an average PSF FWHM, typically measured in the $r$ band, of around 1\farcs{3}, and an approximate magnitude limit of around $r=22.7$ \citep{Blanton_2017_2017AJ....154...28B, Abdurro'uf_2022_2022ApJS..259...35A}. DR17 includes approximately 1.5 million unique spectra for sources, with available spectroscopic redshifts. The SDSS DR17 overlaps solely with the EDF-N, and even then, only a limited number of sources fall within this area. Nevertheless, we conduct a cross-match with the quality-filtered EDF-N catalogue using a fixed radius of $0\farcs5$, based on SDSS' PSF FWHM, and obtain a total of $18$ counterparts.

\subsubsection{Other spectroscopic surveys}\label{sc:spec-wp2}

\gdr{3} announced a sample of 6.6 million quasar candidates \citep[the \texttt{qso\_candidates} table\footnote{The \gdr{3} quasar candidate catalogue is available at the \gaia archive \url{https://gea.esac.esa.int/archive} with table name \texttt{gaiadr3.qso\_candidates}.};][]{GaiaCollaboration_2023_2023A&A...674A...1G,GaiaCollaboration_2023_2023A&A...674A..41G}, which has high completeness thanks to the combination of several different classification modules, including the Discrete Source Classifier (DSC), the Quasar Classifier (QSOC), the variability classification module, the surface brightness profile module, and the \gdr{3} Celestial Reference Frame source table. Nevertheless, the \gdr{3} QSO candidate catalogue has an estimated low purity of quasars (52\%) and a large scatter of redshift estimates.

Instead of using the original \gdr{3} QSO candidates catalogue, we take a purified version to find \Euclid counterparts of the sources. This purified catalogue includes: (i) Quaia \citep{Storey-Fisher2024}, with nearly 1.3 million sources at $G<20.5$; (ii) CatNorth \citep{Fu_2024_2024ApJS..271...54F}, with more than 1.5 million sources down to the \gaia limiting magnitude in the 3$\pi$ sky of the Pan-STARRS1 \citep[PS1;][]{Chambers2016_PanSTARRS} footprint ($\delta>-30\degr$); and (iii) CatSouth (Fu et al., in prep), with 0.9 million sources with $G<21.0$ covered by the fourth data release (DR4) of the SkyMapper Southern Survey \citep[SMSS; $\delta\lesssim 16\degr$;][]{Onken_2024_2024PASA...41...61O}. The compilation of the three catalogues contains more than 1.9 million unique (with unique \gaia \verb|source_id|) quasar candidates in the entire sky. This catalogue from now on will be referred to as the `purified' GDR3 QSO candidate sample (GDR3-QSOs). We cross-match the \Euclid fields with this integrated GDR3-QSOs catalogue using the Q1 provided \gaia IDs and find 647, 811, and 513 matches in quality-filtered EDF-N, EDF-S, and EDF-F, respectively.



\section{Identified populations}
\label{sc:qso}


In this section, we present the two main populations identified in the \Euclid data for this work: stars and AGN. While the primary objective is to investigate various diagnostics for compiling a comprehensive AGN candidate catalogue, stars significantly contribute to the contamination of AGN selection techniques. Therefore, developing an effective selection method to identify the stellar population within the Q1 data is crucial.

To identify AGN, we use traditional colour selection techniques and investigate new colour diagnostics using the limited labelled data obtained after cross-matching the \Euclid catalogues. Additionally, we refer to other Q1 papers that also explore AGN detection techniques. All these approaches include using a combination of \Euclid's photometry, photometric and spectroscopic information from the matched data sets, spectral energy distributions (SEDs) fitting, morphological analysis, and machine-learning techniques.

However, despite the cross-matching between the various data sets and \Euclid, we lack a substantial number of reliably labelled sources, excluding those from DESI. Consequently, we are unable to accurately quantify the purity and completeness of some of these methods. Follow-up work is necessary to further test the methodologies described here and to accurately quantify these parameters.

For the AGN diagnostics based on \Euclid's photometry, we use the template-fit fluxes provided by the Q1 catalogues. These fluxes are colour corrected, following the prescription outlined in \citet{Q1-TP004}. 

Additionally, we examined the impact of correcting the fluxes for Galactic extinction using $E(B-V)$ values from the Galactic dust map from \citet{PlanckCollaboration_2014_2014A&A...571A..11P}. As shown in \citet{Galametz-2017A&A...598A..20G}, the colour of the source SED can lead to significant correction variations. In this work we provide extinction corrections for two extreme cases, blackbody temperatures of 100\,000\,K and 5700\,K. As the Galactic latitudes of the EDFs were selected to be in regions of low Galactic extinction, we find that these corrections have a minimal effect on the template-fit fluxes (approximately 5\% variation). Consequently, we decided not to implement these corrections.

\subsection{Stellar candidates}\label{subsect:star_contaminants}

The proper motion and parallax of an object track its apparent transverse movement over time, as well as its shift in position against a distant background when viewed from different angles. Most stars within our Galaxy show measurable proper motions and parallaxes due to their proximity to us. Vice versa, distant objects like quasars or galaxies have negligible proper motions and parallaxes. Therefore, tracking these two parameters is crucial when attempting to identify stars. 

\gdr3 provides parallaxes and proper motions for around 1.46 billion sources, with a limiting magnitude of about $G\approx21$ and a bright limit of about $G\approx3$ \citep{GaiaCollaboration_2023_2023A&A...674A...1G}. By identifying the \gaia counterparts and utilising the information provided by \gdr{3}, we have the necessary data to identify as stellar candidates those sources with significant proper motion and parallaxes. However, since the detection limit of \gaia ($G<21$) is not as deep as that of \Euclid (\IE$\leq 26.2$), for sources beyond $G \geq 21$, an alternative method is required for detecting stellar candidates.

The Q1 data release also includes catalogues with object classifications that provide the probabilities of an object being a star, a galaxy, or a QSO based on the source's photometry. This is obtained by performing a supervised machine-learning method called \verb|Probabilistic Random Forest| \citep[PRF,][]{Reis_2019_2019AJ....157...16R}. Specifics on the method used can be found in \cite{Q1-TP005}, hereafter referred to as \citetalias{Q1-TP005}. To summarise, the classifiers estimate the probability of objects belonging to a particular class and set a threshold that must be surpassed for an object to be classified into one of the groups. The advised threshold for objects to be considered as stars differs between EDF-N, EDF-S, and EDF-F, with values of 0.58, 0.68, and 0.68, respectively. Upon investigating the data, we decided that a threshold of 0.7 for all three fields provides a purer sample of stellar candidates

By combining information from both \gaia and the Q1 PRF, we are able to construct a comprehensive approach for identifying stellar candidates. However, to refine the star selection and ensure that no extended objects are recorded to have large proper motions and/or parallaxes, or are misclassified by the random forest, we impose an additional condition on the morphology to select only star-like objects. Therefore, only point-like sources (i.e., \verb|MUMAX_MINUS_MAG|$\;<-2.6$) are considered stellar candidates \citep{Q1-TP004}. 

\begin{figure}[htbp!]
\centering
\includegraphics[angle=0,width=1.0\hsize]{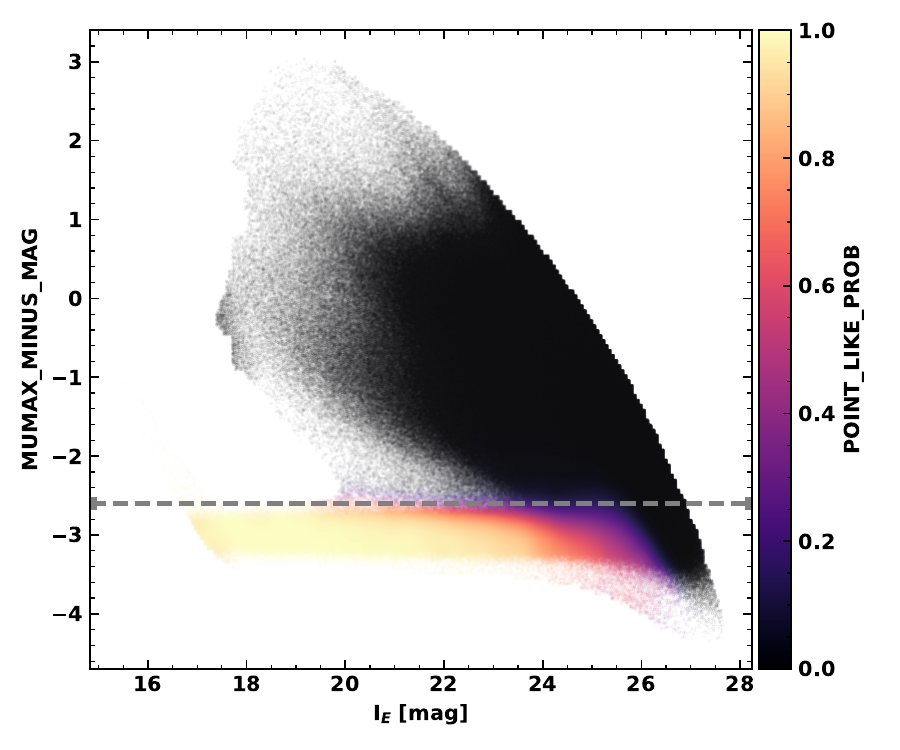}
\caption{\texttt{MUMAX\_MINUS\_MAG} versus \IE for sources in the EDF-N. The colour scale indicates the point like probability of a source. The dotted line indicates the threshold (\texttt{MUMAX\_MINUS\_MAG}$\;<-2.6$) below which most sources appear to be point-like.}
\label{fig:mumax_minus_mag_dataset}
\end{figure}

The parameter \verb|MUMAX_MINUS_MAG| is the difference between two quantities, both available in the Q1 catalogue: a global measure, named \verb|MAG_STARGAL_SEP|; and a local one, named \verb|MU_MAX|. The first is the magnitude used to compute point-like probability, and the second is the peak surface brightness above the background detection level. \verb|MU_MAX| measures the brightest, most concentrated light within a specific area of the source. The \verb|MUMAX_MINUS_MAG| is useful to identify point-sources or nearly point-like sources \citep{Q1-TP004}. \
\Cref{fig:mumax_minus_mag_dataset} indicates how this cut in the data set (\verb|MUMAX_MINUS_MAG|$<-2.6$) is able to capture those sources that have a high probability of being a point-like source.

As a result, we adopt the following prescription to select stellar candidates,
\begin{equation}
\label{eq:stars}
    \begin{aligned}
      & \verb|MUMAX_MINUS_MAG| \leq -2.6 \; \wedge \\
        & \quad
        \begin{cases}
            \sqrt{\displaystyle\left( \frac{\varpi}{\sigma_{\varpi}}\right)^2}>5 \vee 
            \sqrt{\displaystyle\left(\frac{\mu_{\alpha *}}{\sigma_{\mu_{\alpha *}}}\right)^2+\displaystyle\left(\frac{\mu_{\delta}}{\sigma_{\mu_{\delta}}}\right)^2} >5 
            & \text{for } G < 21, \\
            \verb|phz_star_prob| > 0.7 & \text{for } G \geq 21,
        \end{cases}
    \end{aligned}
\end{equation}
where $\varpi$ stands for the parallax of an object, $\sigma_{\varpi}$ the error of this measurement, and ($\mu_{\alpha *}$, $\mu_{\delta}$) are the proper motion measured in the right ascension and declination positions, with their corresponding errors, ($\sigma_{\mu_{\alpha *}}$, $\sigma_{\mu_{\delta}}$). We refer to these selected sources as \verb|stellar_candidates|. \Cref{Table:star} gives a summary of the number of stellar candidates that we identify in each of the three Q1 fields.

\begin{table}
  \caption{\label{Table:star} Number of selected stellar candidates per Q1 fields. } 
  \centering
  \newcolumntype{.}{D{.}{.}{.}}
\begin{tabular}{c r r r}  
    \hline\hline
    \noalign{\vskip 1pt}
    Field & \multicolumn{3}{c}{No. stellar candidates} \\
    &\multicolumn{1}{c}{18< \IE $\leq$21}&\multicolumn{1}{c}{21<\IE$\leq$22}&\multicolumn{1}{c}{22< \IE $\leq$24.5} \\
    \hline
    \noalign{\vskip 1pt}
    EDF-N & 45\,700 & 43\,918 & 84\,156 \\

    EDF-S & 30\,856& 27\,173 & 63\,589\\

    EDF-F & 10\,159& 7\,818 & 17\,163\\
    \hline 
  \end{tabular} 
\tablefoot{
All sources must satisfy the conditions set by \cref{eq:stars}. The numbers reported are based on the quality-filtered catalogues.}
\end{table}

\subsection{AGN candidates: photometric selection}\label{subsec:AGNphot}

We now present different AGN selections applied to the Q1 fields. This initial section examines previously established photometric criteria and introduces novel photometric diagnostics developed for this work. We report the number of AGN candidates recorded per criterion in \cref{Table:AGN_candidates_photometry}.

\begin{sidewaystable*}
\newlength{\tabcs}
\setlength{\tabcs}{4mm}
\caption{\label{Table:AGN_candidates_photometry}Number of selected AGN candidates per criteria.} 
  \centering
 	\scalebox{0.98}{
\begin{tabular}{c@{\hskip\tabcs}r@{\hskip\tabcs}r@{\hskip\tabcs}r@{\hskip\tabcs}r@{\hskip\tabcs}r@{\hskip\tabcs}r@{\hskip\tabcs}r@{\hskip\tabcs}r@{\hskip\tabcs}r@{\hskip\tabcs}r}
    \hline\hline
    \noalign{\vskip 1pt}
    Data& Diagnostic & \multicolumn{3}{c}{EDF-N} & \multicolumn{3}{c}{EDF-S} & \multicolumn{3}{c}{EDF-F} \\
               && \multicolumn{1}{c}{18< \IE $\leq$21}&\multicolumn{1}{c}{21< \IE $\leq$22}&\multicolumn{1}{c}{22< \IE $\leq$24.5} & \multicolumn{1}{c}{18< \IE $\leq$21}&\multicolumn{1}{c}{21< \IE $\leq$22}&\multicolumn{1}{c}{22< \IE $\leq$24.5} & \multicolumn{1}{c}{18< \IE $\leq$21}&\multicolumn{1}{c}{21< \IE $\leq$22}&\multicolumn{1}{c}{22< \IE $\leq$24.5} \\
    \hline
    \noalign{\vskip 1pt}
    Photometry & PRF\tablefootmark{a}  & \multicolumn{1}{c}{\phantom{0\,}138}&\multicolumn{1}{c}{1\,882} &\multicolumn{1}{c}{44\,429}& \multicolumn{1}{c}{\phantom{0\,}161}&\multicolumn{1}{c}{\phantom{0}2\,838} & \multicolumn{1}{c}{37\,869}& \multicolumn{1}{c}{\phantom{0\,0}59}&\multicolumn{1}{c}{1\,476}&\multicolumn{1}{c}{17\,694} \\
    
     &B24A\phantom{\tablefootmark{a}}         & \multicolumn{1}{c}{1\,085}&\multicolumn{1}{c}{2\,038} & \multicolumn{1}{c}{12\,146}&\multicolumn{1}{c}{1\,210} & \multicolumn{1}{c}{\phantom{0}2\,113}&\multicolumn{1}{c}{13\,155}&\multicolumn{1}{c}{\phantom{0\,}480}&\multicolumn{1}{c}{\phantom{0\,}886}&\multicolumn{1}{c}{\phantom{0}4\,690} \\
    
     &B24B\phantom{\tablefootmark{a}}         & \multicolumn{1}{c}{\phantom{0\,}658}&\multicolumn{1}{c}{\phantom{0\,}997} & \multicolumn{1}{c}{\phantom{0}8\,178}&\multicolumn{1}{c}{\dots} & \multicolumn{1}{c}{\dots}&\multicolumn{1}{c}{\dots}&\multicolumn{1}{c}{\dots}&\multicolumn{1}{c}{\dots}&\multicolumn{1}{c}{\dots} \\
    
     &C75\phantom{\tablefootmark{a}}        & \multicolumn{1}{c}{1\,106}&\multicolumn{1}{c}{2\,081} & \multicolumn{1}{c}{10\,888}&\multicolumn{1}{c}{2\,149} & \multicolumn{1}{c}{\phantom{0}3\,379}&\multicolumn{1}{c}{17\,031}&\multicolumn{1}{c}{1\,278}&\multicolumn{1}{c}{1\,639}&\multicolumn{1}{c}{\phantom{0}7\,561} \\
    
     &R90\phantom{\tablefootmark{a}}         & \multicolumn{1}{c}{\phantom{0\,}498}&\multicolumn{1}{c}{\phantom{0\,}196} & \multicolumn{1}{c}{\phantom{00\,}405}&\multicolumn{1}{c}{\phantom{0\,}792} & \multicolumn{1}{c}{\phantom{00\,}347}&\multicolumn{1}{c}{\phantom{00\,}690}&\multicolumn{1}{c}{\phantom{0\,}399}&\multicolumn{1}{c}{\phantom{0\,}204}&\multicolumn{1}{c}{\phantom{00\,}579} \\

     &GDR3-QSOs\phantom{\tablefootmark{a}}          & \multicolumn{1}{c}{\phantom{0\,}606}&\multicolumn{1}{c}{\phantom{0\,0}35} & \multicolumn{1}{c}{\phantom{00\,00}3}&\multicolumn{1}{c}{\phantom{0\,}769} & \multicolumn{1}{c}{\phantom{00\,0}42}&\multicolumn{1}{c}{\phantom{00\,00}0}&\multicolumn{1}{c}{\phantom{0\,}442}&\multicolumn{1}{c}{\phantom{0\,0}68}&\multicolumn{1}{c}{\phantom{00\,00}0} \\

    &$JH$\_$I_{\text{E}}Y$\phantom{\tablefootmark{a}}        & \multicolumn{1}{c}{1\,084}&\multicolumn{1}{c}{1\,789} & \multicolumn{1}{c}{15\,599}&\multicolumn{1}{c}{1\,244} & \multicolumn{1}{c}{\phantom{0}2\,369}&\multicolumn{1}{c}{18\,236}&\multicolumn{1}{c}{\phantom{0\,}509}&\multicolumn{1}{c}{\phantom{0\,}980}&\multicolumn{1}{c}{\phantom{0}6\,888} \\
    
     &$I_{\text{E}}H$\_$gz$\phantom{\tablefootmark{a}}         & \multicolumn{1}{c}{\phantom{0\,}591}&\multicolumn{1}{c}{\phantom{0\,}909} & \multicolumn{1}{c}{\phantom{0}6\,802}&\multicolumn{1}{c}{\phantom{0\,}787} & \multicolumn{1}{c}{\phantom{0}1\,247}&\multicolumn{1}{c}{\phantom{0}9\,340}&\multicolumn{1}{c}{\phantom{0\,}400}&\multicolumn{1}{c}{\phantom{0\,}611}&\multicolumn{1}{c}{\phantom{0}4\,323} \\

    \hline
    \noalign{\vskip 1pt}
& Total &  \multicolumn{1}{c}{2\,547} & \multicolumn{1}{c}{6\,474}&\multicolumn{1}{c}{80\,184} & \multicolumn{1}{c}{3\,410}&\multicolumn{1}{c}{8\,505} & \multicolumn{1}{c}{80\,390}&\multicolumn{1}{c}{1\,697}&\multicolumn{1}{c}{3\,995}&\multicolumn{1}{c}{34\,854}\\
    \hline
    \noalign{\vskip 1pt}
    Spectroscopy & QSO\phantom{\tablefootmark{a}} & \multicolumn{1}{c}{\phantom{0\,}400}&\multicolumn{1}{c}{\phantom{0\,}484} & \multicolumn{1}{c}{\phantom{00\,}673}&\multicolumn{1}{c}{\dots} & \multicolumn{1}{c}{\dots}&\multicolumn{1}{c}{\dots}&\multicolumn{1}{c}{\dots}&\multicolumn{1}{c}{\dots}&\multicolumn{1}{c}{\dots}\\
                & Broad-line QSO\phantom{\tablefootmark{a}} & \multicolumn{1}{c}{\phantom{0\,}366}&\multicolumn{1}{c}{\phantom{0\,}460} & \multicolumn{1}{c}{\phantom{00\,}603}&\multicolumn{1}{c}{\dots} & \multicolumn{1}{c}{\dots}&\multicolumn{1}{c}{\dots}&\multicolumn{1}{c}{\dots}&\multicolumn{1}{c}{\dots}&\multicolumn{1}{c}{\dots}\\
    & Broad-line Galaxy\phantom{\tablefootmark{a}} & \multicolumn{1}{c}{\phantom{0\,0}57}&\multicolumn{1}{c}{\phantom{0\,0}48} & \multicolumn{1}{c}{\phantom{00\,}113}&\multicolumn{1}{c}{\dots} & \multicolumn{1}{c}{\dots}&\multicolumn{1}{c}{\dots}&\multicolumn{1}{c}{\dots}&\multicolumn{1}{c}{\dots}&\multicolumn{1}{c}{\dots}\\
                & NII BPT\phantom{\tablefootmark{a}}& \multicolumn{1}{c}{\phantom{0\,0}93}&\multicolumn{1}{c}{\phantom{0\,00}1} & \multicolumn{1}{c}{\phantom{00\,00}1}&\multicolumn{1}{c}{\dots} & \multicolumn{1}{c}{\dots}&\multicolumn{1}{c}{\dots}&\multicolumn{1}{c}{\dots}&\multicolumn{1}{c}{\dots}&\multicolumn{1}{c}{\dots}\\
                & SII BPT\phantom{\tablefootmark{a}} & \multicolumn{1}{c}{\phantom{0\,0}69}&\multicolumn{1}{c}{\phantom{0\,00}0} & \multicolumn{1}{c}{\phantom{00\,00}5}&\multicolumn{1}{c}{\dots} & \multicolumn{1}{c}{\dots}&\multicolumn{1}{c}{\dots}&\multicolumn{1}{c}{\dots}&\multicolumn{1}{c}{\dots}&\multicolumn{1}{c}{\dots}\\
                & OI BPT\phantom{\tablefootmark{a}} & \multicolumn{1}{c}{\phantom{0\,0}72}&\multicolumn{1}{c}{\phantom{0\,00}3} & \multicolumn{1}{c}{\phantom{00\,0}38}&\multicolumn{1}{c}{\dots} & \multicolumn{1}{c}{\dots}&\multicolumn{1}{c}{\dots}&\multicolumn{1}{c}{\dots}&\multicolumn{1}{c}{\dots}&\multicolumn{1}{c}{\dots}\\
                & WHAN\phantom{\tablefootmark{a}} & \multicolumn{1}{c}{1\,792}&\multicolumn{1}{c}{\phantom{0\,0}12} & \multicolumn{1}{c}{\phantom{00\,0}14}&\multicolumn{1}{c}{\dots} & \multicolumn{1}{c}{\dots}&\multicolumn{1}{c}{\dots}&\multicolumn{1}{c}{\dots}&\multicolumn{1}{c}{\dots}&\multicolumn{1}{c}{\dots}\\
                & Blue\phantom{\tablefootmark{a}} & \multicolumn{1}{c}{\phantom{0\,}104}&\multicolumn{1}{c}{\phantom{0\,0}47} & \multicolumn{1}{c}{\phantom{00\,}103}&\multicolumn{1}{c}{\dots} & \multicolumn{1}{c}{\dots}&\multicolumn{1}{c}{\dots}&\multicolumn{1}{c}{\dots}&\multicolumn{1}{c}{\dots}&\multicolumn{1}{c}{\dots}\\
                &  KEX\phantom{\tablefootmark{a}} & \multicolumn{1}{c}{\phantom{0}188}&\multicolumn{1}{c}{\phantom{0}110} & \multicolumn{1}{c}{\phantom{0\,0}467}&\multicolumn{1}{c}{\dots} & \multicolumn{1}{c}{\dots}&\multicolumn{1}{c}{\dots}&\multicolumn{1}{c}{\dots}&\multicolumn{1}{c}{\dots}&\multicolumn{1}{c}{\dots}\\

    \hline
    \noalign{\vskip 1pt}
    & Total &  \multicolumn{1}{c}{2\,250} & \multicolumn{1}{c}{\phantom{0\,}636}&\multicolumn{1}{c}{\phantom{0}1\,350} &\multicolumn{1}{c}{\dots} & \multicolumn{1}{c}{\dots}&\multicolumn{1}{c}{\dots}&\multicolumn{1}{c}{\dots}&\multicolumn{1}{c}{\dots}&\multicolumn{1}{c}{\dots}\\
    \hline
    \noalign{\vskip 1pt}
    Other Q1 selections & RW25\tablefootmark{b}         & \multicolumn{1}{c}{\phantom{0\,0}62}&\multicolumn{1}{c}{\phantom{0\,0}72} & \multicolumn{1}{c}{\phantom{00\,}210}&\multicolumn{1}{c}{\phantom{0\,}625} & \multicolumn{1}{c}{\phantom{00\,}425}&\multicolumn{1}{c}{\phantom{000\,}609}&\multicolumn{1}{c}{\phantom{0\,}447}&\multicolumn{1}{c}{\phantom{0\,}520}&\multicolumn{1}{c}{\phantom{0}1\,973} \\
    
    & SG25\tablefootmark{c} & \multicolumn{1}{c}{1\,831}&\multicolumn{1}{c}{2\,078} & \multicolumn{1}{c}{\phantom{0}1\,326}&\multicolumn{1}{c}{3\,435} & \multicolumn{1}{c}{\phantom{0}2\,678}&\multicolumn{1}{c}{\phantom{00}1\,609}&\multicolumn{1}{c}{1\,400}&\multicolumn{1}{c}{1\,012}&\multicolumn{1}{c}{\phantom{00\,}571} \\
                
    & MB25\tablefootmark{d} & \multicolumn{1}{c}{\phantom{0}494}&\multicolumn{1}{c}{2\,813} & \multicolumn{1}{c}{12\,779}&\multicolumn{1}{c}{\phantom{0}650} & \multicolumn{1}{c}{\phantom{0}3\,361}&\multicolumn{1}{c}{\phantom{0}15\,043}&\multicolumn{1}{c}{\phantom{0\,}232}&\multicolumn{1}{c}{1\,464}&\multicolumn{1}{c}{\phantom{0}6\,721} \\ 
    \hline
    \noalign{\vskip 1pt}
    & Total &  \multicolumn{1}{c}{2\,303} & \multicolumn{1}{c}{4\,566}&\multicolumn{1}{c}{13\,700} & \multicolumn{1}{c}{4\,179}&\multicolumn{1}{c}{\phantom{0}5\,691} & \multicolumn{1}{c}{\phantom{0}16\,169}&\multicolumn{1}{c}{1\,759}&\multicolumn{1}{c}{2\,471}&\multicolumn{1}{c}{\phantom{0}7\,325}\\
    \hline
    \noalign{\vskip 1pt}
    SED fitting & LB25\tablefootmark{e} & \multicolumn{1}{c}{2\,172}&\multicolumn{1}{c}{1\,218} & \multicolumn{1}{c}{\phantom{0}4\,166}&\multicolumn{1}{c}{\dots} & \multicolumn{1}{c}{\phantom{0}\dots}&\multicolumn{1}{c}{\phantom{00}\dots}&\multicolumn{1}{c}{\dots}&\multicolumn{1}{c}{\dots}&\multicolumn{1}{c}{\phantom{00\,}\dots} \\
    \noalign{\vskip 1pt}

    \hline

  \end{tabular}
  }
\tablefoot{
\tablefoottext{a}{PRF selected QSOs: \citet{Q1-TP005}.}
\tablefoottext{b}{X-ray selected AGN candidates: \citet{Q1-SP003}.}
\tablefoottext{c}{AGN candidates from diffusion models trained on VIS data: \citet{Q1-SP009}.}
\tablefoottext{d}{AGN candidates from deep learning trained on IllustrisTNG: \citet{Q1-SP015}.}
\tablefoottext{e}{AGN candidates with spectroscopic redshift from SED fitting: Euclid Collaboration: Laloux et al., in prep}}
\end{sidewaystable*}

\subsubsection{Probabilistic random forest}\label{subsec:random_forest}

As outlined in \cref{subsect:star_contaminants}, Q1 provides object classification catalogues with the probabilities of an object being a star, galaxy, or QSO \citepalias{Q1-TP005}. They also provide the probability threshold to use in order to select different populations of sources. In particular, the recommended thresholds for QSOs are 0.67, 0.85, and 0.85 for the EDF-N, EDF-S, and EDF-F, respectively. When investigating the data we decided that a threshold of 0.85 for the EDF-N, and 0.95 for the EDF-S and EDF-F created an overall more refined sample of QSOs. However, since the PRF is trained using photometry only, to avoid contaminants from stellar objects that might have QSO-like colours, we exclude the stars identified by \cref{subsect:star_contaminants}. We refer to this `purified' version of the \citetalias{Q1-TP005} catalogue as the PRF candidates.

The total number of identified QSO candidates with this recipe in the quality-filtered catalogues is 180\,666, and the breakdown per field is reported in \cref{Table:AGN_candidates_photometry}.

\subsubsection{Bisigello+24 selections} \label{subsect:B24_selection}

In preparation for the \Euclid mission, \citet{EP-Bisigello}, from now on referred to as \citetalias{EP-Bisigello}, identified various colour-colour selection criteria for AGN using \Euclid photometry alone, and combinations of \Euclid photometry with additional external photometric bands. Their study was carried out for both, the EWS and the EDS, using simulated data from the Spectro-Photometric Realisations of IR-Selected Targets at all-z \citep[\texttt{Spritz},][]{Bisigello2021}. Selection criteria were identified by maximising the F1-score -- the harmonic mean of the completeness ($C$, the fraction of true AGN recovered from the original AGN sample) and the purity ($P$, the fraction of true AGN among all AGN candidates). 

In this work, due to the depth of the MER catalogues, we use the diagnostics derived for the EWS. It is worth noting that all selection criteria developed in \citetalias{EP-Bisigello} assumed that stars had been previously selected and removed from the sample. Therefore, before applying any of their selections, we remove all stellar candidates identified in \cref{subsect:star_contaminants}.

We then apply their purest selection criterion in the EWS to identify QSO candidates based on three \Euclid filters. This selection provided them with an F1 score of $0.224 \pm 0.001$, and was derived from a low purity ($P = 0.166 \pm 0.015$), as well as a low completeness ($C = 0.347 \pm 0.004$). This particular selection follows the prescription 
\crefformat{equation}{#2Eq.~(#1)#3}
\begin{equation}
\label{eq:B24_2}
\begin{aligned}
    \left(\IE-\YE  <0.5 \right) & \land \left(\IE-\JE<0.7 \right) \\
    & \land \left[\IE-\JE<-2.1 (\IE-\YE)+0.9 \right].
\end{aligned}
\end{equation}
This criterion identifies a total of 521\,252 QSO candidates in the EDF-N, 755\,032 in the EDF-S, and 315\,683 in the EDF-F. 

In addition to stars, based on the large number of candidates obtained and the low purity of this selection, it is possible that other contaminants also affect this selection. Therefore, to improve the purity of this diagnostic, which was specifically designed to identify unobscured AGN, also referred to as Type I AGN, we impose an additional requirement of point-likeness to eliminate potential extended contaminants (i.e., \verb|MUMAX_MINUS_MAG| $\leq -2.6$). This criterion removes approximately 87\% of candidates in each field, indicating that many of the initial candidates were extended sources. However, by applying this additional morphological filter, we exclude several AGN at $z\leq 1.2$ that may appear as extended sources due to \Euclid's resolution.

\begin{figure}[htbp!]
\centering
\includegraphics[angle=0,width=1.0\hsize]{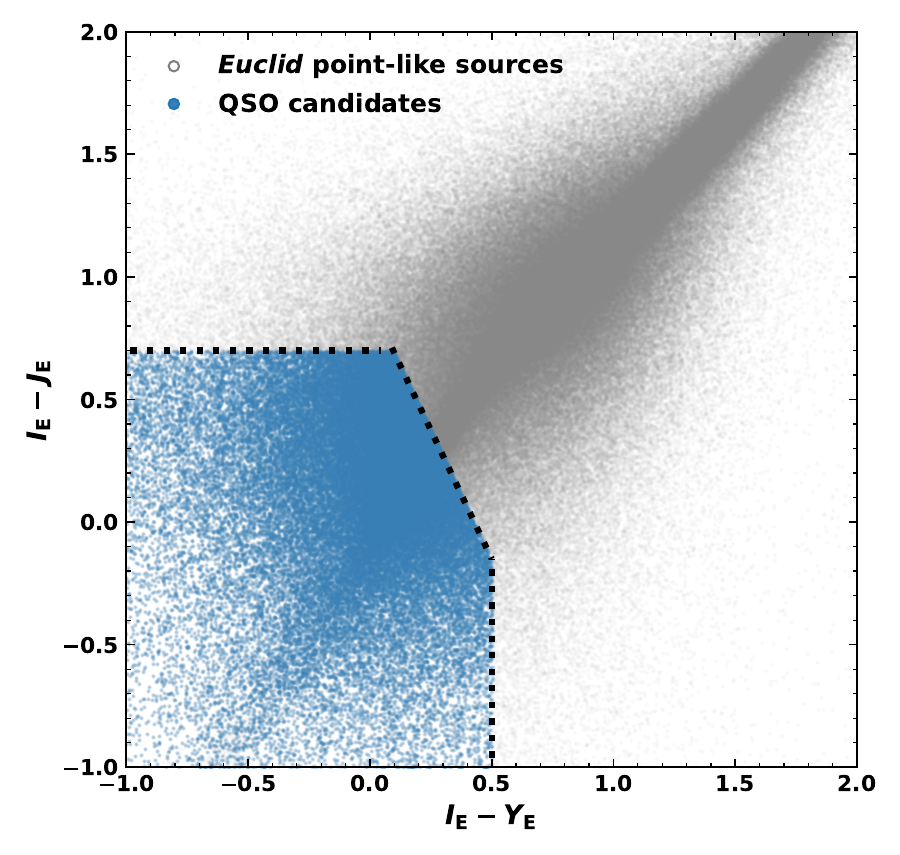}
\caption{B24A selection criteria applied to the EDF-N. In grey we show all quality-filtered \Euclid compact sources, while the QSO candidates are shown in blue. We also showcase, with the black dotted line, the limit of the B24A selection.}
\label{fig:bisigello+24_N}
\end{figure}

Although it is challenging to assess how this additional condition might enhance the purity of the selection, this approach is a straightforward method to further clean the candidate sample without deviating from the prescription established by \citetalias{EP-Bisigello}. The combination of selecting sources with \cref{eq:B24_2} and \verb|MUMAX_MINUS_MAG| $\leq -2.6$ will henceforth be referred to as selection `B24A'. The total number of QSO candidates obtained in the quality-filtered catalogues after applying the B24A selection is 211\,797, the breakdown of which is reported in \cref{Table:AGN_candidates_photometry}. Furthermore, \cref{fig:bisigello+24_N} illustrates the QSO candidates selected with B24A in the EDF-N, showcasing the colours of all quality-filtered compact sources in this colour-colour plane and highlighting the QSOs selected as candidates.

\crefformat{equation}{#2Eq.~#1#3}

\citetalias{EP-Bisigello} also included some additional diagnostics designed to combine the \Euclid photometry with that of other surveys. Specifically focussing on future surveys, such as Rubin/LSST, they characterised a series of criteria using the $ugriz$ bands. Since the \Euclid photometry is accompanied by ancillary ground-based optical photometry taken with these bands (details in \citetalias{Q1-TP004}), we opt to test the selection criteria derived using the $u$ and $z$ bands.

This specific selection follows the prescription
\begin{equation}\label{eq:B24_uz}
\begin{aligned}
    \left(\IE-\HE  <1.1\right) & \land \left(u-z <1.2\right)\\
    & \land \left[\IE-\HE<-1.3(u-z)+1.9\right], \\
\end{aligned}
\end{equation}
\crefformat{equation}{#2Eq.~(#1)#3}
and was designed to identify Type I AGN, assuming all stellar candidates had already been removed from the samples. For this selection, \citetalias{EP-Bisigello} obtained F1 $= 0.861 \pm 0.004$, with $C=0.813\pm0.011$ and $P=0.922\pm0.017$, making it the purest and most complete selection of their work.

The EDF-N is the only Q1 field that contains the $u$ band as part of the \Euclid ancillary data. Therefore we can only apply this selection to the EDF-N, obtaining a total of 1\,092\,763 QSO candidates.

\begin{figure}[htbp!]
\centering
\includegraphics[angle=0,width=1.0\hsize]{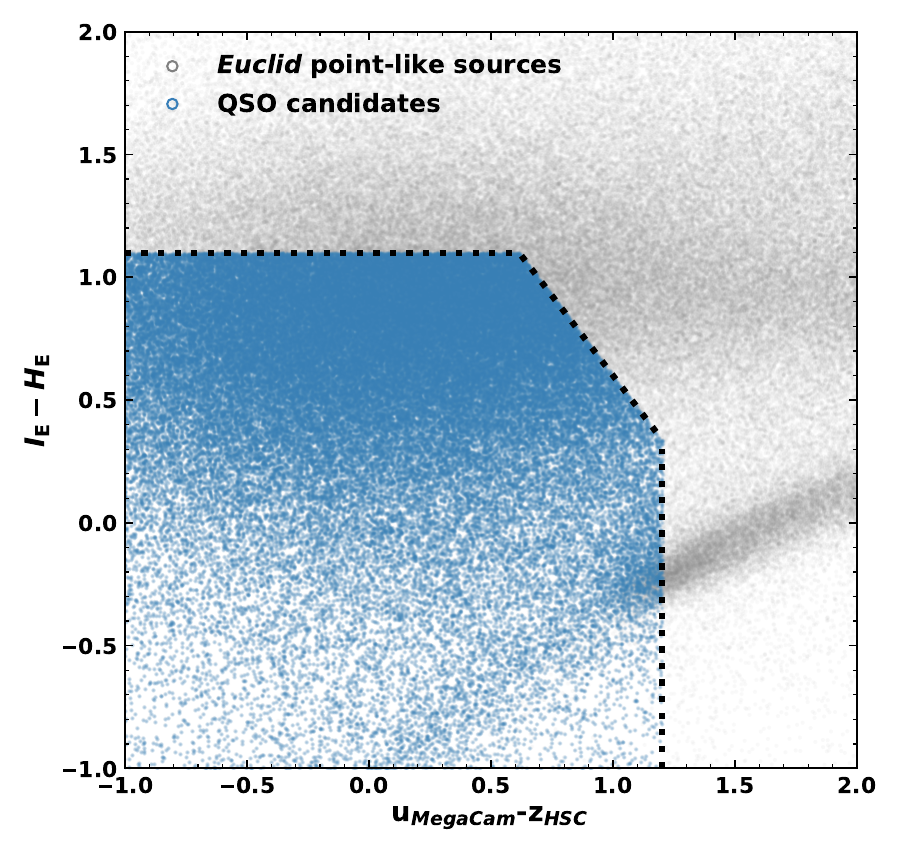}
\caption{B24B selection criteria applied to the EDF-N. In grey we show all quality-filtered \Euclid point-like sources, while the QSO candidates are shown in blue. We also showcase, with the black dotted line, the limit of the B24B selection.}
\label{fig:B24_ur}
\end{figure}

Similarly to B24A, we suspect that the large number of candidates might be attributed to potential contaminants infiltrating this selection. Consequently, driven by these other contaminants, we apply the same morphological cut to this criterion. We refer to the combination of \cref{eq:B24_uz} and \verb|MUMAX_MINUS_MAG| $\leq -2.6$ as the B24B selection. Applying this combination reduces the number of selected QSO candidates to 114\,145. We report this number, split into magnitude bins, in \cref{Table:AGN_candidates_photometry}. Additionally, \cref{fig:B24_ur} illustrates the QSO candidates selected with B24B in the EDF-N, showcasing the colours of all quality-filtered point-like sources and highlighting the QSOs selected as candidates. Once again, it is not possible to assess how the additional condition on morphology might impact the purity of the selection, but this approach provides a way of further cleaning the candidate sample without having to alter the \citetalias{EP-Bisigello} criterion.

\subsubsection{WISE-AllWISE selection}\label{subsect:WISE_selection}

Utilising the supplementary photometry derived from the WISE-AllWISE counterparts, we implement the selection criteria established by \citet{Assef_2018_2018ApJS..234...23A}, hereinafter called \citetalias{Assef_2018_2018ApJS..234...23A}. In their work, two distinct diagnostics, C75 and R90, were examined. As their names suggest, these criteria were designed to generate catalogues with 75\% completeness and 90\% reliability, whereby reliability measures the purity of the selection. The completeness-optimised AGN diagnostic is defined by
\begin{equation}
\label{eq:c75}
\centering
    \rm{W}1 - \rm{W2} > 0.71,
\end{equation}
\crefformat{equation}{#2Eq.~(#1)#3}
where the completeness fractions for a given $W1-W2$ colour cut are independent of magnitude. Meanwhile, the reliability driven AGN selection takes the form,
\begin{equation}
\label{eq:r90}
\begin{aligned}
    \rm{W}1 - \rm{W}2 > \begin{cases}
      0.65\, \mathrm{exp}[{0.153(W2-13.86)^2]}&,\, \text{for }\, W2>13.86,\\
      0.65 &,\,\text{for }\, W2 \leq 13.86.\\
    \end{cases}
\end{aligned}
\end{equation}
\crefformat{equation}{#2Eq.~(#1)#3}
Additionally, to maintain the completeness and reliability of these diagnostics, it is essential to impose extra conditions, such as \rm{W}1>8$, \rm{W}2>9$, SNR$_{\rm{W}2}>5$, and the WISE-AllWISE quality flags \verb|cc_flags|=0. Both of these selections are established for the Vega magnitude system.

\begin{figure}[htbp!]
\centering
\includegraphics[angle=0,width=1.0\hsize]{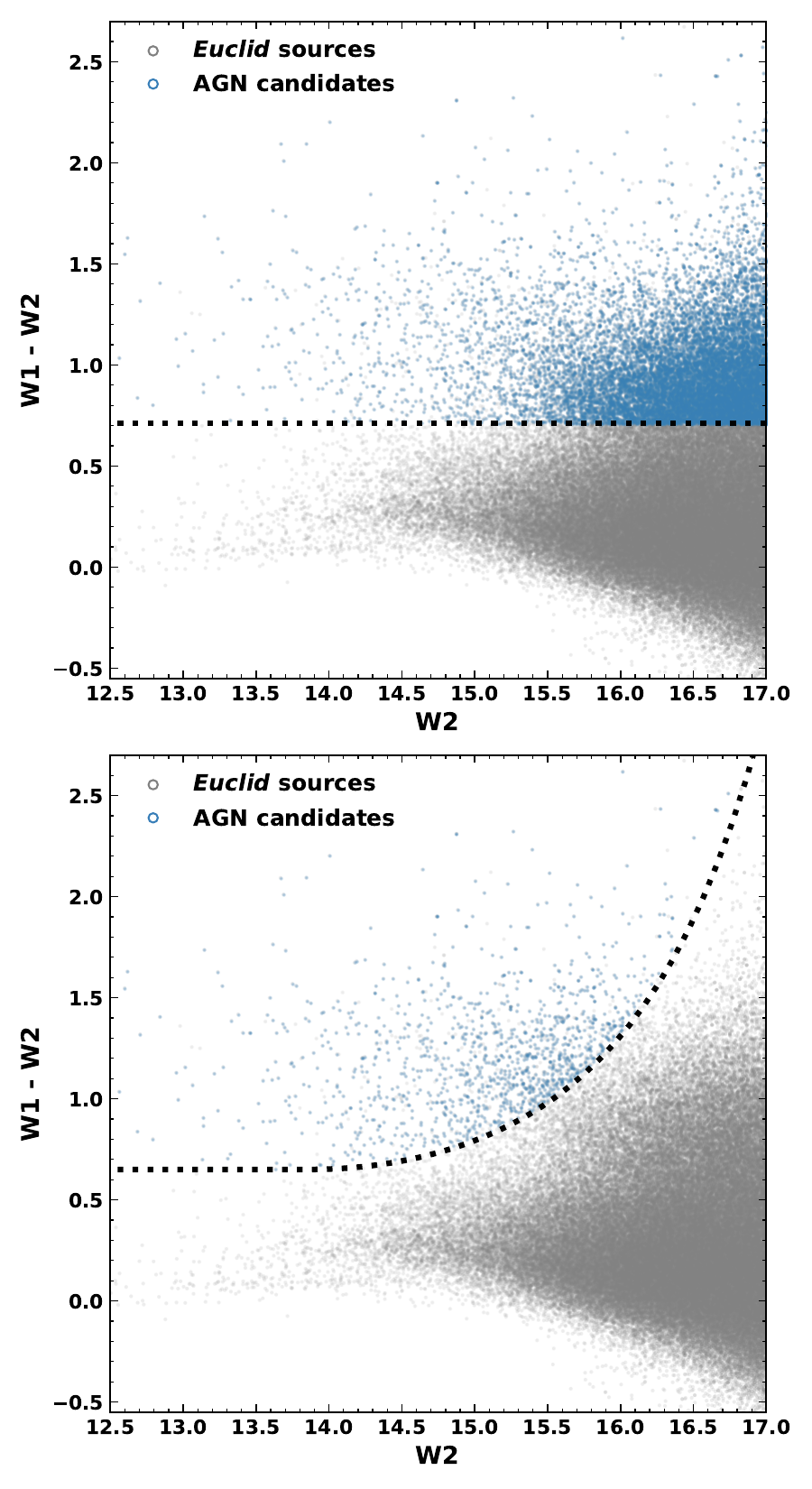}
\caption{WISE-AllWISE AGN candidates in the EDF-N defined by \cref{eq:c75}, top panel, and \cref{eq:r90}, bottom panel. In grey we show all \Euclid sources with WISE-AllWISE counterparts, while the blue sources represent the selected AGN candidates. The black dotted lines represent the limits of the C75 and R90 selections. The stellar candidates have been removed from the samples of AGN candidates. }
\label{fig:wise_c75_r90}
\end{figure}

Before applying either one of these selections, we remove the stellar candidates identified with the \cref{subsect:star_contaminants} prescription. The total numbers of C75 and R90 AGN candidates is 65\,083 and 4\,688, respectively, and the numbers of AGN candidates per field are reported in \cref{Table:AGN_candidates_photometry}. 
\Cref{fig:wise_c75_r90} shows both of these selections applied to EDF-N sources, where we include all \Euclid sources matched to WISE-AllWISE and we highlight those sources that are selected as AGN candidates by C75 or R90.

\subsubsection{\gdr{3}}\label{subsect:GR3-QSOs}

Using the \gdr{3} cross-match data, we identify sources marked as QSO candidates from the GR3-QSOs sample. Specifically, we identify 647, 811, and 513 candidates in the quality-filtered catalogues of EDF-N, EDF-S, and EDF-F, respectively. To address potential stellar contamination within this sample, we exclude the very few sources classified as stellar candidates according to \cref{subsect:star_contaminants}. The refined number of sources in the quality-filtered catalogues is presented in \cref{Table:AGN_candidates_photometry}.

\subsubsection{New \Euclid-only colour cut: $JH$\_$I_{\text{E}}Y$}\label{subsect:jhvisjy}

\Euclid provides us with a myriad pieces of information that can be utilised for selecting QSO candidates. Motivated to obtain a purer QSO selection, we investigate a new diagnostic using all four \Euclid bands, starting with a morphological cut. Focusing on QSOs, we consider only point-like sources with \verb|MUMAX_MINUS_MAG| $\leq -2.6$. Imposing this restriction allows us to exclude all extended sources that could act as contaminants in our selection, to the expense of detectable AGN within extended host galaxies (see \cref{subsect:WISE_selection,subsec:AGNspec,subsec:AGNmorph}.

Subsequently, we create a smaller `ground truth' catalogue where we include sources that have been matched to DESI and have been classed as DESI QSOs while possessing broad-line detection and $z_\mathrm{spec}$ (see \cref{subsect:desi}). In this way, we are able to differentiate between objects that have a good reliability of being QSOs, galaxies, or stellar candidates.

Imposing good photometry on all sources (i.e., working with the quality-filtered catalogues), we explore the colour--colour space $\JE-\HE$ versus $\IE-\YE$. 
We identify a cut that excludes the stellar locus, and obtain the following:
\begin{equation}
\label{eq:mumax}
\begin{aligned}
\left[-0.1\leq (\IE-\YE) <1.0\right] & \land [-0.5 \leq (\JE - \HE) < 0.6] \\ & \land \{ [(\JE - \HE) > 0.5(\IE-\YE) - 0.20] \\ & \lor [(\JE-\HE) > 0.13] \}.
\end{aligned}
\end{equation}
\crefformat{equation}{#2Eq.~(#1)#3}

We then investigate the confusion matrix, completeness and purity in two $z_\mathrm{spec}$ bins separated by $z_\mathrm{spec} = 1.6$. We limit this analysis to \IE$<21$, since at fainter magnitudes galaxy contamination from DESI (especially at $z>1.6$) is largely unknown. 
The confusion matrix is computed applying the following labels to the `ground truth' sample:
\begin{itemize}
    \item true; DESI QSO labelled objects with a detected broad-line and spec-$z$ are broad-line QSOs (BLQSOs); or
    \item false; anything that is not a broad-line QSO acts as contaminants (galaxies, Type II AGN, stars).
\end{itemize}
Then, depending on whether or not an object is within our colour-colour selection we can assign:
\begin{itemize}
    \item positive; is compact and is within the selection; or
    \item negative; is not point-like or is outside the selection.
\end{itemize}
We opt to count stellar contaminants across all redshift bins, given that stars can interfere with selection processes at both low and high $z$. Using the aforementioned prescription, we achieve $P=0.92$ with $C=0.63$ for the \IE$<21$ $\land$ $z_\mathrm{spec}<1.6$ bin, and $P=0.95$ with $C=0.90$ for the \IE$<21$ $\land$ $z_\mathrm{spec}>1.6$ bin. Nevertheless, these values should be taken with caution since they can not be straightforwardly extrapolated to fainter magnitudes, as the presence of numerous contaminants, especially compact galaxies, could significantly reduce the purity and completeness of this selection at fainter magnitudes. Given the lack of sufficiently reliable galaxy labels at fainter magnitudes, assessing the impact of contaminants is challenging. Therefore, we consider this selection method particularly effective for our two brightest \IE bins, while the statistics for the faintest bin remain less constrained. In \cref{app:a}, we illustrate the number of candidates picked up by this selection, split into our three magnitude bins, where it is evident that at fainter magnitudes, the number of candidates is considerably larger, hinting at higher contamination rates.

\begin{figure}[htbp!]
\centering
\includegraphics[angle=0,width=1.0\hsize]{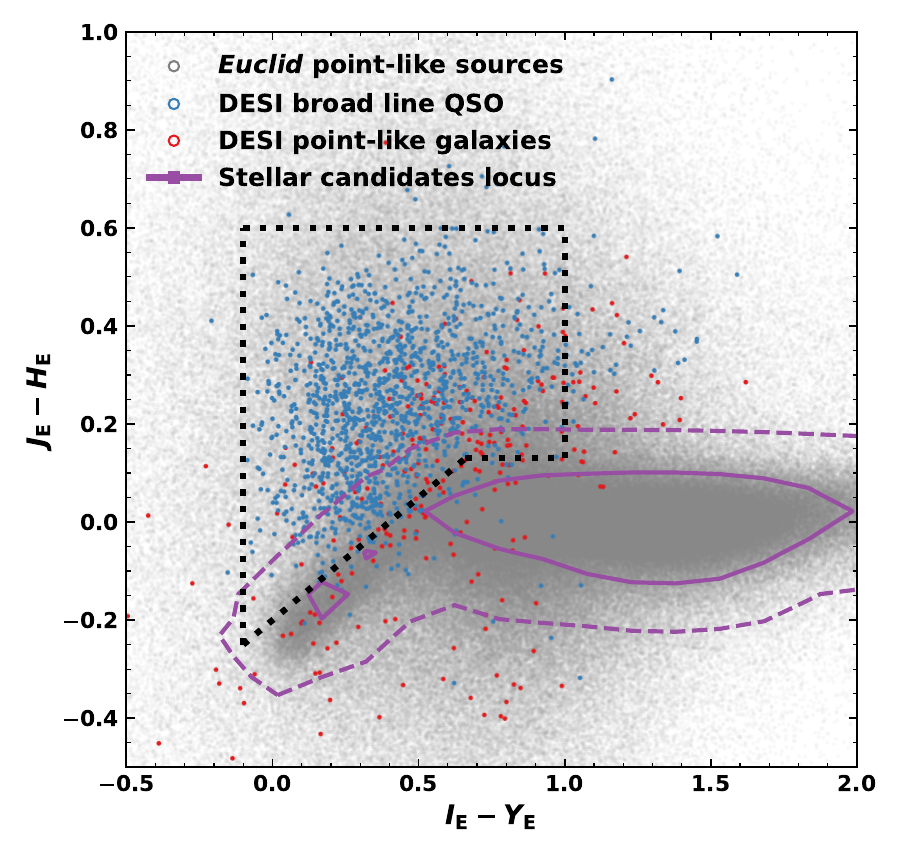}
\caption{New colour-cut criteria defined by \cref{eq:mumax}, black dotted line, applied to EDF-N. In grey we show all \Euclid compact sources that have a DESI counterpart. The blue coloured points represent the \Euclid compact sources selected as DESI BLQSO candidates and the red ones represent the \Euclid compact objects selected as galaxy candidates by DESI. Moreover, the purple lines represent the 68\% (solid) and 95\% (dashed) contours of the stellar candidates found in \cref{subsect:star_contaminants}. }
\label{fig:mumax}
\end{figure}

With this specific criterion, we identify a total of 313\,714 QSO candidates, the breakdown of which can be found in \cref{Table:AGN_candidates_photometry}. Additionally, \cref{fig:mumax} illustrates the application of this colour cut to the EDF-N, highlighting the corresponding point-like sources that served as DESI counterparts and were used to derive this QSO selection criterion, as well as the stellar locus that we identify in \cref{subsect:star_contaminants}.

\subsubsection{New \Euclid and ancillary photometry colour cut: $I_{\text{E}}H$\_$gz$}\label{sc:vishgz}

Employing \Euclid's ancillary data from Q1, we explore an additional diagnostic using the $\IE-\HE$ vs. $g-z$ colour space. Similarly to \cref{subsect:jhvisjy}, we create a smaller \enquote*{ground truth} catalogue including sources that have been matched to DESI classed as DESI broad-line QSOs. We impose the same morphology cut as in \cref{subsect:jhvisjy} and we then identify the area occupied by the stellar locus, as well as the locations that the DESI broad-line QSOs and galaxies populate. Based on this initial realisation, it becomes apparent that it is possible to separate these objects in this colour space with the following prescription:
\begin{equation}
\label{eq:vishgz}
\begin{aligned}
(g-z<1) & \land  [(g-z<0.5) \land (\IE-\HE>0.1)]\\
& \lor [(g-z>0.5) \land (\IE-\HE>(g-z)-0.4)].
\end{aligned}
\end{equation}
\crefformat{equation}{#2Eq.~(#1)#3}

We then follow the same guidelines as the ones presented in \cref{subsect:jhvisjy} and investigate the confusion matrix, $C$ and $P$, in two $z_\mathrm{spec}$ bins separated by $z_\mathrm{spec} = 1.6$, again limiting the analysis to \IE$<21$. We obtain $P=0.93$ with $C=0.60$ for \IE$<21$ $\land$ $z_\mathrm{spec}<1.6$, and a $P=0.97$ with $C=0.77$ for \IE$<21$ $\land$ $z_\mathrm{spec}>1.6$. Once again, given the limited number of labels, particularly at fainter magnitudes, we refrain to assess the performance of this selection at \IE > 21. \cref{app:a} illustrates this selection split into the three \IE bins to show the increasing number of candidates and therefore contaminants with fainter magnitudes.

\begin{figure}[htbp!]
\centering
\includegraphics[angle=0,width=1.0\hsize]{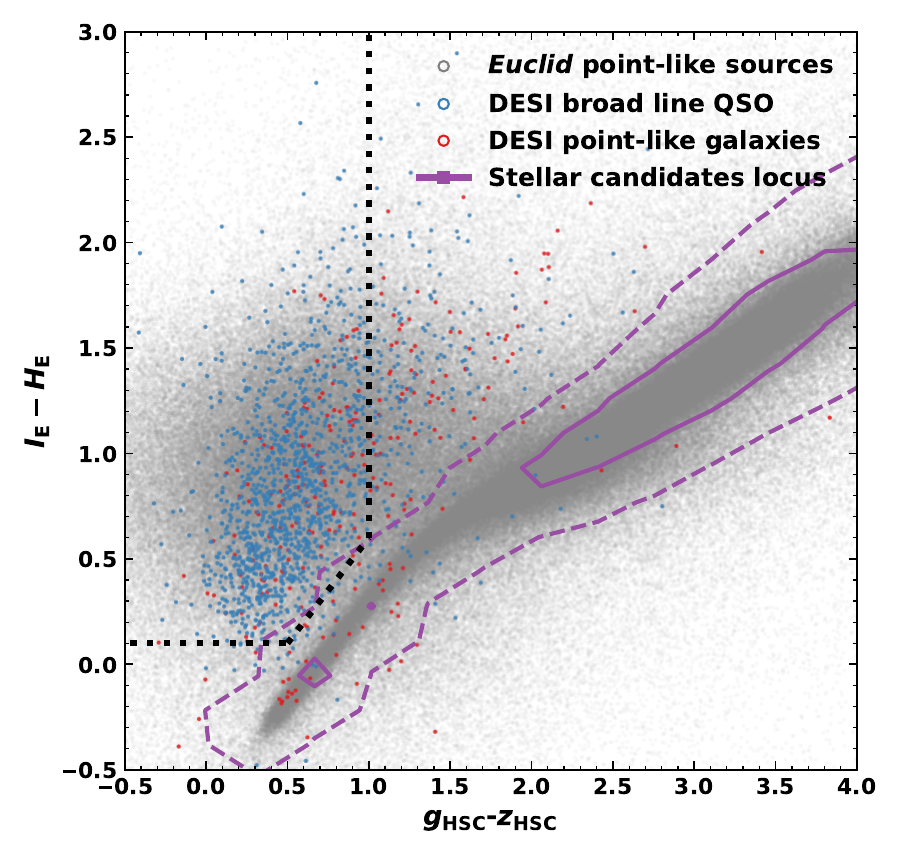}
\caption{New colour-cut criteria defined by \cref{eq:vishgz}, black dotted line, applied to EDF-N. In grey we show all \Euclid compact sources with that contained a DESI counterpart. The blue coloured points represent the \Euclid compact sources selected as DESI BLQSO candidates and the red ones represent the \Euclid compact objects selected as galaxy candidates by DESI. Moreover, the purple lines represent the 68\% (solid) and 95\% (dashed) contours of the stellar candidates found in \cref{subsect:star_contaminants}.}
\label{fig:desi_onvishgz}
\end{figure}

With this specific criterion, we identify a total of 267\,513 QSO candidates, its breakdown shown in \cref{Table:AGN_candidates_photometry}. \Cref{fig:desi_onvishgz} shows the application of this colour cut to the EDF-N, highlighting the corresponding point-like sources that serve as DESI counterparts and are used to derive this QSO selection criteria. 

\subsection{AGN candidates: spectroscopic selection}\label{subsec:AGNspec}

We present the various AGN selections applied to the Q1 fields based on spectroscopy obtained from the DESI counterparts. We report the number of AGN candidates recorded per criterion on \cref{Table:AGN_candidates_photometry}.

\subsubsection{DESI selection} \label{subsect:desi}

\begin{figure*}[htbp!]
\centering
\includegraphics[angle=0,width=\textwidth]{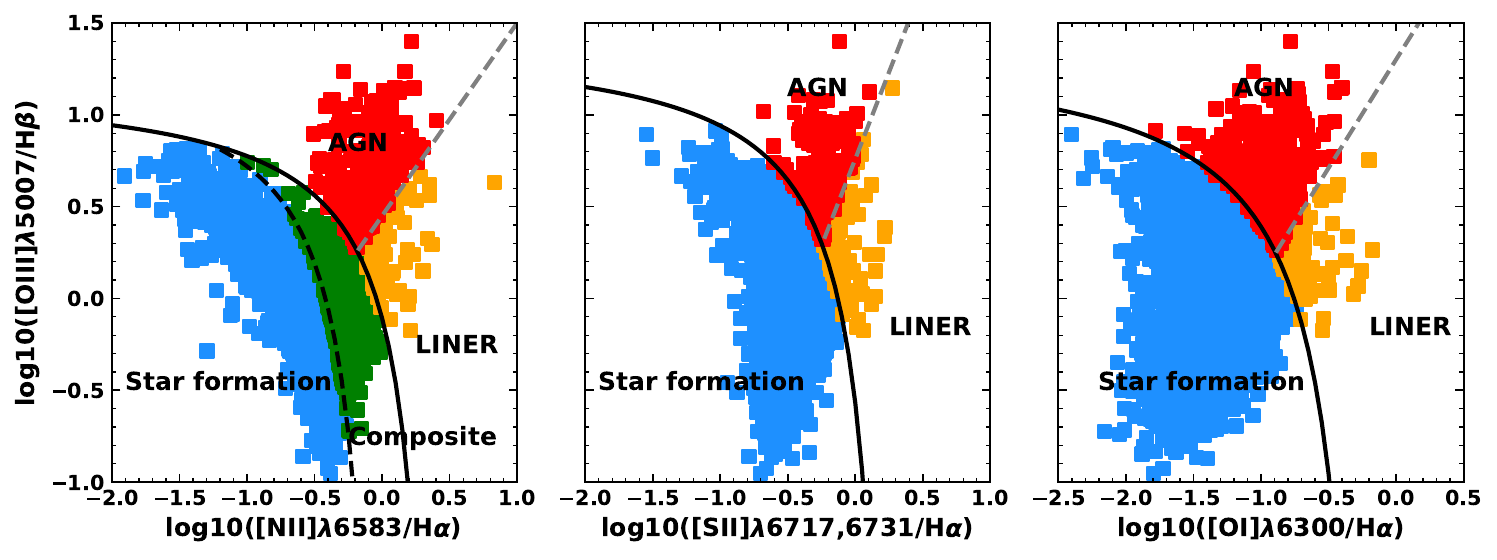}
\caption{\ion{N}{ii} (left), \ion{S}{ii} (centre), and \ion{O}{i} (right) emission line diagnostic diagram for those \Euclid sources with a DESI spectroscopic counterpart.}
\label{fig:desi_bpt}
\end{figure*}

We investigated the presence of QSOs and AGN within the 64\,039 matches of DESI EDR to the \Euclid matches. We began by creating a subsample of extragalactic objects that only included targeted objects with positive redshifts that had not been spectroscopically selected as DESI stars. To do so, we implemented the following criteria:

\begin{itemize}
    \item \texttt{z $>$ 0.001};
    \item \texttt{spectype $\ne$ STAR}.
\end{itemize}
From this subsample, the simplest method for selecting the QSO candidates is by using the DESI spectral type classification \citep[\texttt{SPECTYPE=QSO},][]{DESI_EDR}. Additionally, for sources classified as galaxies \citep[\texttt{SPECTYPE=GALAXY},][]{DESI_EDR}, we make use of multiple AGN diagnostics based on emission line fluxes, widths, and equivalent widths measured with \texttt{FastSpecFit} \citep{2023ascl.soft08005M}. These measurements are available for 40\,274 of the DESI EDR-\Euclid MER sources.

\begin{itemize}
    \item The detection of broad H\,$\alpha$, H\,$\beta$, \ion{Mg}{ii} or \ion{C}{IV} emission lines with a FWHM $\geq$ 1200\,\kms.
    
    \item An AGN classification in either the \ion{N}{ii} ([\ion{O}{III}]$\lambda$5007/H$\beta$ versus [\ion{N}{II}]$\lambda$6583/H$\alpha$), \ion{S}{ii} ([\ion{O}{III}]$\lambda$5007/H$\beta$ versus [\ion{S}{II}]$\lambda$6717,6731/H$\alpha$) or \ion{O}{i} ([\ion{O}{III}]$\lambda$5007/H$\beta$ versus [\ion{O}{I}]$\lambda$6300) emission line diagnostic diagrams \citep[or `BPT diagrams' after][see \cref{fig:desi_bpt}]{Baldwin1981}. For the \ion{N}{ii} BPT, we make use of \citet{Kewley2001}, \citet{Kauffmann2003}, and \citet{Schawinski2007} to distinguish between AGN, Low-ionization nuclear emission-line region (LINER), composite, and star formation (SF) ionisation. For the \ion{S}{ii} and \ion{O}{i} BPTs, we make use of the \citet{Kewley2006} and \citet{Law2021} demarcations to distinguish between AGN, LINER and star-formation ionisation. 

    \item A strong AGN or weak AGN classification in the WHAN diagram of \cite{CidFernandes2010}, which makes use of the equivalent width of the H\,$\alpha$ emission line.

    \item An AGN classification in the BLUE diagram of \cite{Lamareille2010}, which makes use of the equivalent width of the H\,$\beta$ and [\ion{O}{II}]$\lambda$3727 emission lines.

    \item An AGN classification in the KEX diagnostic diagram of \cite{ZhangHao2018}, which makes use of the [\ion{O}{III}]$\lambda$5007 emission line width.
    
\end{itemize}
Moreover, to identify QSOs containing broad emission lines, we also perform a test to detect broad H\,$\alpha$, H\,$\beta$, \ion{Mg}{ii}, or \ion{C}{IV} emission lines with a FWHM $\geq$ 1200\,\kms\ for those sources classed with \verb|SPECTYPE=QSO|. 

This results in a total of 4392 AGN candidates in the quality-filtered catalogue, the breakdown of which is shown in \cref{Table:AGN_candidates_photometry}. Additionally, \cref{fig:desi_bpt} shows an example of the \ion{N}{ii}, \ion{S}{ii}, and \ion{O}{i} emission-line diagnostics performed for the DESI counterparts.

\subsection{AGN candidates: other AGN selections}\label{subsec:AGNmorph}

The following AGN diagnostics were developed in other works conducted in preparation for the Q1 data release. We report the number of AGN candidates recorded per criterion on \cref{Table:AGN_candidates_photometry}.

\subsubsection{X-rays}\label{subsect:xrays}

\citet{Q1-SP003}, subsequently referred to as \citetalias{Q1-SP003}, present the Q1 counterparts to X-ray point sources, starting from the 4XMM-DR14 \citep{Webb2020}, the \textit{Chandra} Source Catalog (CSC) Release 2 Series \citep[][]{Weisskopf_2002, Evans_2024} and the eROSITA \citep{Predehl_2021_2021A&A...647A...1P}  first Data Release \citep[DR1;][]{Merloni2024}. For each of the Q1 fields and each of the X-ray surveys they first identify the best \Euclid counterpart by means of the Bayesian algorithm \texttt{NWAY} \citep{Salvato_2018}, which assigns the probability of a good association considering a) the separations between sources, their positional uncertainties, and their number density, and b) the similarity between the SED of a candidate counterpart and the SED of a typical X-ray emitter, regardless whether the source is Galactic or extragalactic. The latter information is provided by a prior externally defined using a random forest on a set of \Euclid-only features (i.e., no features from ground-based photometry) extracted from the Q1 catalogues for a training sample of secure X-ray emitters and secure field sources. The same procedure is repeated, randomizing the coordinates of the X-ray catalogues so that the probability of a chance association can be determined (see \citetalias{Q1-SP003} for details).

After the determination of the counterparts, the authors then assign to each source a probability of being Galactic (star, compact object) or extragalactic (galaxy, AGN, QSO). This is again done using a training sample of secure Galactic and extragalactic sources and \Euclid-only features from the Q1 catalogues.

Finally, for the sources that have a probability larger than 50\% of being extragalactic, the authors provide either photometric redshifts computed using {\sc{PICZL}} \citep{Roster24} on Legacy Survey DR10 \citep{Dey_2019} images (and thus limited to the sources detected in that survey), or spectroscopic redshifts from literature.

The released catalogues enable users to refine their samples based on specific scientific needs, balancing purity and completeness through the \texttt{NWAY} output parameters. In total, they report 12\,645 AGN candidates, though some sources have multiple counterparts; when considering only the best match for each unique X-ray source, the sample reduces to 11\,286 candidates. They identify 949 in EDF-N, 3789 in EDF-S, and 6548 in EDF-F. 

From their catalogue, in order to identify those candidates that have the highest probability of being an AGN candidate, we select a subsample of sources with $P_{\textrm{Gal}} < 0.2$. Implementing our `cleaning' on these candidates we obtain 434, 1812, and 3813 X-ray candidates in the quality-filtered EDF-N, EDF-S, EDF-F catalogues (see \cref{Table:AGN_candidates_photometry}).

\subsubsection{Diffusion models }
\citet{Q1-SP009}, from here on called \citetalias{Q1-SP009}, use the reconstruction error of a diffusion model, a type of generative model, to select AGN candidates. Through training on VIS images, the model is able to learn a bias for the light profile at the centres of galaxies. Since AGN are rare, the bright pixel and steep fall off of light they exhibit is converted to one that is significantly flatter, leading to a high reconstruction error for suspected AGN. They obtain a total of 15\,940 AGN candidates across the three EDFs (see \cref{Table:AGN_candidates_photometry}).

\subsubsection{Deep learning }

\citet{Q1-SP015}, hereafter called \citetalias{Q1-SP015}, present a deep learning (DL) method to quantify the AGN contribution ($f_{\rm PSF}$) of a galaxy using VIS imaging. The DL model is trained with a sample of mock images generated from the IlllustrisTNG simulations, designed to mimic \Euclid VIS observations, with different levels of AGN contributions artificially injected as PSFs. The DL model is trained to estimate the level of the injected PSF, achieving a root mean square error (RMSE) of 0.052 on the test set. After applying this model to the Q1 data, they find 48\,840 galaxies across the EDFs that are classified as AGN based on this AGN contribution, that is, $f_{\rm PSF} > 0.2$. Adopting a less conservative threshold of $f_{\rm PSF} > 0.1$ increases the number to $158\,711$ AGN. This method allows the identification of AGN even when the AGN component is not the primary contributor to the host galaxy’s luminosity. The resulting number of AGN candidates found per field and magnitude bin is in \cref{Table:AGN_candidates_photometry}.

\subsection{AGN fraction from SED fitting}\label{subsec:AGNsed}

SED-template fitting is a powerful method for measuring physical properties of galaxies and AGN by reproducing the observed photometry with a combination of theoretical and empirical SED models that account for the different AGN and galaxy emission processes. The result of the multi-component SED fitting constrains a variety of physical properties, notably the AGN fraction, defined as the ratio of the AGN flux to the total flux in the MIR band, used for AGN identification \citep[see e.g.,][]{Dale_2014, Thorne_2022}. By decomposing the emission from the galaxy and the potential AGN components, SED fitting permits the identification of fainter AGN compared to colour-colour approaches. Moreover, since the IIR emission is not significantly impacted by AGN obscuration, SED fitting can reliably identify obscured AGN missed by optical or X-ray methods \citep{Pouliasis_2020, Andonie_2022}.

Since accurate redshift measurements are required for reliable SED fitting, we first restricted the analysis to sources with spectroscopic redshift in EDF-N, computing the AGN fraction as the ratio between the AGN flux and the total flux over the 5--20\,\micron\ wavelength range \citep[following][]{Dale_2014, Thorne_2022}. These results can be used to define an accurate selection threshold for AGN that could later be applied to the entire sample with photometric redshift. We used the SED fitting algorithm \texttt{CIGALE} \citep{Boquien_2019, Yang_2022} to fit the UV-to-mid-IR photometry of our sources, including: GALEX FUV/NUV, {\it ugriz}, \gaia-G/BP/RP, \Euclid \IE/\YE/\JE/\HE, WISE \rm{W}1/2/3/4.
Our modelling consists of a delayed star-formation history with a simple stellar population from \cite{Bruzual_2003} and the \cite{Chabrier_2003} initial mass function, a galactic dust attenuation \citep{Calzetti_2000} and emission \citep{Draine_2014}, nebular lines \citep{Inoue_2011}, and an AGN model \citep{Fritz_2006}. For further details, we refer to Euclid Collaboration: Laloux et al., in prep, from now on referred to as LB25, which presents the physical properties of the AGN candidates in the three EDFs.

\begin{figure}
    \centering
    \includegraphics[width=1\hsize]{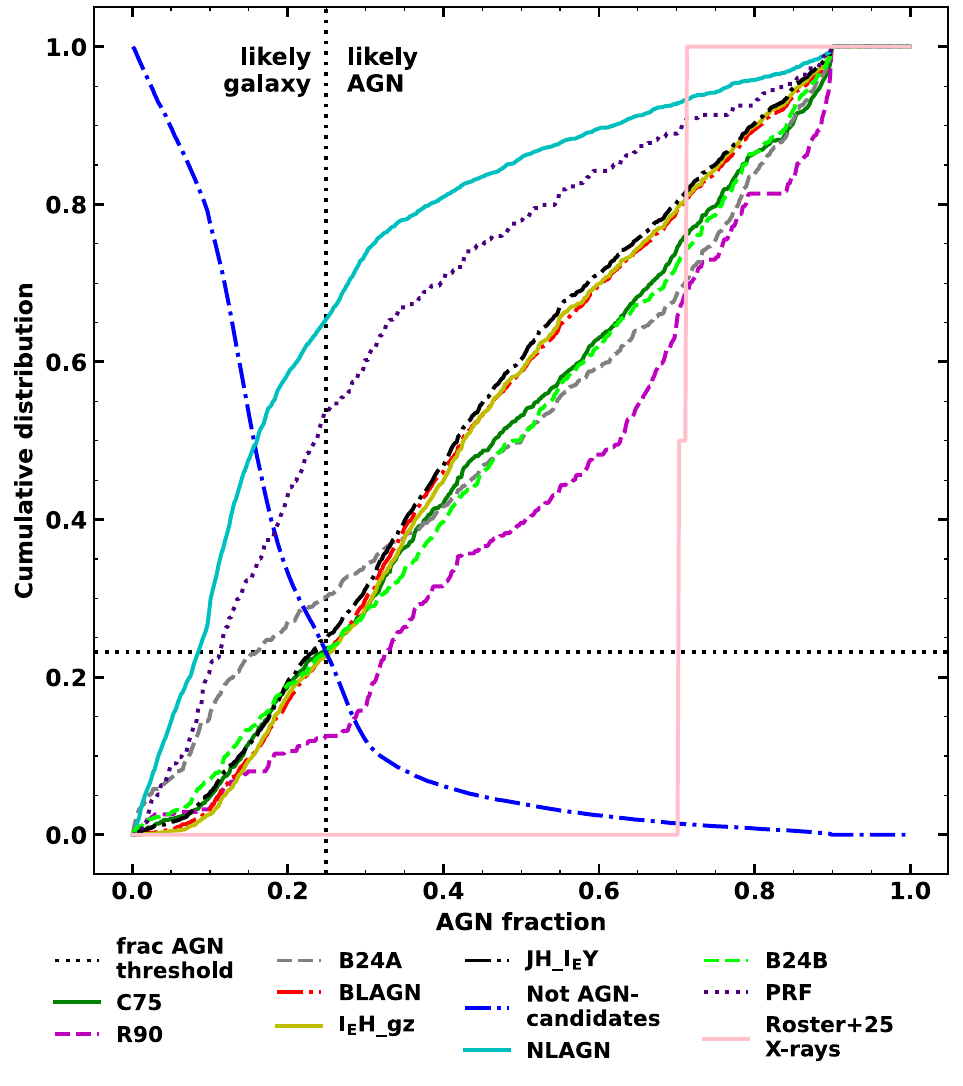}
    \caption{Normalised cumulative distribution of the AGN fraction measured by SED fitting for the different AGN selection methods in EDFN: C75/R90 (green solid/purple dashed), B24A/B (grey/lime dashed), spectroscopically-confirmed broad/narrow line AGN (red dash-dotted/cyan solid), $I_{\text{E}}H$\_$gz$ (yellow solid), $JH$\_$I_{\text{E}}Y$ (black dash-dotted), PRF (dark blue dotted) and X-ray (pink solid). 
    The blue dashed-dotted line is the inverse cumulative distribution of all non-AGN candidates. 
    The vertical black solid line represents the proposed AGN fraction threshold ($f_{\rm AGN}=0.25$) discriminating AGN and non-AGN sources, while the horizontal black dotted line shows the corresponding false-positive probability (23\%). Note that only two X-ray sources have counterparts with $z_\mathrm{spec}$ in the quality-filtered EDF-N catalogue.
    }
    \label{fig:fracAGN_cumulative_distribution}
\end{figure}

The results are shown in \cref{fig:fracAGN_cumulative_distribution}, where the normalised cumulative distribution of the AGN fraction for normal galaxies is compared to the different AGN samples. 
As indicated by the vertical black dotted line, we define our AGN fraction threshold as the intersection between the normal galaxy distribution (dash-dotted blue line) and the broad-line AGN one (BLAGN, dash-dotted red line), whereby BLAGN refers to those sources classified as QSOs or galaxies that exhibit broad emission lines of H$\alpha$, H$\beta$, \ion{Mg}{ii}, or \ion{C}{IV}. An AGN fraction threshold at $f_{\rm AGN}=0.25$ is a compromise between purity and completeness, since it correctly selects 77\% of the spectroscopically confirmed BLAGN while only 23\% of the non-AGN candidates are misclassified as AGN. With this threshold we obtain a total of 7766 AGN candidates within the quality-filtered catalogues, the breakdown of which can be found in \cref{Table:AGN_candidates_photometry}.
A lower AGN fraction threshold, such as $f_{\rm AGN}=0.1$ \citep[][]{Thorne_2022}, would lead to an increased AGN completeness (97\%), but with a much higher false-detection probability (78\%). Conversely, a higher AGN fraction threshold of $f_{\rm AGN}=0.5$ improves the purity, with a false-detection probability of 5\%, at the expense of the completeness (44\%). 

In \cref{fig:fracAGN_cumulative_distribution}, we also compare the AGN fraction distribution for different AGN-selection methods. 
We find that AGN candidates selected by C75, B24B, $I_{\text{E}}H$\_$gz$, and $JH$\_$I_{\text{E}}Y$ show a similar AGN fraction distribution to the BLAGN, while the R90 candidates tend to have higher AGN fractions. This suggests that the more reliable methods select sources where the AGN emission dominates over its host, potentially missing weaker AGN. Conversely, the B24A method tends to select sources with lower AGN fractions, some of which are likely to be contaminants, as indicated by the curve of the non-AGN candidates. Additionally, the cyan curve, representing the narrow-line AGN (NLAGN) from the DESI emission-line diagnostics, is significantly shifted to lower AGN fraction values compared to other methods. This demonstrates that while SED fitting is efficient at identifying unobscured AGN, achieving both high completeness and purity for obscured AGN is more challenging. Nevertheless, it still manages to reach 32\% completeness for these elusive AGN.

In this Section we discussed only qualitatively the comparison of the results of the SED fitting analysis for different selection approaches applied to sub-samples with spectroscopic redshift. Not necessarily the same results will hold for samples which have only photometric redshifts as the availability of a spectroscopic redshift could impact the selections and introduce a bias in the AGN fraction distributions. The AGN fraction from SED fitting for the three deep fields for spectroscopic and photometric redshift sources will be provided in LB25.



\section{Discussion}\label{sc:discussion}

\subsection{Contaminants}

Although \Euclid offers exciting prospects for AGN selection, similar to all AGN criteria, several contaminants must be considered when developing these methods. 
As already discussed in \cref{subsect:star_contaminants}, stars are a big contaminant when it comes to colour-colour diagnostics. However, other types of stellar objects also make it hard to create a pure AGN-selection technique. 

For instance, young stellar objects (YSOs), asymptotic giant branch (AGB) stars, and \ion{H}{ii}-regions can show very similar colours to those of AGN, and therefore slip through the selection criteria as potential candidates. 

Another form of contaminants, related to low mass products of star formation would be that of brown dwarfs \citep{Davy_Kirkpatrick_2011} whose feature at 1\,\micron\  can sometimes resemble a Ly\,$\alpha$ break at redshifts $z=6$--7.5 \citep{Wilkins_2014_2014MNRAS.439.1038W, hainline2024browndwarfcandidatesjades, Langeroodi_2023}.

Compact normal galaxies can also play a big role as contaminants in AGN diagnostics \citep{Kouzuma_2010_2010A&A...509A..64K}. This is because, at higher redshifts, distant galaxies may appear as point-like sources and show similar colours to that of AGN, therefore contaminating the AGN locus in colour-colour plots. 

Additionally, dwarf irregulars (Irr) have a 4000\,\AA\, break, meaning that, at some redshifts ($z<1$), their SEDs are very similar to those of QSOs \citepalias{EP-Bisigello}. 

Moreover, high-$z$ star-forming galaxies (SFGs) can sometimes also be considered a contaminant, since their SEDs resemble those of AGN. However, since their 4000\,\AA\ break lies within \Euclid's bands at $z>1$, it might be easier to separate between AGN and SFGs at these higher redshifts.

These contaminants are likely affecting the colour selections analysed in this work. With our current knowledge and understanding of the Q1 data, we are unable to identify these sources, making it difficult to distinguish between them and potential AGN candidates. Consequently, we acknowledge that, especially at fainter magnitudes, these sources are definitely contaminating our AGN sample, which worsens our calculated values for purity and completeness. Addressing this issue would require further work, such as combining our data with other photometric catalogues that include bands that would facilitate the separation of these objects, or obtaining more spectroscopic data to test for specific emission lines, like those mentioned in \cref{subsect:desi}.

\subsection{Comparison to expectations}

Prior to this work, \citet{EP-Selwood}, from now on addressed as \citetalias{EP-Selwood}, examined the AGN surface density expected for the \Euclid mission. Starting from an X-ray luminosity function \citep{Fotopoulou_2016_2016A&A...587A.142F}, they predicted the observational expectations for AGN with $z < 7$ in the EWS and EDS. They generated volume-limited samples covering $0.01\leq z \leq7$ and $43\leq \text{log}_{10}(L_{\text{bol}}/\text{erg s}^{-1})\leq 47$. Each AGN was assigned an SED based on its X-ray luminosity and redshift. Dust extinction was applied, and once assigned and scaled, they performed mock observations of each AGN SED in their sample, convolving with the \Euclid bands and an assortment of ancillary photometric bands to explore the observable population of $z < 7$ AGN in the \Euclid surveys. They concluded that \Euclid should be able to detect significantly more AGN in the EWS and EDS compared to those identified in other surveys covering similar regions. In the EDS they predicted an unobscured AGN surface density of 346\,deg$^{-2}$ based on the \citetalias{EP-Bisigello} \cref{eq:B24_2} selection, which is comparable to the densities obtained by \textit{Spitzer} \citep{Lacy_2020_2020NatAs...4..352L} and the {XXL--3XLSS} \citep{Chiappetti_2018_2018A&A...620A..12C} survey. Similarly, the EWS was estimated to recover an AGN surface density of 331\,deg$^{-2}$, surpassing the AGN densities obtained from both ground-based optical and space-based MIR missions.

\begin{figure}[htbp!]
\centering
\includegraphics[angle=0,width=1.0\hsize]{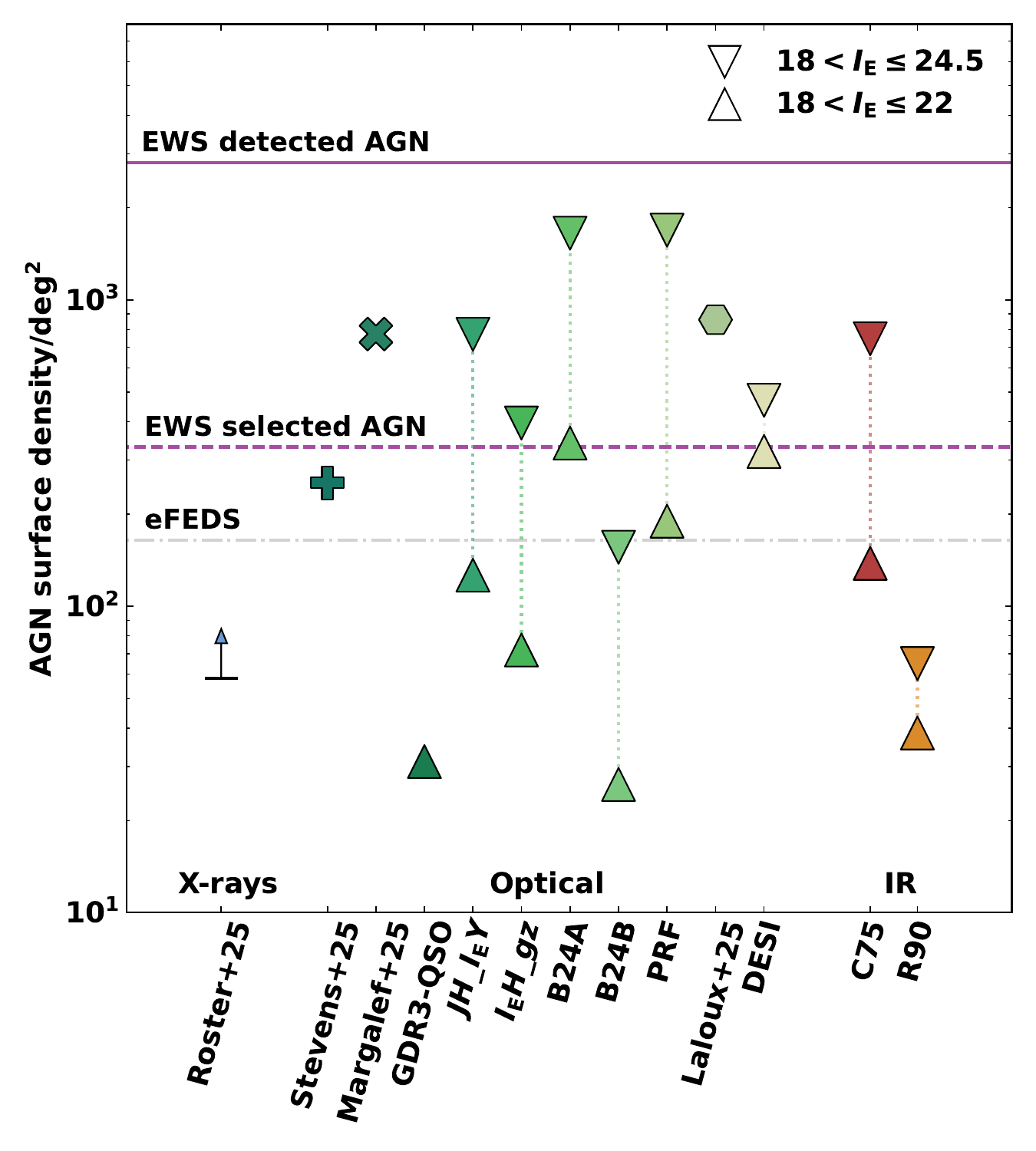}
\caption{Comparison of AGN surface densities obtained from the selection methods discussed in this work, divided into energy bands: X-ray selections (blue;, optical selections (shades of green); and IR selections (shades of orange). For the selections B24A, B24B, $JH$\_$I_{\text{E}}Y$, $I_{\text{E}}H$\_$gz$, C75, R90, DESI, and PRF the AGN surface densities are split into $18<\IE\leq 24.5$ (upside down triangles); and $18<\IE\leq 22$ (upside up triangles). For the GDR3-QSOs, only the $18<\IE\leq 22$ is shown due to the limiting magnitude of \gaia. Individual markers indicate AGN surface densities from other Q1 related works, including X-rays from \citetalias{Q1-SP003} (with the lower limit indicated by an arrow pointing up); morphology-based selections from \citetalias{Q1-SP009} (plus sign); and \citetalias{Q1-SP015} (cross); and SED fitting from LB25 (hexagon). The predictions for the detectable AGN (purple horizontal solid line) and identifiable AGN (purple dashed line) in the EWS from \citetalias{EP-Selwood} are included. The grey horizontal dashed line represents recovered AGN surface density by eFEDS \citep{Liu_2022_2022A&A...661A...2L}.}
\label{fig:selwoodpredictions}
\end{figure}

We compare the AGN surface densities recovered in $18<\IE\leq 24.5$ and $18<\IE\leq 22$ for each one of the selections investigated in this work against the EWS predictions from \citetalias{EP-Selwood}, which were based on a $5\,\sigma$ limit in all four \Euclid bands, corresponding to a limiting magnitude of \IE$\approx$26.2.

Starting with the \citetalias{EP-Bisigello} candidates, in \cref{fig:selwoodpredictions} we observe differences in the surface densities recovered for B24A and B24B. Given that B24B could only be applied to the EDF-N, the area used to calculate its AGN surface density is that of the EDF-N (23.9\,deg$^{2}$), as opposed to the Q1 area (63.1\,deg$^{2}$), which is used for most of the other selections. B24A/B have distinct statistics, with B24A having $P\approx0.17$ and $C\approx0.35$, while B24B has $P\approx0.92$ and $C\approx0.81$. Based on these numbers, the AGN surface densities for both selections align with their expectations. B24A retrieves a greater number of AGNs, likely due to its low purity, indicating that a notable amount of contaminants could be identified as candidates. The notably lower AGN surface density obtained for B24B can be attributed to the high purity of this selection, which indicates a smaller number of contaminants. Both selections reveal a significant difference in AGN surface densities recovered in $18<\IE\leq 24.5$ and $18<\IE\leq 22$, highlighting that at fainter magnitudes, we encounter a higher number of contaminants that are difficult to distinguish from actual AGN.

From \cref{fig:selwoodpredictions} it is apparent that the number of AGNs selected by the PRF is comparable to those selected by B24A in the range of $18<\IE \leq 24.5$, but it is significantly smaller in the range of $18<\IE\leq 22$. This variation is likely due to the PRF potentially misclassifying compact galaxies with QSOs at fainter magnitudes. The larger number of candidates at fainter magnitudes indicates that, even when we increase the thresholds beyond the recommended values, many contaminants still affect this selection, highlighting that the advised Q1 PRF thresholds are too lenient and should be revised.

The observed numbers for both the C75 and R90 criteria align with what we expected for both of these catalogues. For the C75 criteria, with a completeness of 75\%, we anticipated a larger number of candidates, as illustrated in \cref{fig:selwoodpredictions}. However, this expectation comes with the potential for an increase in contaminants. By excluding the stellar candidates, we have likely marginally increased the reliability of this selection, though we currently lack the tools to precisely quantify these effects. In contrast, R90 aims to create a more reliable, or purer, catalogue, resulting in a smaller AGN candidate size, as illustrated in \cref{fig:selwoodpredictions}. Notably, C75 recovers almost over an order of magnitude more candidates than R90 in both $18<\IE\leq 24.5$ and $18<\IE\leq 22$. Another observation from R90 is that the AGN surface density for $18<\IE\leq 24.5$ and $18<\IE\leq 22$ are not very different from each other, whereas other methods show a significantly higher higher AGN surface density in the faintest bin. The relatively bright limiting magnitude of this WISE-AllWISE selection (\rm{W1}<17.0 (Vega) = 19.7 (AB)), together with its high purity, makes this selection highly incomplete at faint \IE magnitudes, since the number of very red selected candidates (with faint \IE magnitude) is small. This effect produces the small increase of the number of candidates going from $\IE<22.0$ to $\IE<24.5$.

Both of the new selections, $JH$\_$I_{\text{E}}Y$ and $I_{\text{E}}H$\_$gz$, recover similar AGN surface densities, which in the $18<\IE\leq 24.5$ appear to surpass the expectations of what is identifiable as an AGN according to \citetalias{EP-Selwood}. However, the high purity achieved by these selections was only quantified for $\IE<21$. Therefore, we consider the AGN surface densities recovered in $18<\IE\leq 22$ to be more reliable since we lack the means to assess how the $P$ and $C$ of these selections might change with increasing magnitude. This raises questions about the reliability of the candidates obtained for both selections at $\IE>21$--22. Further work involving additional galaxy labels or spectra may be necessary to verify whether these AGN surface densities are accurate or inflated by contaminants.

The GDR3-QSO sample was originally quite small compared to the other samples of candidates (1971), so it was expected that the recovered AGN surface density was going to be orders of magnitude lower than that of other surveys. In \cref{fig:selwoodpredictions}, we only show GDR3-QSO's number density for $18<\IE\leq 22$. This is due to the lack of GDR3-QSO candidates at the faintest magnitudes, which is linked to \gaia's detection limit at $G<21$. Additionally, DESI only covers a reduced area of the EDF-N, which we calculate to be approximately 9 deg$^2$. We use this area to obtain the corresponding AGN surface density for the sources selected with DESI, which include QSOs, galaxies with detected broad-lines, and AGN selected via BPT or other narrow-line emission diagnostics. We find that the results are consistent with the expectations outlined by \citetalias{EP-Selwood} within the range $18<\IE\leq 22$, and exceed the expectations for $18<\IE\leq 24.5$. This is indicative of potential synergies between \Euclid and DESI in the future when more data from both survey are available. Moreover, we also assess the overall AGN surface density of the X-ray AGN identified by \citetalias{Q1-SP003}, which is lower than that recovered from eFEDS \citep{Liu_2022_2022A&A...661A...5L}. However, we note that the X-ray catalogue created in \citetalias{Q1-SP003} is a combination of different X-ray catalogues with varying depths (see Figure 1 of \citetalias{Q1-SP003}). Therefore, the surface density reported in \cref{fig:selwoodpredictions} should only be taken as a lower limit, where to estimate the surface area, we generated a multi-order coverage map with a resolution of 6\farcs87, yielding a total area of 36.72 deg$^2$. We also note that the SED fitting AGN surface density from the LB25 candidates surpasses the limit of the identifiable AGN set by \citetalias{EP-Selwood}. This could be due to the latter's use of Type I specific diagnostics to obtain their predictions, while SED fitting is more efficient, though not perfect, at identifying both, obscured and unobscured AGN. However, SED fitting is still affected by contaminants, and the quality of the available photometry will dictate the quality of the SED fit.

For most of the selections applied in this work, there seems to be a consistent trend. The magnitude range $18<\IE\leq 24.5$ recovers a larger AGN surface density, likely filled with contaminants, while the brightest $18<\IE\leq 22$ bin shows lower densities, usually below the predictions of \citetalias{EP-Selwood}. Despite the goal of achieving a reliable surface density across all magnitudes, further work is required to assess and enhance the reliability of these selections in the faintest magnitudes. Therefore, we consider the $18<\IE\leq$22 range to provide a `purer' catalogue of AGN candidates.

The resulting AGN surface density of our selections B24A, B24B, $JH$\_$I_{\text{E}}Y$, $I_{\text{E}}H$\_$gz$, GDR3-QSOs, PRF, SED-fitting, and DESI candidates is 3641\,deg$^{-2}$ for 18 < \IE $\leq$ 24.5. By applying this magnitude cut, we eliminate saturated sources from the brightest magnitudes as well as the faintest sources in the Q1 catalogues. We acknowledge that this approach results in missing a population of \Euclid sources detectable at a 5$\sigma$ limit of \IE, for which we currently lack the tools to study comprehensively. Given the potential contamination even within this cut, we propose that the purest sample of AGN lies in the magnitude range 18 < \IE $\leq$ 22, resulting in an AGN surface density of 482\,deg$^{-2}$. Even after narrowing the magnitude range of our selection, the AGN surface density recovered remains higher than what was expected to be identifiable with \Euclid, yet falls short of the expected number of detectable AGN. This indicates that further refinement of our selection criteria is necessary to bring the number of reliable AGN candidates closer to what should be observable. Notably, this need for improvement is especially critical for Type II AGN, which have been largely excluded from most of the photometric selections evaluated in this study (not including the MIR, X-ray, SED-fitting, and spectroscopic diagnostics). With machine-learning and contributions such as those conducted by \citetalias{Q1-SP009} and \citetalias{Q1-SP015} we hope to be able to bridge the gap between detected and selected AGN, thereby reducing the bias against Type II AGN that typically arises from most colour-colour selections. In fact, the recovered AGN surface densities from \citetalias{Q1-SP009} and \citetalias{Q1-SP015}, which are limited to $\IE<22$ and $\IE<24.5$ respectively, already highlight how effective machine learning can be in this regard.

\subsection{Comparison among AGN selections}

In this work, we have examined a variety of AGN diagnostics to construct the first \Euclid multi-wavelength catalogue of AGN candidates. However, as expected, most of these AGN selections are incomplete and biased.

Focusing on the photometric selections applied to the \Euclid photometry, which include B24A, B24B, $JH$\_$I_{\text{E}}Y$, and $I_{\text{E}}H$\_$gz$, it is the latter diagnostic that achieves the highest purity, with $P\approx0.93$ for $z_\mathrm{spec}<1.6$ and $P\approx0.97$ for $z_\mathrm{spec}>1.6$. However, $JH$\_$I_{\text{E}}Y$ ($P\approx0.92$ for $z_\mathrm{spec}<1.6$ and $P\approx0.95$ for $z_\mathrm{spec}>1.6$) and B24B ($P\sim0.92$) are only slightly lower. For the new selections, $JH$\_$I_{\text{E}}Y$ and $I_{\text{E}}H$\_$gz$, the calculation of $P$ and $C$ is limited to \IE < 21, thus considering only the brightest sources. This limitation is problematic at the faintest magnitudes, where we can not assess the $P$ and $C$ values and the effects of different contaminants due to the lack of reliable labels (see \cref{fig:jhvisy_magbins,fig:vishgz_magbins}).

\begin{figure}[htbp!]
\centering
\includegraphics[angle=0,width=1.0\hsize]{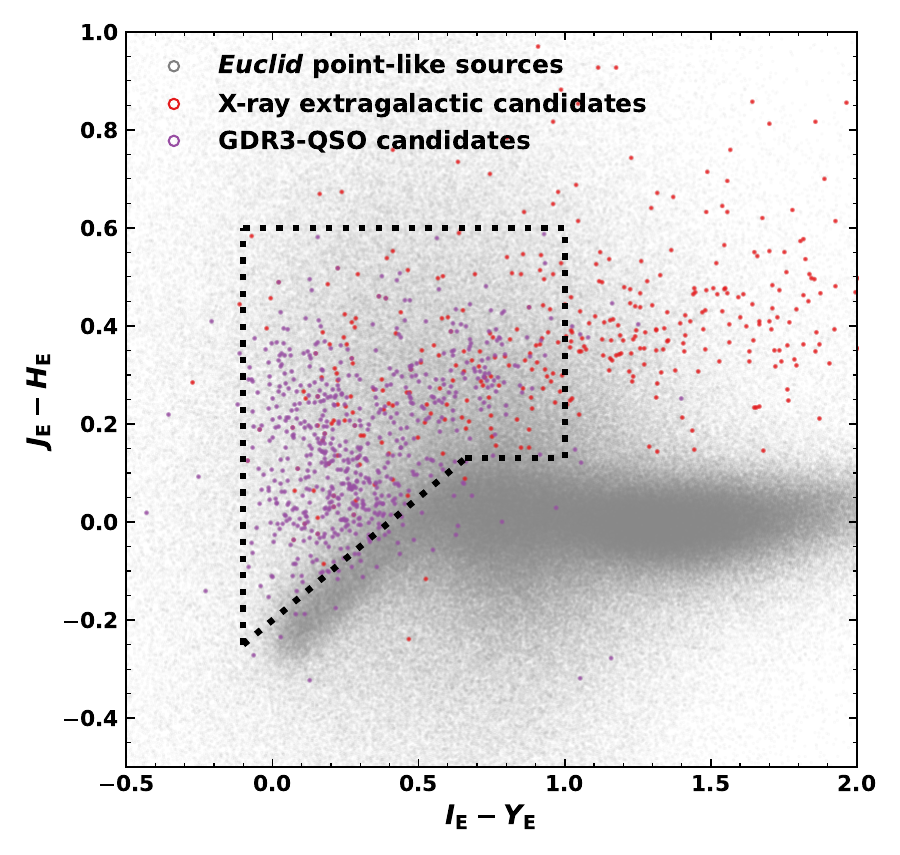}
\caption{Selection $JH$ versus $I_{\text{E}}Y$ (black dotted line) applied to the EDF-N. In grey we show all \Euclid compact sources. The red coloured dots represent the X-ray selected AGN candidates from \citetalias{Q1-SP003}, while the purple dots indicate the GDR3-QSOs, which mainly lie within the selection.}
\label{fig:jhvisy_xrays}
\end{figure}

\begin{figure}[htbp!]
\centering
\includegraphics[angle=0,width=1.0\hsize]{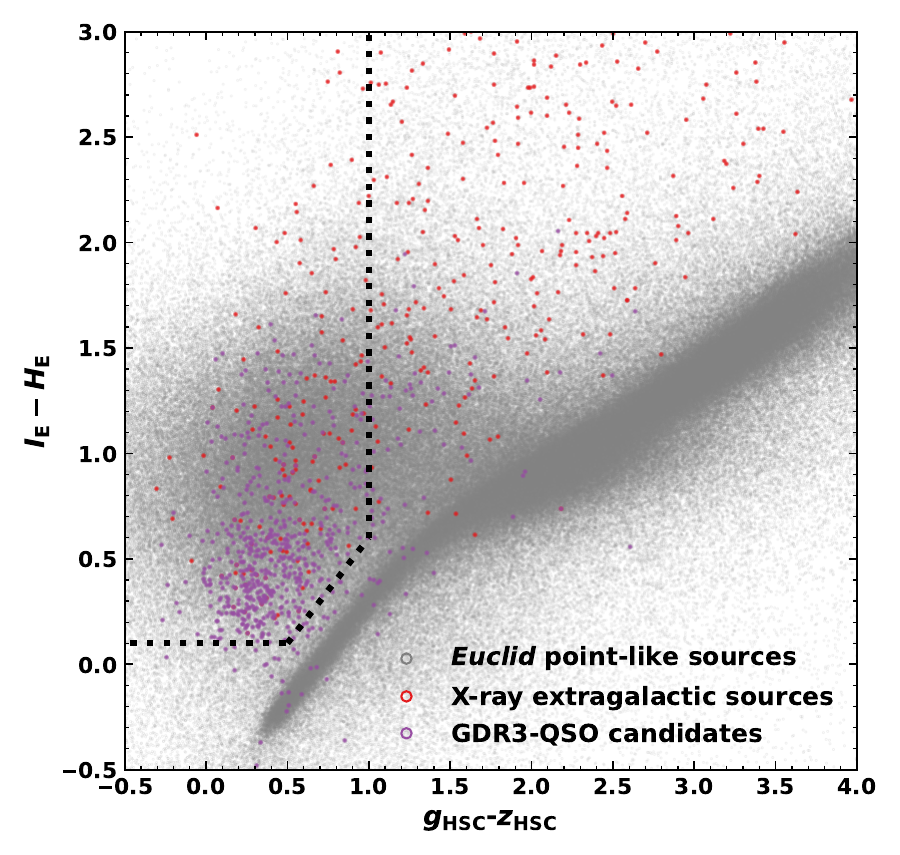}
\caption{Selectio $I_{\text{E}}H$ versus $gz$ (black dotted line) applied to the EDF-N. In grey we show all \Euclid compact sources. The red coloured dots represent the X-ray selected AGN candidates from \citetalias{Q1-SP003}, while the purple dots indicate the GDR3-QSOs, which mainly lie within the selection.}
\label{fig:vishgz_xrays}
\end{figure}

\begin{figure*}[htbp!]
\centering
\includegraphics[angle=0,width=1.0\hsize]{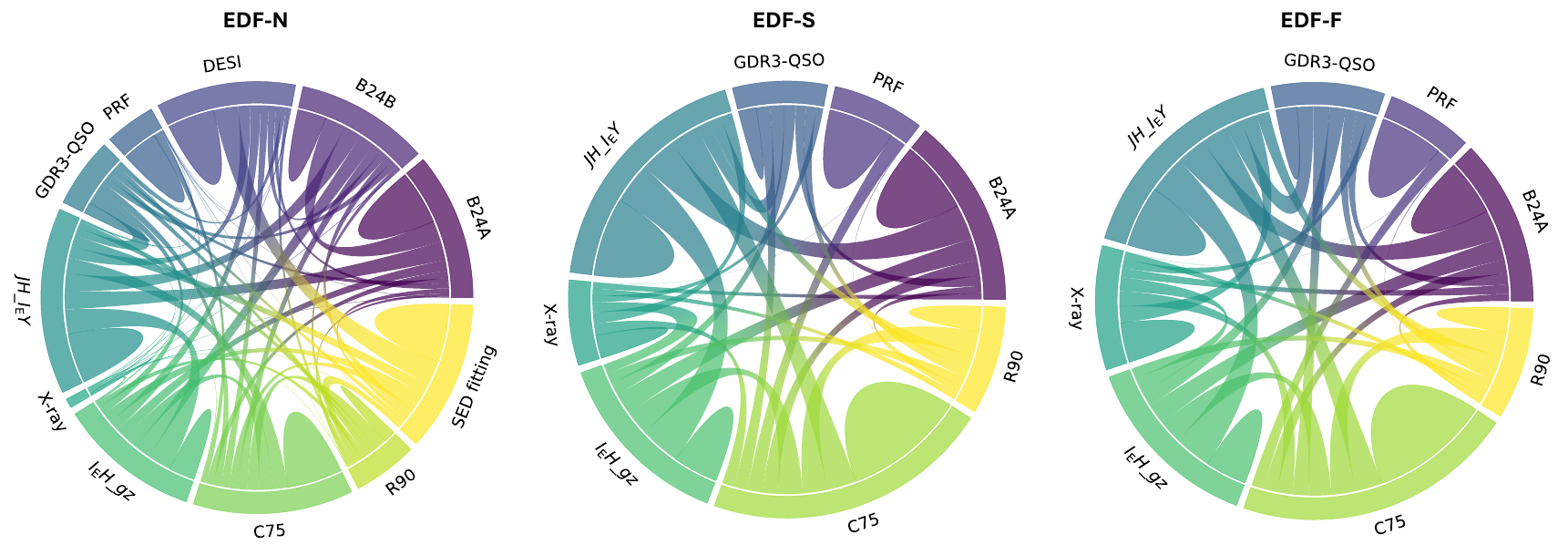}
\caption{Comparison among the number of sources selected as AGN candidates for the different criteria investigated in this work per EDF. We exclude the morphology-based AGN candidates due to the significant differences in their methodologies compared to other techniques explored in this work. We set the detection limit to the range of 18 < \IE $\leq$ 22, ensuring all selections match in depth. Note that EDF-N shows more selection criteria than EDF-S and EDF-F.}
\label{fig:chords}
\end{figure*}

To determine whether the new selections align with other QSO and AGN candidates, we examine the agreement of these methods with the GDR3-QSO and X-ray AGN candidates. From \cref{fig:jhvisy_xrays,fig:vishgz_xrays}, which illustrate the GDR3-QSO and X-ray candidates in the selections $JH$\_$I_{\text{E}}Y$ and $I_{\text{E}}H$\_$gz$ colour spaces, it is evident that both selections agree with the GDR3-QSOs, capturing most of them, while only selecting some of the X-ray sources. This outcome is expected as both of the new selections are designed to select Type I AGN, i.e., QSOs, hence the agreement with GDR3-QSOs, while X-ray samples contain a significant fraction of Type II AGN, which would occupy a different region of the colour space. The agreement with the GDR3-QSOs can also be attributed to the fact that these sources are selected using \gaia information, which has a detection limit of $G<21$, therefore somewhat agreeing with the $\IE < 21$ cut we use to calculate the $P$ and $C$ values of selections A and B.

The issue of contaminants at fainter magnitudes also affects the B24A and B24B selections. At the faintest magnitudes, these selections introduce a large number of candidates that are likely compact galaxies or other types of contaminants, such as brown dwarfs or stellar objects, which cannot be disentangled using colour cuts alone (see \cref{fig:b24a_magbins,fig:b24b_magbins}).

The DESI-selected QSO and AGN candidates exhibit higher reliability since they utilise DESI spectra from \Euclid counterparts to assess the population a source may belong to. We particularly trust those DESI BLQSO candidates, since these are indicative of Type I AGN activity, due to the high velocities of the ionised clouds within the broad-line region (BLR) of an AGN \citep{Antonucci_1993_1993ARA&A..31..473A, Veilleux_2002}. Moreover, the NLAGN candidates tend to have high reliability because BPT diagnostics use a combination of nebular emission lines to differentiate between various ionisation mechanisms in gas \citep{Baldwin1981}. This helps distinguish between AGN, LINERs, SFGs, and composite objects, which encompass starburst-AGN objects \citep{Kewley_2006}. The WHAN, KEX, and BLUE diagrams are similar in the sense that they also use specific line ratios and compare these to source qualities in order to identify different populations. However, although these may be more reliable AGN diagnostics at times, they are still incomplete and biased, and the sources identified with them should still be considered candidates.

For the specifics on the purity and completeness of the GDR3-QSOs, C75, R90, and X-ray selected AGN candidates, as well as those obtained using morphology and machine-learning information, we point the readers to the corresponding papers \citet{Storey-Fisher2024}, \citet{Fu_2024_2024ApJS..271...54F}, Fu et al., in prep, \citetalias{Assef_2018_2018ApJS..234...23A}, \citetalias{Q1-SP003}, \citetalias{Q1-SP009}, \citetalias{Q1-TP005}, \citetalias{Q1-SP015}, and LB25.

To assess the overlap among the various selection methods investigated (excluding the morphology-based ones, since their methodologies differ significantly from the other techniques explored in this work), \cref{fig:chords} visualizes the portion of AGN candidates identified by multiple selections simultaneously per EDF. To enable a better comparison, we set the detection limit to the range of $18<\IE\leq 22$, ensuring all selections match in depth.

We find that although most selections overlap to some extent in AGN candidates, there is a large number of candidates that do not co-exist in the different AGN samples. Notably, the PRF, B24A, and $JH$\_$I_{\text{E}}Y$ selections have substantial portions of their QSO candidate populations that are not selected by other diagnostics, potentially indicating that these sources are contaminants. Similarly, this behaviour is observed for some C75 candidates. However, the C75 selection is designed to identify AGN in general, which suggests that these sources may be Type II AGN, not detectable by the other QSO-specific diagnostics. Potential future work combining \Euclid's photometry with that of WISE-AllWISE could be a promising approach to reduce the bias against Type II AGN. Nevertheless, this lies outside the scope of the current work.

We observe that most of the DESI AGN, GDR3-QSO, and X-ray extragalactic candidates appear to be consistently identified by the other diagnostics.
\cref{ap:b} provides a numerical representation of \cref{fig:chords} to quantify the agreements between selections.

In total, and including the AGN candidates identified in this work, which include B24A, B24B, $JH$\_$I_{\text{E}}Y$, $I_{\text{E}}H$\_$gz$, DESI, PRF, SED-fitting and GDR3-QSOs, our current catalogue includes 229\,779  AGN candidates in 18 < \IE $\leq$ 24.5, which is equivalent to an AGN surface density of 3\,641\,deg$^{-2}$. However, due to contamination, we believe the purest sample of AGN is in the magnitude range $18 < \IE \leq 22$, resulting in a total of 30\,422 AGN candidates i.e.\ 482\,deg$^{-2}$. This sample, although primarily composed of Type I AGN, also includes some Type II AGN identified through the DESI diagnostics.

\subsubsection{Obscured vs.\ unobscured AGN}

\begin{figure}[htbp!]
\centering
\includegraphics[angle=0,width=1.0\hsize]{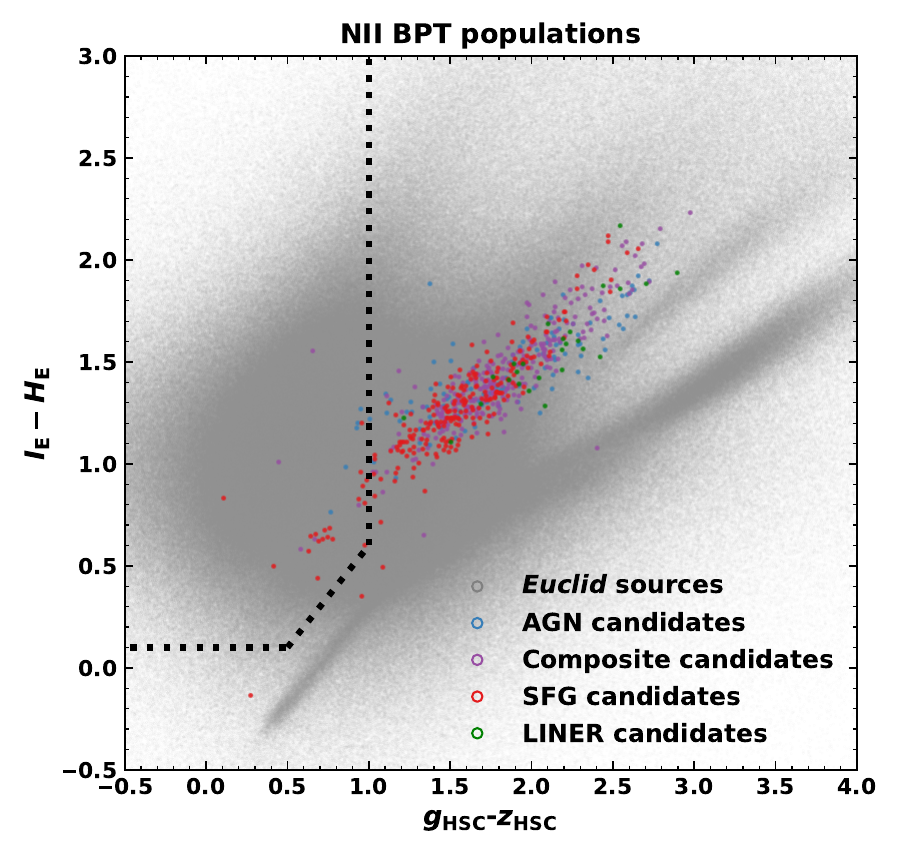}
\caption{Selection $I_{\text{E}}H$ versus $gz$ (black dotted line) applied to the EDF-N. In grey we show all \Euclid sources. The blue dots represent the \ion{N}{ii} selected AGN candidates, the composite galaxies are shown in purple, the SFGs in red and the LINERs in green.}
\label{fig:desi_bpt_vishgz}
\end{figure}

\begin{figure*}[htbp!]
\centering
\includegraphics[angle=0,height=0.55\textheight]{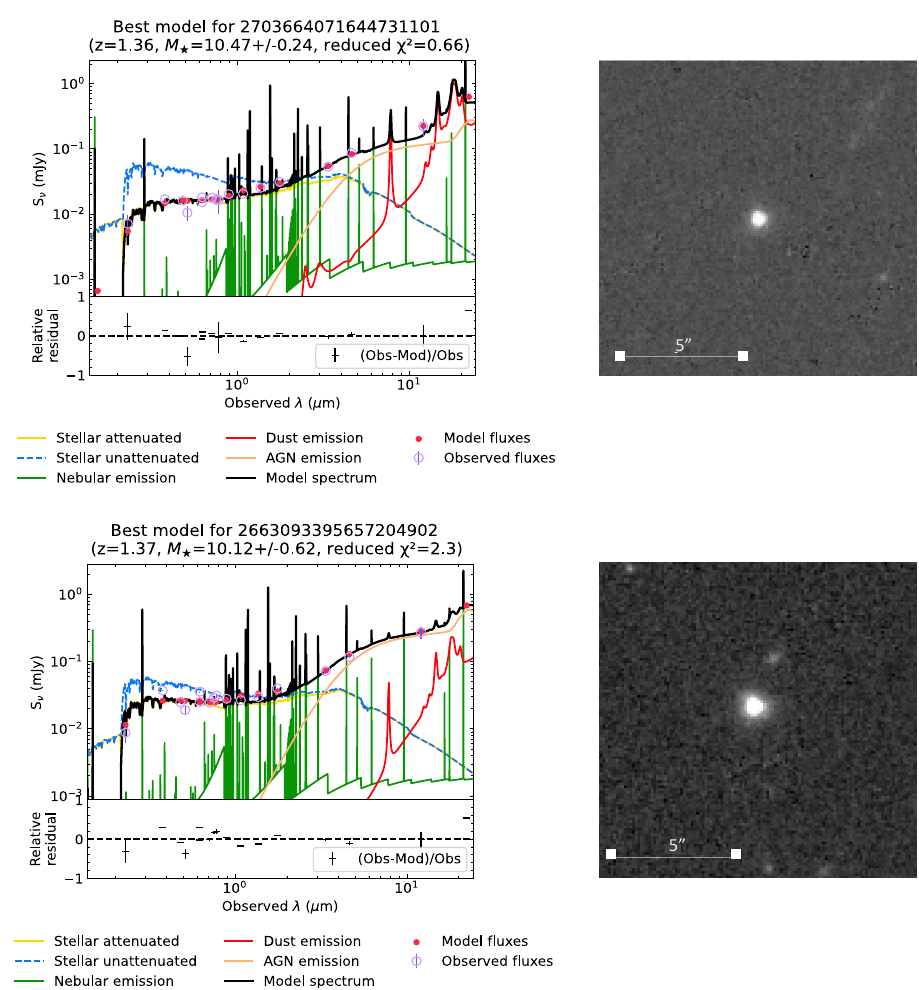}
\caption{SEDs and corresponding VIS cutouts of two QSO candidates identified by the selections $JH$\_$I_{\text{E}}Y$, $I_{\text{E}}H$\_$gz$, B24A, B24B and simultaneously classified as broad-line QSOs by DESI.}
\label{fig:seds_type1}
\end{figure*}

\begin{figure*}[htbp!]
\centering
\includegraphics[angle=0,height=0.55\textheight]{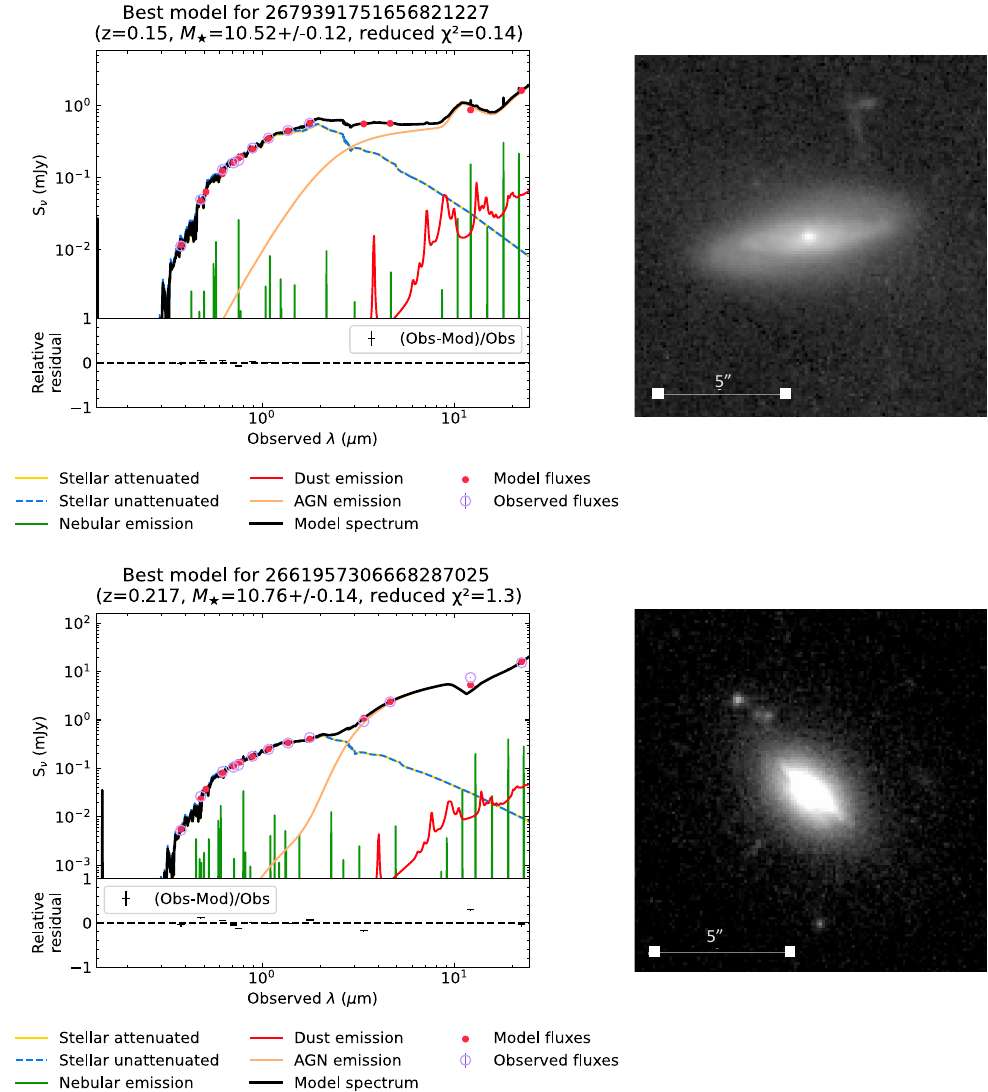}
\caption{SEDs and corresponding VIS cutouts of two narrow line AGN candidates identified via BPT diagnostics.}
\label{fig:seds_type2}
\end{figure*}

The majority of the AGN selection methods used in this work are specifically designed to select QSOs. These sources are easier to detect due to their distinct colours and point-like appearance. Additionally, as they are face-on AGN, they are minimally affected by dust, meaning their observed fluxes have not been significantly attenuated. The main drawback of optical photometric selections is that they are heavily affected by dust, making them easily optimised for Type I AGN, while being heavily biased against Type II AGN. Despite \Euclid having NIR filters, it was predicted that selecting all AGN, including optically obscured AGN and composite systems, would be challenging with \Euclid filters alone or supplemented by optical or other bands \citepalias{EP-Bisigello}.

To ascertain this, we investigate the regions that the DESI spectroscopically selected narrow line AGN populate in different colour spaces. \Cref{fig:desi_onvishgz} showcases an example of one of the DESI spectroscopic tests, the \ion{N}{ii} BPT diagnostic, plotted on the $\IE-\HE$ versus\ $g-z$ space. It is apparent that the area populated by the AGN is significantly entangled with that of composite, SFGs, and LINER galaxies, highlighting that obscured AGN do not occupy a specific and distinct region of the colour space. This behaviour was observed for all spectroscopic diagnostics across the different \Euclid colour combinations.

Moreover we note that out of the total number of DESI broad-line QSOs (1\,434) 91\% are selected by our QSO diagnostics, while from the total of NLAGN selected emission line diagnostics (2\,761) only 8\% are detected by other selections, therefore highlighting the still existing bias against Type II AGN.
Additionally, to further demonstrate the differences in populations identified by the spectroscopic and QSO selections, we use the  ancillary photometry from the multi-wavelength catalogue to explore the SEDs of the AGN selected by these methods.

We first create a subsample of sources simultaneously identified as QSO candidates by B24A, B24B, $JH$\_$I_{\text{E}}Y$, $I_{\text{E}}H$\_$gz$, and DESI BLQSO, resulting in a total of 279 candidates. For these sources, we perform SED fitting with \texttt{CIGALE} using ancillary optical-to-IR photometric information and DESI's $z_{\text{spec}}$ and examine their best-fit models and corresponding VIS cutouts. \Cref{fig:seds_type1} provides an example of the SED fitting and VIS images for two of these selected Type I QSO. 

The SEDs show a notably strong AGN component in the mid-IR, even dominating the SF emission at longer wavelengths for source \verb|2663093395657204902|. Linked to the SF, we also notice moderate dusty absorption on the stellar emission.
The corresponding AGN fractions and stellar masses obtained for these sources are approximately 0.46 and $M_{\ast}\approx 10^{10.5}$\,M$_{\odot}$ for source \verb|2703664071644731101| and 0.69 with $M_{\ast}\approx 10^{10.1}$\,M$_\odot$ for source \verb|2663093395657204902|.
The VIS cutouts reveal the point-like appearance of these candidates.

We then create a second subsample of sources identified as AGN candidates by either one of the DESI narrowline spectroscopic selections, and similarly fitted their optical-to-IR SED. \Cref{fig:seds_type2} showcases the SEDs and VIS cutouts for two of these AGN candidates. 
The corresponding AGN fractions and stellar masses obtained for these sources are approximately 0.31 and $M_{\ast} \approx^{10.5}$\,M$_\odot$ for source \verb|2679391751656821227| and 0.97 with $M_{\star} \approx 10^{10.8}$\,M$_\odot$ for source \verb|2661957306668287025|. It should be noted that the absence of a WISE-AllWISE counterpart, and therefore photometry, for source \verb|2679391751656821227| results in a large uncertainty ($\pm 0.16$) in its AGN fraction.
The VIS cutouts reveal extended sources with bright centres and, in the case of \verb|2661957306668287025|, a dust lane.

This test highlights the difference between the average unobscured sources selected with our current QSO diagnostics and the obscured sources identified with spectroscopy. This opens an exciting path for future work to exploit similar types of spectroscopic diagnostics on the \Euclid spectra to verify if more obscured AGN can be identified using \Euclid's spectroscopic capabilities, notably in extended sources, currently excluded of most QSO-selection approaches.

\subsection{Galaxy major mergers with AGN contributions}

\citet{Q1-SP013} make use of a convolutional neural network (CNN) to perform morphological classification of a stellar mass-complete sample of Q1 galaxies in the redshift range $0.5\leq z \leq 2.0$. The CNN is trained with \Euclid \IE mock observations created from the IllustrisTNG simulations, with different levels of AGN contributions injected in 20\% of the sample. The authors classify $113\,155$ galaxies as mergers and $269\,933$ as non-mergers. Then, they utilise the AGN catalogue presented here to select AGN in four different ways (X-ray detections, optical spectroscopy, through the $f_{\rm{PSF}}$ parameter, and with MIR colours) to study the possible connection of mergers with each AGN type. \citet{Q1-SP013} observe a larger fraction of AGN in mergers compared to non-mergers, with the largest AGN excess seen in MIR-selected AGN, and a dependence of the merger fraction on the $f_{\rm{PSF}}$ parameter and the AGN luminosity. Their analysis supports the scenario in which mergers are most closely connected to the most powerful and dust-obscured AGN.


\section{The AGN catalogue}\label{sc:agn_catalogue}

The catalogue created through this work and made public through Zenodo contains the counterparts to the \Euclid sources from GALEX, \gaia, WISE-AllWISE, DESI, SDSS, DES, and \textit{Spitzer}, with their corresponding IDs, RA, and Dec (columns 1--25). We also include the flags to clean the data similarly to our work (column 26), split the data into our magnitude bing (27--29), and identify the stellar candidates (30--32). Additionally, columns 33--48 flag the sources that have been selected as AGN candidates via the various tests conducted in this study, including B24A, B24B, $JH$\_$I_{\text{E}}Y$, $I_{\text{E}}H$\_$gz$, C75, R90, PRF, QDR3-QSOs, and the different DESI diagnostics. Finally, we also include the results from the SED fitting explored alongside this work, which includes columns for the AGN fraction of those sources with DESI redshifts, their corresponding errors, and the resulting selected AGN candidates (columns 49--51). A detailed description of the columns included in this catalogue can be found in \cref{app:c}.


\section{Conclusions}

In this paper, we have created and presented a multi-wavelength AGN candidate catalogue, incorporating ancillary photometric and spectroscopic data from surveys such as \gaia, WISE-AllWISE, DES, SDSS, DESI, and \textit{Spitzer}. We summarise the most important results as follows.

\begin{itemize} 
\item Counterpart associations are performed using a nearest-neighbour approach with \verb|STILTS|, deciding the best fixed error radius for each survey based on their angular resolution and PSF FWHM.

\item Two QSO diagnostics derived by \citetalias{EP-Bisigello} are applied to the Q1 data. Upon investigating and refining these selections with a morphology cut (\verb|MUMAX_MINUS_MAG|<$-$2.6), we obtain a total of 211\,797 QSO candidates using \Euclid photometry only (B24A), and 114\,145 QSO candidates using \Euclid plus ancillary bands (B24B). 

\item We apply the C75 and R90 \citetalias{Assef_2018_2018ApJS..234...23A} diagnostics to the WISE-AllWISE counterparts and obtain a total of 65\,083 and 4\,688 AGN candidates respectively.

\item Labelled sources from the DESI counterparts are used to create two new QSO diagnostics: one tailored for \Euclid-only photometry ($JH$\_$I_{\text{E}}Y$), achieving $P\approx.92$ and $C\approx.63$ for $z_{\text{spec}}<1.6$ and $P\approx.95$ and $C\approx.9$ for $z_{\text{spec}}>1.6$ ; and another using \Euclid plus ancillary data ($I_{\text{E}}H$\_$gz$), obtaining $P\approx.93$ and $C\approx.60$ for $z_{\text{spec}}<1.6$ and $P\approx.97$ with $C\approx.77$ for $z_{\text{spec}}>1.6$, both for \IE < 21.

\item The spectra of DESI counterparts are utilised to test for broad-line detection in the DESI QSOs and galaxies. Additionally, spectroscopic tests (\ion{N}{ii} BPT, \ion{S}{ii} BPT, \ion{O}{i} BPT, WHAN, BLUE, and KEX) are performed to identify a sample of obscured narrow line AGN. In total we obtain 4392 DESI AGN candidates.

\item Our catalogue is matched to a ‘purified’ version of the \gdr3 QSO catalogue (GDR3-QSO) with 1971 QSO candidates.

\item By conducting SED fitting on a subset of sources with available $z_{\text{spec}}$, we are able to determine their AGN fraction and consequently identify a total of 7766 AGN candidates.

\item Results are compared to other Q1 AGN-related works \citet{Q1-TP005}, \citet{Q1-SP003}, \citet{Q1-SP009}, \citet{Q1-SP015}, and Euclid Collaboration: Laloux et al., in prep, assessing the differences and strengths of each selection.

\item The purity and completeness of our selections are discussed, acknowledging the need for more labels in order to assess the impact of contaminants at the faintest magnitudes.

\item A total of 229\,779  AGN candidates are identified at 18 < \IE $\leq$ 24.5, with a refined sample of 30\,422    AGN candidates within the magnitude bin of 18 < \IE $\leq$ 22.

\item The AGN surface density expected from \citetalias{EP-Selwood} in the EWS, 331\,deg$^{-2}$, is compared to our catalogue, 3\,641\,deg$^{-2}$, which reaches a higher AGN surface densities, most probably due to contaminants in the faintest magnitudes. Even when limiting AGN to 18 < \IE $\leq$ 22, with 482\,deg$^{-2}$, we surpass the expected number of selected AGN, although we still fall short of the expected detected AGN. This gap could be bridged by future machine-learning studies.

\item The AGN catalogue is presented, containing a wealth of information, including the data needed to replicate the numbers obtained in this work, as well as flags to easily identify different types of selected AGN. \end{itemize}

\begin{acknowledgements}
\AckQone\\
\AckEC\\

Based on data from UNIONS, a scientific collaboration using
three Hawaii-based telescopes: CFHT, Pan-STARRS, and Subaru
\url{www.skysurvey.cc}\,.

Based on data from the Dark Energy Camera (DECam) on the Blanco 4-m Telescope
at CTIO in Chile \url{https://www.darkenergysurvey.org}\,.

This work uses results from the ESA mission {\it Gaia},
whose data are being processed by the Gaia Data Processing and
Analysis Consortium \url{https://www.cosmos.esa.int/gaia}\,.

This publication makes use of data products from the Wide-field Infrared Survey Explorer, which is a joint project of the University of California, Los Angeles, and the Jet Propulsion Laboratory/California Institute of Technology, funded by the National Aeronautics and Space Administration.

This publication is partially based on observations made with the Spitzer Space Telescope, which is operated by the Jet Propulsion Laboratory, California Institute of Technology under a contract with NASA, and has made use of the NASA/IPAC Infrared Science Archive, which is funded by the National Aeronautics and Space Administration and operated by the California Institute of Technology. 

This research is based on observations made with the Galaxy Evolution Explorer, obtained from the MAST data archive at the Space Telescope Science Institute, which is operated by the Association of Universities for Research in Astronomy, Inc., under NASA contract NAS 5–26555. We thank the MAST team for providing the GALEX catalogue in a convenient format.


DESI construction and operations is managed by the Lawrence Berkeley National Laboratory. This research is supported by the U.S. Department of Energy, Office of Science, Office of High-Energy Physics, under Contract No. DE–AC02–05CH11231, and by the National Energy Research Scientific Computing Center, a DOE Office of Science User Facility under the same contract. Additional support for DESI is provided by the U.S. National Science Foundation, Division of Astronomical Sciences under Contract No. AST-0950945 to the NSF’s National Optical-Infrared Astronomy Research Laboratory; the Science and Technology Facilities Council of the United Kingdom; the Gordon and Betty Moore Foundation; the Heising-Simons Foundation; the French Alternative Energies and Atomic Energy Commission (CEA); the National Council of Science and Technology of Mexico (CONACYT); the Ministry of Science and Innovation of Spain, and by the DESI Member Institutions. The DESI collaboration is honored to be permitted to conduct astronomical research on Iolkam Du’ag (Kitt Peak), a mountain with particular significance to the Tohono O’odham Nation
Funding for the Sloan Digital Sky 
Survey IV has been provided by the 
Alfred P. Sloan Foundation, the U.S. 
Department of Energy Office of 
Science, and the Participating 
Institutions. 

SDSS-IV acknowledges support and 
resources from the Center for High 
Performance Computing  at the 
University of Utah. The SDSS 
website is www.sdss4.org.

SDSS-IV is managed by the 
Astrophysical Research Consortium 
for the Participating Institutions 
of the SDSS Collaboration including 
the Brazilian Participation Group, 
the Carnegie Institution for Science, 
Carnegie Mellon University, Center for 
Astrophysics | Harvard \& 
Smithsonian, the Chilean Participation 
Group, the French Participation Group, 
Instituto de Astrof\'isica de 
Canarias, The Johns Hopkins 
University, Kavli Institute for the 
Physics and Mathematics of the 
Universe (IPMU) / University of 
Tokyo, the Korean Participation Group, 
Lawrence Berkeley National Laboratory, 
Leibniz Institut f\"ur Astrophysik 
Potsdam (AIP),  Max-Planck-Institut 
f\"ur Astronomie (MPIA Heidelberg), 
Max-Planck-Institut f\"ur 
Astrophysik (MPA Garching), 
Max-Planck-Institut f\"ur 
Extraterrestrische Physik (MPE), 
National Astronomical Observatories of 
China, New Mexico State University, 
New York University, University of 
Notre Dame, Observat\'ario 
Nacional / MCTI, The Ohio State 
University, Pennsylvania State 
University, Shanghai 
Astronomical Observatory, United 
Kingdom Participation Group, 
Universidad Nacional Aut\'onoma 
de M\'exico, University of Arizona, 
University of Colorado Boulder, 
University of Oxford, University of 
Portsmouth, University of Utah, 
University of Virginia, University 
of Washington, University of 
Wisconsin, Vanderbilt University, 
and Yale University.


This work has benefited from the support of Royal Society Research Grant RGS{\textbackslash}R1\textbackslash231450.
This research was supported by the International Space Science Institute (ISSI) in Bern, through ISSI International Team project \#23-573 ``Active Galactic Nuclei in Next Generation Surveys''.
F.~R., L.~B.,V.~A, B.~L.,J.~C.,F. ~LF., and A.~B. acknowledge the support from the INAF Large Grant ``AGN \& Euclid: a close entanglement'' Ob.\ Fu.\ 01.05.23.01.14. A.~F.\ acknowledges the support from project ``VLT- MOONS'' CRAM 1.05.03.07, INAF Large Grant 2022, ``The metal circle: a new sharp view of the baryon cycle up to Cosmic Dawn with the latest generation IFU facilities'' and INAF Large Grant 2022 ``Dual and binary SMBH in the multi-messenger era''

\end{acknowledgements}

\bibliography{Euclid, Q1, Euclid_official}

\begin{appendix}

\section{QSO candidates in magnitude bins}\label{app:a}

To highlight the increase in contaminants with increasing magnitudes, we plot the selected QSO candidates across three magnitude bins for the \Euclid-based photometric selections. 

\begin{figure}[htbp!]
\centering
\includegraphics[height=0.8\textheight]{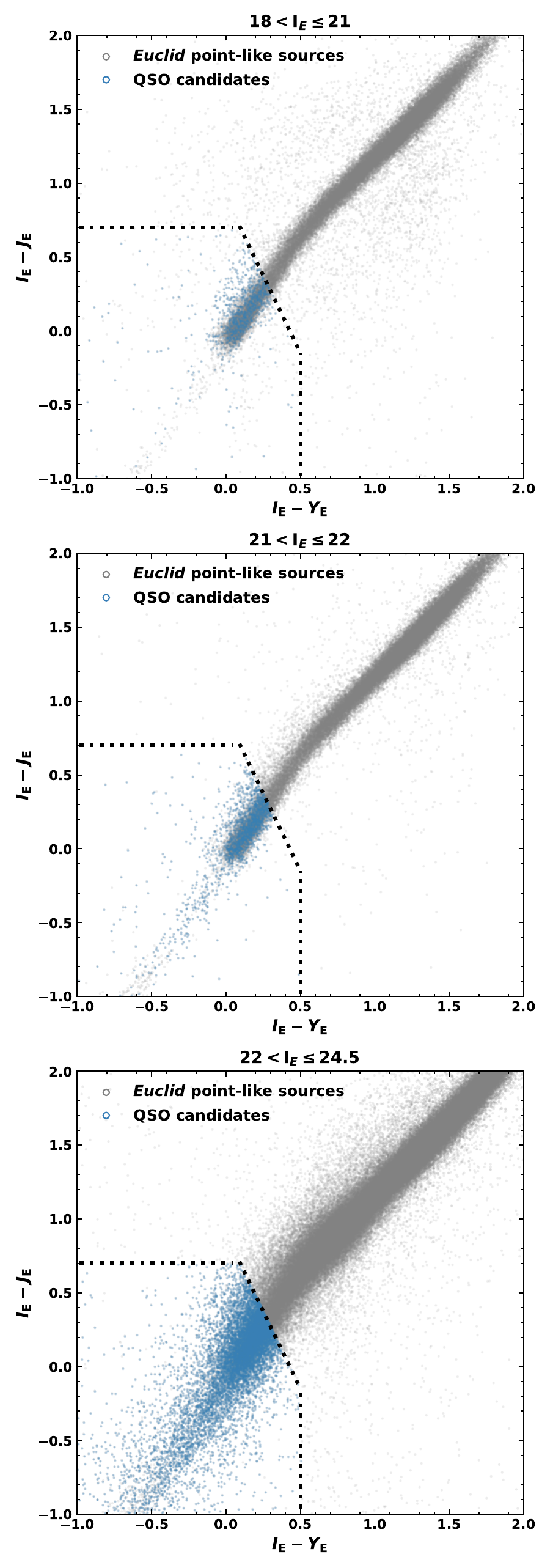}
\caption{Comparison between the number of B24A QSO (blue) candidates per magnitude bin in the EDF-N. In grey we show all \Euclid compact sources in the corresponding magnitude bin. Because the magnitude bins go from brighter colours (top plot) to faintest (bottom plot) the number of sources and QSO candidates increase.}
\label{fig:b24a_magbins}
\end{figure}

\begin{figure}[htbp!]
\centering
\includegraphics[height=0.8\textheight]{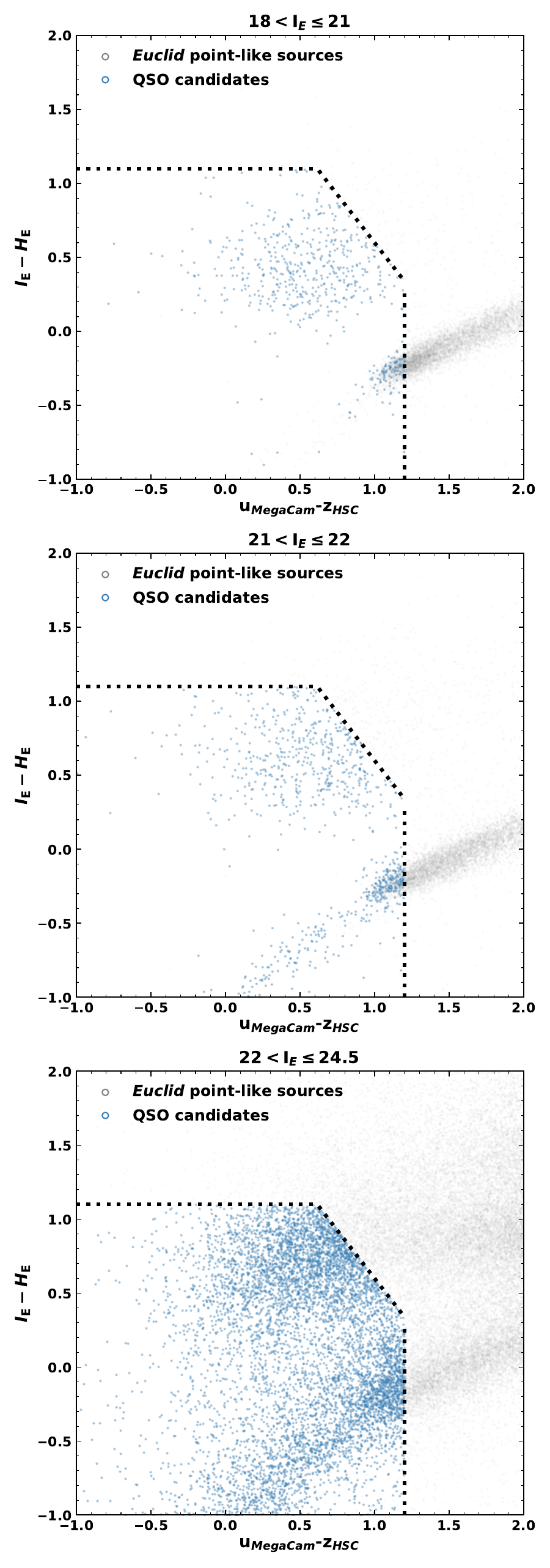}
\caption{The same as \cref{fig:b24a_magbins}, but for B24B}
\label{fig:b24b_magbins}
\end{figure}

\begin{figure}[htbp!]
\centering
\includegraphics[height=0.8\textheight]{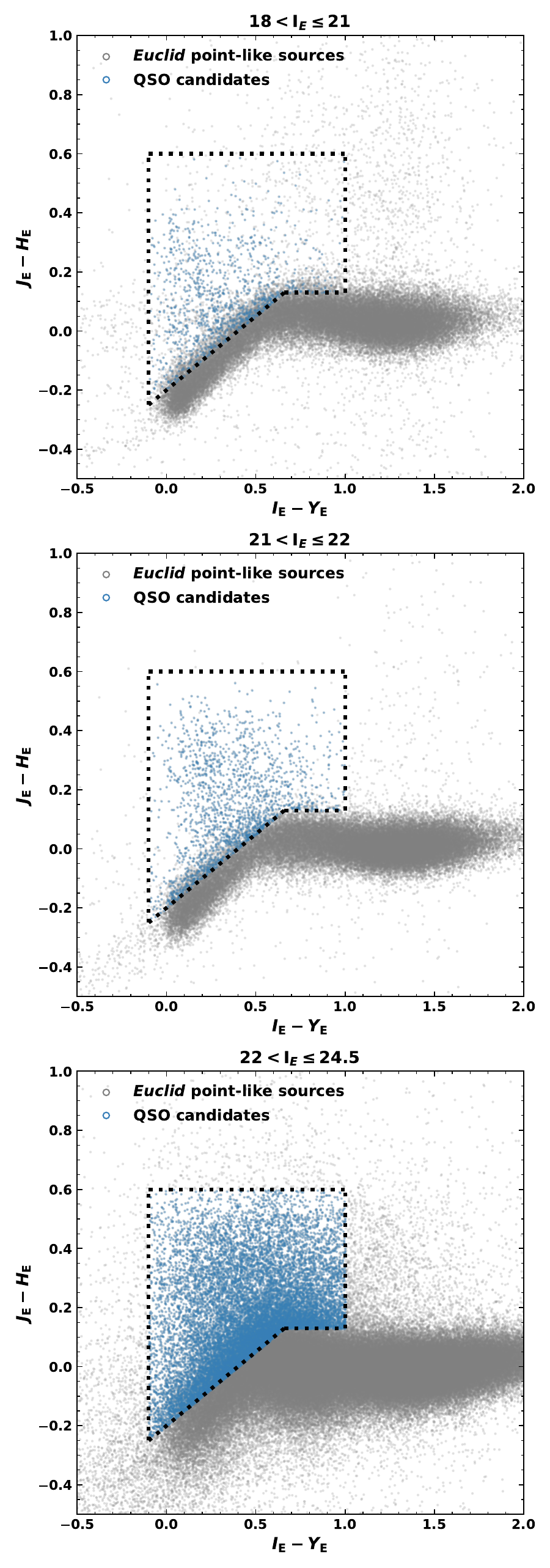}
\caption{Comparison between the number of QSO (blue) candidates per magnitude bin in the EDF-N for the new \Euclid-only colour cut, $JH$\_$I_{\text{E}}Y$. In grey we show all \Euclid compact sources in the corresponding magnitude bin. As the magnitude bins go from brighter colours (top plot) to faintest (bottom plot) the number of sources and QSO candidates increase.}
\label{fig:jhvisy_magbins}
\end{figure}

\begin{figure}[htbp!]
\centering
\includegraphics[height=0.8\textheight]{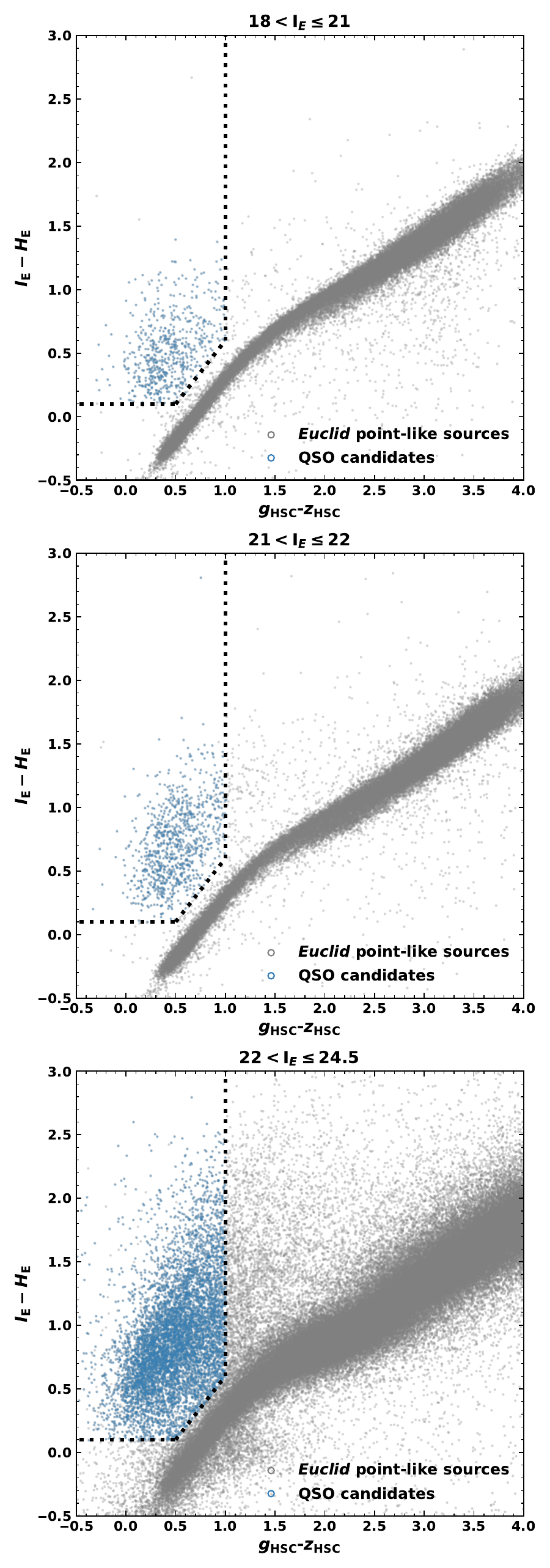}
\caption{Comparison between the number of QSO (blue) candidates per magnitude bin in the EDF-N for the new \Euclid and ancillary data colour cut, $I_{\text{E}}H$\_$gz$. In grey we show all \Euclid compact sources in the corresponding magnitude bin. As the magnitude bins go from brighter colours (top plot) to faintest (bottom plot) the number of sources and QSO candidates increase.}
\label{fig:vishgz_magbins}
\end{figure}

\Cref{fig:b24a_magbins} showcases selection B24A, \cref{fig:b24b_magbins} B24B, \cref{fig:jhvisy_magbins} $JH$\_$I_{\text{E}}Y$, and \cref{fig:vishgz_magbins} $I_{\text{E}}H$\_$gz$. It is evident that each selection is impacted by the higher number of candidates at fainter magnitudes. This highlights the necessity to conduct further work at \IE>21, either by refining our selections or devising methods to identify potential contaminants in this region.

\newpage
\section{Comparison between AGN selections}\label{ap:b}

\begin{table*}[htp!]
\caption{EDF-N intersection matrix between the different AGN selection methods investigated in this work, excluding the morphology based ones, and limited to $18<\IE\leq 22$.}
\centering
\begin{tabular}{crrrrrrrrrrr}
\hline\hline
 & B24A & B24B & DESI & PRF & GDR3-QSO & $JH$\_$I_{\text{E}}Y$ & X-ray & $I_{\text{E}}H$\_$gz$ & C75   &   R90 & SED fitting\\
\hline
B24A & 3123 & \dots & \dots & \dots & \dots & \dots & \dots & \dots & \dots & \dots & \dots \\
B24B & 1209 & 1655 & \dots & \dots & \dots & \dots & \dots & \dots & \dots & \dots & \dots \\
DESI & 299 & 433 & 2886 & \dots & \dots & \dots & \dots & \dots & \dots & \dots & \dots \\
PRF & 26 & 1 & 43 & 2020 & \dots & \dots & \dots & \dots & \dots & \dots & \dots \\
GDR3-QSO & 298 & 381 & 281 & 15 & 641 & \dots & \dots & \dots & \dots & \dots & \dots \\
$JH$\_$I_{\text{E}}Y$ & 1091 & 1000 & 714 & 97 & 490 & 2873 & \dots & \dots & \dots & \dots & \dots \\
X-ray & 28 & 41 & 43 & 3 & 24 & 59 & 134 & \dots & \dots & \dots & \dots \\
$I_{\text{E}}H$\_$gz$ & 622 & 927 & 635 & 73 & 457 & 1400 & 54 & 1500 & \dots & \dots & \dots \\
C75   & 400 & 563 & 518 & 295 & 437 & 870 & 46 & 769 & 3187 & \dots & \dots \\
  R90 & 185 & 238 & 256 & 17 & 318 & 369 & 23 & 311 & 685 & 694 & \dots \\
SED fitting & 294 & 409 & 1\,505 & 30 & 272 & 635 & 40 & 572 & 488 & 251 & 3390 \\
\hline
\end{tabular}
\label{tab:intersection_matrix_N}
\end{table*}

\begin{table*}[htp!]
\caption{EDF-S intersection matrix between the different AGN selection methods investigated in this work, excluding the morphology based ones, and limited to $18<\IE\leq 22$.}
\centering
\begin{tabular}{crrrrrrrr}
\hline\hline
 & B24A & PRF & GDR3-QSO & $JH$\_$I_{\text{E}}Y$ & X-ray & $I_{\text{E}}H$\_$gz$ & C75 & R90 \\
\hline
B24A & 3323 & \dots & \dots & \dots & \dots & \dots & \dots & \dots \\
PRF & 0 & 2999 & \dots & \dots & \dots & \dots & \dots & \dots \\
GDR3-QSO & 465 & 0 & 811 & \dots & \dots & \dots & \dots & \dots \\
$JH$\_$I_{\text{E}}Y$ & 1607 & 0 & 616 & 3593 & \dots & \dots & \dots & \dots \\
X-ray & 292 & 5 & 315 & 520 & 1050 & \dots & \dots & \dots \\
$I_{\text{E}}H$\_$gz$ & 992 & 0 & 557 & 1857 & 452 & 2034 & \dots & \dots \\
C75 & 708 & 700 & 623 & 1326 & 524 & 1110 & 5528 & \dots \\
R90 & 322 & 25 & 457 & 566 & 300 & 456 & 1123 & 1139 \\
\hline
\end{tabular}
\label{tab:intersection_matrix_S}
\end{table*}

\begin{table*}[htp!]
\caption{EDF-F intersection matrix between the different AGN selection methods investigated in this work, excluding the morphology based ones, and limited to $18<\IE\leq 22$.}
\centering
\begin{tabular}{crrrrrrrr}
\hline\hline
 & B24A & PRF & GDR3-QSO & $JH$\_$I_{\text{E}}Y$ & X-ray & $I_{\text{E}}H$\_$gz$ & C75 & R90 \\
\hline
B24A & 1366 & \dots & \dots & \dots & \dots & \dots & \dots & \dots \\
PRF & 0 & 1535 & \dots & \dots & \dots & \dots & \dots & \dots \\
GDR3-QSO & 273 & 0 & 510 & \dots & \dots & \dots & \dots & \dots \\
$JH$\_$I_{\text{E}}Y$ & 720 & 0 & 401 & 1489 & \dots & \dots & \dots & \dots \\
X-ray & 225 & 4 & 213 & 399 & 967 & \dots & \dots & \dots \\
$I_{\text{E}}H$\_$gz$ & 519 & 0 & 381 & 911 & 341 & 1011 & \dots & \dots \\
C75 & 327 & 307 & 396 & 642 & 366 & 549 & 2917 & \dots \\
R90 & 175 & 19 & 282 & 311 & 180 & 262 & 596 & 603 \\
\hline
\end{tabular}
\label{tab:intersection_matrix_F}
\end{table*}

We present the intersection table for the different selection methods investigated in this work, excluding the morphology-based methods due to their distinct methodologies (see \cref{fig:chords}). We limit the depth of every selection to $\IE<22$ to ensure comparability without bias towards the \Euclid-based selections, which are the only ones capable of reaching the faintest magnitudes.

\section{Column description of AGN catalogue}\label{app:c}

We list column descriptions for the three EDFs catalogues below. 

\begin{enumerate}
    \item \verb|object_id_euclid|: \Euclid unique source identifier.
    \item \verb|right_ascension_euclid|: \Euclid source barycent right ascension coordinate in decimal degrees.
    \item \verb|declination_euclid|: \Euclid source barycenter declination coordinate in decimal degrees.
    \item \verb|source_id_gaia|: \gaia unique source identifier.
    \item \verb|right_ascension_gaia|: \gaia barycentric right ascension in ICRS at the reference epoch.
    \item \verb|declination_gaia|: \gaia barycentric declination in ICRS at the reference epoch.
    \item \verb|id_galex|: GALEX merged ID. Extraction identification number.
    \item \verb|alpha_j2000_galex|: GALEX right ascension in degrees (0 to 360) in a J2000 reference frame. 
    \item \verb|delta_j200_galex|: GALEX declination in degrees (-90 to 90) in a J2000 reference frame.
    \item \verb|source_id_allwise|: WISE-AllWISE unique source ID.
    \item \verb|right_ascension_allwise|: WISE-AllWISE J2000 right ascension with respect to the 2MASS PSC reference frame from the non-moving source extraction.
    \item \verb|declination_allwise|: WISE-AllWISE J2000 declination with respect to the 2MASS PSC reference frame from the non-moving source extraction.
    \item \verb|object_id_spitzer|: \textit{Spitzer} unique source identifier.
    \item \verb|brick_desi|: DESI brick ID from tractor input.
    \item \verb|brickname_desi|: DESI brick name from \texttt{tractor} input.
    \item \verb|brick_objid_desi|: DESI imaging surveys OBJID on that brick.
    \item \verb|right_ascension_desi|: DESI barycentric right ascension in ICRS.
    \item \verb|declination_desi|: DESI barycentric declination in ICRS.
    \item \verb|z_desi|: DESI redshift measured by Redrock.
    \item \verb|specobjid_sdss|: SDSS object identification number.
    \item \verb|right_ascension_sdss|: SDSS right ascension in decimal degrees.
    \item \verb|declination_sdss|: SDSS declination in decimal degrees.
    \item \verb|coadd_object_id_des|: DES unique identifier for the coadded objects.
    \item \verb|right_ascension_des|: DES right ascension, with quantised precision for indexing.
    \item \verb|declination_des|:  DES declination, with quantised precision for indexing.
    \item \verb|good_flags|: cleaning implemented in the work to keep only those sources with `good flags'.
    \item \verb|bright_vis_mag_bin|: bright \IE magnitude bin: $18<$\IE $\leq 21$.
    \item \verb|medium_vis_mag_bin|: medium \IE magnitude bin: $21<$\IE $\leq 22$.
    \item \verb|faint_vis_mag_bin|: faint \IE magnitude bin: $22<$\IE $\leq 24.5$.
    \item \verb|star_candidate_gaia|: identifier for stars based on \gaia's proper motion and parallax.
    \item \verb|star_candidate_prf|: identifier for stars based on PRF probability > 0.7.
    \item \verb|star_candidate_all|: identifier for stars combining \gaia's proper motion and parallax, and PRF probability. 
    \item \verb|prf_qso_candidate|: identifier for QSOs based on PRF probability>$0.85$ in the EDF-N and >$0.95$ in the EDF-S and EDF-F.
    \item \verb|B24a_qso_candidate|: identifier for QSO candidates based on B24A.
    \item \verb|B24b_qso_candidate|: identifier for QSO candidates based on B24B.
    \item \verb|C75_agn_candidate|:  identifier for AGN candidates based on C75.
    \item \verb|R90_agn_candidate|:  identifier for AGN candidates based on R90.
    \item \verb|GDR3_qso_candidate|: identifier for QSO candidates based on GDR3-QSO.
    \item \verb|JH_IeY_qso_candidate|: identifier for QSO candidates based on $JH$\_$I_{\text{E}}Y$.
    \item \verb|IeH_gz_qso_candidate|:  identifier for QSO candidates based on $I_{\text{E}}H\_gz$.
    \item \verb|DESI_broad_qso_candidate|: identifier for QSO candidates based on DESI \verb|SPECTYPE| and presence of broad emission lines.
    \item \verb|DESI_broad_galaxy_candidate|: identifier for AGN candidates based on DESI \verb|SPECTYPE| and presence of broad emission lines.
    \item \verb|DESI_niibpt_agn_candidate|: identifier for AGN candidates based on DESI spectra and \ion{N}{ii} BPT diagnostic.
    \item \verb|DESI_siibpt_agn_candidate|: identifier for AGN candidates based on DESI spectra and \ion{S}{ii} BPT diagnostic.
    \item \verb|DESI_oibpt_agn_candidate|: identifier for AGN candidates based on DESI spectra and \ion{O}{i} BPT diagnostic.
    \item \verb|DESI_whan_agn_candidate|: identifier for AGN candidates based on DESI spectra and WHAN diagnostic.
    \item \verb|DESI_blue_agn_candidate|: identifier for AGN candidates based on DESI spectra and Blue diagnostic.
    \item \verb|DESI_kex_agn_candidate|: identifier for AGN candidates based on DESI spectra and KEX diagnostic.
    \item \verb|AGN_fraction|: AGN fraction derived from SED fitting.
    \item \verb|AGN_fraction_err|: AGN fraction error derived from SED fitting.
    \item \verb|AGN_sed_candidate|: AGN candidate based on AGN fraction threshold.
    
\end{enumerate}

\end{appendix}

\end{document}